\documentclass[final,1p,times]{elsarticle}
\usepackage{graphicx}
\usepackage{amssymb}
\usepackage{amsthm}

\usepackage{dcolumn}% Align table columns on decimal point
\usepackage{bm}% bold math
\usepackage{color}% bold math
\usepackage{amsmath}%,amsthm,amssymb}
\usepackage{mathrsfs}
\usepackage{comment}

\newcommand{\Rb}{^{\text{87}}\text{Rb}}

\newcommand{\Hamil}[1]{\hat{\mathcal{H}}_{#1}}

\newcommand{\annih}[1]{\hat{\psi}_{#1}}
\newcommand{\creat}[1]{\hat{\psi}^\dagger_{#1}}
\newcommand{\ann}[1]{\hat{\delta}_{#1}}
\newcommand{\cre}[1]{\hat{\delta}^\dagger_{#1}}

\newcommand{\eps}[1]{\epsilon^0_{#1}}

\newcommand{\ahat}{\hat{a}}
\newcommand{\adagg}{\hat{a}^\dagger}

\newcommand{\R}{\mathbf{r}}
\newcommand{\K}{\mathbf{k}}

\newcommand{\Q}{\mathbf{q}}
\newcommand{\p}{\mathbf{p}}

\newcommand{\integral}{\int\text{d}\mathbf{r}}

\newcommand{\integralq}{\int\frac{\text{d}^4 q}{(2\pi)^4}\,}
\newcommand{\integralQ}{\int\frac{\text{d}^3 \mathbf{q}}{(2\pi)^3}\,}

\journal{Annals of Physics}

\begin{document}

\begin{frontmatter}



\title{Beliaev theory of spinor Bose-Einstein condensates}


\author[DOP]{Nguyen Thanh Phuc\corref{cor1}}
\ead{thanhphuc\_85@cat.phys.s.u-tokyo.ac.jp}

\author[DOP]{Yuki Kawaguchi}

\author[DOP,ERATO]{Masahito Ueda}

\address[DOP]{Department of Physics, University of Tokyo, 7-3-1 Hongo, Bunkyo-ku, Tokyo 113-0033, Japan}
\address[ERATO]{ERATO Macroscopic Quantum Control Project, 7-3-1 Hongo, Bunkyo-ku, Tokyo 113-0033, Japan}

\cortext[cor1]{Corresponding author}

\begin{abstract}
By generalizing the Green's function approach proposed by Beliaev~\cite{Beliaev1,Beliaev2}, we investigate the effect of quantum depletion on the energy spectra of elementary excitations in an $F=1$ spinor Bose-Einstein condensate, in particular, of $\Rb$ atoms in an external magnetic field. We find that quantum depletion increases the effective mass of magnons in the spin-wave excitations with quadratic dispersion relations. The enhancement factor turns out to be the same for both ferromagnetic and polar phases, and also independent of the magnitude of the external magnetic field. The lifetime of these magnons in a $\Rb$ spinor BEC is shown to be much longer than that of phonons. We propose an experimental setup to measure the effective mass of these magnons in a spinor Bose gas by exploiting the effect of a nonlinear dispersion relation on the spatial expansion of a wave packet of transverse magnetization. This type of measurement has practical applications, for example, in precision magnetometry.
\end{abstract}


\end{frontmatter}

% \linenumbers

\section{Introduction}
\label{section: Introduction}
Since the experimental realization of Bose-Einstein condensates (BECs)~\cite{Anderson95, Davis95, Bradley95}, the Bogoliubov theory of weakly interacting dilute Bose gases has been successfully applied to describe a variety of phenomena in these systems~\cite{Pethickbook, Pitaevskiibook, Leggett01}. The Bogoliubov theory was originally invented to describe bosonic systems at absolute zero~\cite{Bogoliubov47}, and then extended to finite temperature~\cite{Popovbook, Griffin96, Andersen04, Proukakis08}. It gives the leading-order values of physical observables of a system in thermodynamic equilibrium. The second-order correction to the Bogoliubov result is usually relatively small for weakly interacting dilute Bose gases. At absolute zero, this correction is a consequence of a small fraction of quantum depleted noncondensed atoms~\cite{Lee57,Braaten97}. The second-order correction to the Bogoliubov energy spectrum was given by Beliaev~\cite{Beliaev1}, who developed a diagrammatic Green's function approach to describe the energy spectrum of elementary excitations at absolute zero~\cite{Beliaev2}. Afterwards, finite-temperature theories based on the Beliaev technique were developed for weakly interacting Bose gases~\cite{Popovbook, Popov65, Shi98, Capogrosso10}. With the rapid development of techniques for precise measurements of physical observables, the small effect of quantum depletion is no longer beyond the scope of experimenters. Furthermore, by using a Feshbach resonance or optical lattices, the effective interatomic interaction can be manipulated to cover both weakly and strongly interacting systems~\cite{Xu06, Papp08, Pollack09, Navon11}. In particular, it has been shown that up to a moderate strength of interaction, by taking the second-order correction to the mean-field (Bogoliubov) calculation, the obtained results for spinless condensates agree excellently with both the results of experiment and those of quantum Monte-Carlo simulation~\cite{Navon11}. Therefore, the second-order correction to the Bogoliubov result, which can be obtained in an analytic form, can be used as an important check for any calculation or measurement of a strongly correlated system. Furthermore, the Beliaev theory also predicts the so-called Beliaev damping which quantitatively shows a finite lifetime of Bogoliubov quasiparticles (phonons) due to their collisions with condensed particles. The Beliaev damping of quasi-particles under various conditions has been a subject of active study~\cite{Halley78, Matthias00, Hodby01, Katz02, Mizushima03}.

Recently, Bose-Einstein condensates with spin degrees of freedom (spinor BECs) have been extensively studied (see, for example,~\cite{UedaarXiv}). These atomic systems simultaneously exhibit superfluidity and magnetism, and the combination of atoms' motional and spin degrees of freedom gives rise to various interesting phenomena in the study of thermodynamic properties and quantum dynamics. Due to the competition between spin-dependent interatomic interactions and the coupling of atoms to an external magnetic field, the system can exist in various quantum phases with different spinor order parameters~\cite{Ho98, Ohmi98, Murata07}. In contrast to spinless BECs, there exist spin-wave excitations in spinor BECs in addition to the conventional density-wave excitations. These are excitations of atoms from the condensate to the other magnetic sublevels, and the corresponding magnons have quadratic dispersion relations at low momenta as opposed to the linear dispersion relations of phonons. Furthermore, in spinor Bose gases the collisions of atoms in different spin channels give rise to spin-conserving and spin-exchange interactions. Particularly, in some atomic species such as $\Rb$, the ratio of the spin-conserving to spin-exchange interactions is so large that it can compensate for the small noncondensate fraction. That is, the mean field caused by noncondensed atoms with spin-conserving interaction can have the same order of magnitude as that caused by condensed atoms with spin-exchange interaction. Consequently, a small number of noncondensed atoms can, in principle, give an appreciable effect on the physical properties of the system, for example, by shifting the phase boundary between different quantum phases~\cite{Phuc11}.

In this study, we apply the Beliaev theory to spin-1 Bose gases to investigate the effect of quantum depletion at absolute zero on the energy spectra of elementary excitations. In the presence of an external magnetic field, the ground state can be in several quantum phases, depending on the strength of the quadratic Zeeman energy relative to the spin-exchange interatomic interaction. In contrast to the work in~\cite{Phuc11}, we do not consider phase transitions between different quantum phases. Instead, we assume that the magnitude of the external magnetic field is chosen so that the system is stable in a certain quantum phase. Here, we consider two characteristic phases of $F=1$ spinor Bose gases: the fully spin-polarized ferromagnetic phase and the unmagnetized polar phase. In the calculation of second-order corrections for ultracold atomic systems like $\Rb$, the spin-conserving interaction must be taken into account while the spin-exchange interaction is neglected because of its much smaller value. (The spin-exchange interaction is, of course, taken into account in the calculation of first-order values.) We find that for both the ferromagnetic and polar phases, the quantum depletion leads to an increase in the effective mass of magnons, while it does not alter the energy gap to the leading order. Although the effective mass is different between the ferromagnetic and polar phases, it is enhanced by the same factor for these quantum phases. This factor is also independent of the magnitude of the external magnetic field. This implies a physical mechanism whereby the quantum depletion affects the motion of quasiparticles in spinor Bose gases in a universal manner under some certain conditions. In the case of $\Rb$, where the spin-conserving interaction is much larger than the spin-exchange one, the lifetime of magnons becomes much longer than that of phonons. We show that this agrees with the mechanism of Beliaev damping which is caused by collisions between quasiparticles and the condensate. To measure the effective mass of magnons in spinor Bose gases, we propose an experimental scheme which exploits the effect of a nonlinear dispersion relation on the spatial expansion of a spinor wave packet during its time evolution. This type of measurement can be used for several applications: to probe the effect of quantum depletion, to identify spinor quantum phases, or to be used for precision magnetometry in a way different from the method given in \cite{Vengalattore07}. 

This paper is organized as follows: Section \ref{section: Green's function formalism for a spinor Bose-Einstein condensate} formulates the diagrammatic Green's function approach for spin-1 spinor BECs, which is the generalization of Beliaev theory to systems with spin degrees of freedom. The explicit forms of the matrices of self-energies for both the ferromagnetic and polar phases are given. The T-matrix which plays the role of an effective interaction potential in dilute Bose gases is also introduced in this section. Section~\ref{section: First-order approximation} summarizes the results of energy spectra of elementary excitations at the first order in the interaction. It is the rederivation of the Bogoliubov energy spectra by using the Green's function approach~\cite{Uchino10}. Section~\ref{section: Second-order approximation} deals with the self-energies to the second order in the interaction and gives the leading-order corrections to the Bogoliubov energy spectra due to the effect of quantum depletion. Section~\ref{section: Spin-density waves} shows that the elementary excitations with quadratic dispersion relations are spin waves. An experimental scheme using spinor wave packets is proposed to measure the effective mass of magnons. An order-of-magnitude estimation of the time evolution of these wave packets is also given in this section. Section~\ref{section: Conclusion} concludes the paper by discussing the application of the measurement to some practical purposes. The detailed calculations are given in the Appendices to avoid digressing from the main subject.
    
%############################
\section{Green's function formalism for a spinor Bose-Einstein condensate}
\label{section: Green's function formalism for a spinor Bose-Einstein condensate}
%&&&&&&&&&&&&&&&&&&&&&&&&&&&&
\subsection{Hamiltonian}
\label{subsection: Hamiltonian}
We consider a homogeneous system of identical bosons with mass $M$ in the $F=1$ hyperfine spin manifold that is subject to a magnetic field in the $z$-direction. The single-particle part of the Hamiltonian is given in the form of a matrix by
\begin{align}
(h_0)_{jj'}= \left[-\frac{\hbar^2\nabla^2}{2M}+q_Bj^2\right]\delta_{jj'},
\label{eq:h0}
\end{align}
where the subscripts $j,j'=0,\pm 1$ refer to the magnetic sublevels, and $q_B$ is the coefficient of the quadratic Zeeman energy. Because of the conservation of the system's total longitudinal magnetization, the linear Zeeman term vanishes. The total Hamiltonian of the $F=1$ spinor Bose gas is then given in the second-quantized form by
\begin{align}
\Hamil{}=&\integral\sum_{jj'}\creat{j}(\R)(h_0)_{jj'} \annih{j'}(\R)+\hat{\mathcal{V}},
\label{eq: Spin-1 spinor BEC's Hamiltonian}
\end{align}
where $\annih{j}(\R)$ is the field operator that annihilates an atom in magnetic sub-level $j$ at position $\R$, and the interaction energy $\hat{\mathcal{V}}$ is given by
\begin{align}
\hat{\mathcal{V}}=\frac{1}{2}\integral\integral' \sum_{j,j',m,m'}\creat{j}(\R)\creat{m}(\R')V_{jm,j'm'}(\R-\R')\annih{m'}(\R')\annih{j'}(\R).
\label{eq:interaction energy}
\end{align}
Here, the matrix element $V_{jm,j'm'}(\R-\R')$ can be written as a sum of interactions in two spin channels $\mathcal{F}=0$ and 2 ($\mathcal{F}$ denotes the total spin of two colliding atoms) as follows:
\begin{align}
V_{jm,j'm'}(\R-\R')=&\,\langle j,m|\mathcal{F}=0\rangle\langle \mathcal{F}=0|j',m'\rangle V_0(\R-\R')\nonumber\\
&+\langle j,m|\mathcal{F}=2\rangle\langle \mathcal{F}=2|j',m'\rangle V_2(\R-\R'),
\end{align}
where quantum statistics prohibits bosons from interacting via the spin channel $\mathcal{F}=1$.

In the presence of a condensate, the field operator $\annih{j}(\R)$ is decomposed into the condensate part, which can be replaced by a classical field $\sqrt{n_0}\xi_j$, and the noncondensate part $\ann{j}(\R)$:
\begin{align}
\annih{j}(\R)=\sqrt{n_0}\xi_j+\ann{j}(\R).
\label{eq:field operator decomposition}
\end{align}
For a homogeneous system, the condensate is characterized by the condensate number density $n_0$ and the spinor order parameter $\xi_j (j=0,\pm1)$, which is normalized to unity:
\begin{align}
\sum_j |\xi_j|^2=1.
\end{align}

\begin{figure}[tbp] % float placement: (h)ere, page (t)op, page (b)ottom, other (p)age
  \centering
  % file name: E:/Spinor Beliaev (June 27, 2011)/Figures/interaction.eps
  \includegraphics[width=5in,keepaspectratio]{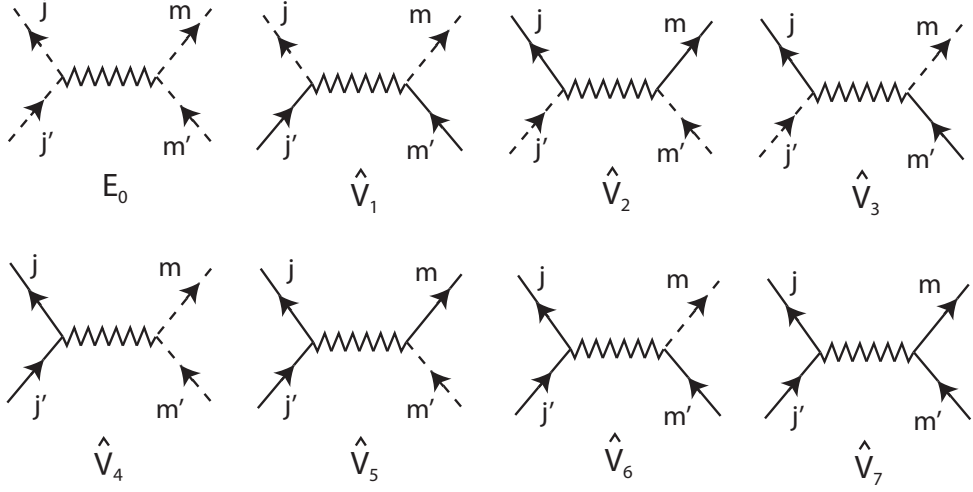}
  \caption{Two-particle interactions involving different numbers of condensed and noncondensed atoms. The dashed, solid, and wavy lines represent a condensed atom, a noncondensed atom, and the interaction, respectively.}
  \label{fig:interaction}
\end{figure}

Substituting Eq.~\eqref{eq:field operator decomposition} into Eq.~\eqref{eq:interaction energy}, we can decompose the interaction energy as
\begin{align}
\hat{\mathcal{V}}=E_0+\sum_{n=1}^{7}\hat{V}_n,
\label{eq:interaction energy 2}
\end{align}
where
\begin{subequations}
\label{eq: seven-types of interactions}
\begin{align}
E_0=&\frac{1}{2}n_0^2\integral\integral' \xi^*_j\xi^*_m V_{jm,j'm'}(\R-\R') \xi_{m'}\xi_{j'}, \label{eq:E0}\\
\hat{V}_1=&\frac{1}{2}n_0\integral\integral' \xi^*_j\xi^*_m V_{jm,j'm'}(\R-\R')\ann{m'}(\R')\ann{j'}(\R),\\
\hat{V}_2=&\frac{1}{2}n_0\integral\integral' \cre{j}(\R)\cre{m}(\R')V_{jm,j'm'}(\R-\R')\xi_{m'}\xi_{j'},\\
\hat{V}_3=&2(\frac{1}{2}n_0)\integral\integral' \xi^*_j\cre{m}(\R')V_{jm,j'm'}(\R-\R')\xi_{m'}\ann{j'}(\R),\\
\hat{V}_4=&2(\frac{1}{2}n_0)\integral\integral' \cre{j}(\R)\xi^*_m V_{jm,j'm'}(\R-\R')\xi_{m'}\ann{j'}(\R),\\
\hat{V}_5=&2(\frac{1}{2}n_0^{1/2})\integral\integral'\cre{j}(\R)\cre{m}(\R')V_{jm,j'm'}(\R-\R')\xi_{m'}\ann{j'}(\R),\\
\hat{V}_6=&2(\frac{1}{2}n_0^{1/2})\integral\integral'\cre{j}(\R)\xi^*_m V_{jm,j'm'}(\R-\R')\ann{m'}(\R')\ann{j'}(\R),\\
\hat{V}_7=&\frac{1}{2}\integral\integral'\cre{j}(\R)\cre{m}(\R')V_{jm,j'm'}(\R-\R')\ann{m'}(\R')\ann{j'}(\R).
\end{align}
\end{subequations}
These interactions are illustrated by the Feynman diagrams in Fig. \ref{fig:interaction}.

We consider a grand canonical ensemble of the above atomic system, and introduce the operator 
\begin{align}
\hat{\mathcal{K}}\equiv \hat{\mathcal{H}}-\mu\hat{\mathcal{N}},
\label{eq:operator K definition}
\end{align}
where $\mu$ denotes the chemical potential and $\hat{\mathcal{N}}$ is the total number operator:
\begin{align}
\hat{\mathcal{N}}=\integral \sum_j \creat{j}(\R)\annih{j}(\R).
\end{align}
Using Eqs.~\eqref{eq: Spin-1 spinor BEC's Hamiltonian},\eqref{eq:field operator decomposition},\eqref{eq:interaction energy 2}, and \eqref{eq:operator K definition}, we have
\begin{align}
\hat{\mathcal{K}}=E_0+\left(\sum_j q_Bj^2|\xi_j|^2-\mu\right) N_0+\hat{\mathcal{K}}',
\end{align}
where $E_0$, given in Eq.~\eqref{eq:E0}, is the interaction energy between condensed atoms, $N_0=Vn_0$ is the total number of condensed atoms with $V$ being the volume of the system, and
\begin{align}
\hat{\mathcal{K}}'\equiv&\, \hat{\mathcal{K}}_0+\hat{\mathcal{K}}_1
\label{eq: operator K'}
\end{align}
is the corresponding operator for the noncondensate part with
\begin{align}
\hat{\mathcal{K}}_0\equiv& \sum_{\K\not=0,j}(\epsilon_\K^0-\mu+q_Bj^2)\adagg_{j,\K}\ahat_{j,\K},\label{eq: operator K0}\\
\hat{\mathcal{K}}_1\equiv& \sum_{n=1}^{7}\hat{V}_n. \label{eq: operator K1}
\end{align}
Here, $\epsilon_\K^0=\hbar^2\K^2/(2M)$ is the kinetic energy of a particle with momentum $\hbar\K$, and $\ahat_{j,\K}$ is related to the noncondensate field operator $\ann{j}(\R)$ via a Fourier transform:
\begin{align}
\ahat_{j,\K}=\frac{1}{\sqrt{V}}\integral e^{-i\K\cdot\R} \ann{j}(\R).
\label{eq: Fourier transform of the field operator}
\end{align}
In the following sections, $\hat{\mathcal{K}}_0$ and $\hat{\mathcal{K}}_1$ are referred to as the noninteracting and interacting parts of operator $\hat{\mathcal{K}}'$ in Eq.~\eqref{eq: operator K'}, respectively. For a weakly interacting system, $\hat{\mathcal{K}}_1$ can be treated as a perturbation to $\hat{\mathcal{K}}_0$.

%&&&&&&&&&&&&&&&&&&&&&&&&&&&&
\subsection{Green's functions}
\label{subsection: Green's functions}
In the presence of the condensate, the Green's function is given by \cite{Beliaev2, Fetterbook}
\begin{align}
iG^\mathrm{total}_{jj'}(x,y)=n_0\xi_j\xi_{j'}^*+iG_{jj'}(x,y),
\end{align}
where $j,j'=0,\pm1$ indicate the spin components, and $x=(\mathbf{r}, t), y=(\mathbf{r}', t')$ are four-vectors in the time-coordinate space. The noncondensate part of the Green's function is defined as
\begin{align}
iG_{jj'}(x,y)\equiv\, \frac{\langle\mathbf{O}|\mathcal{T}{\ann{j,\mathrm{H}}(x)\cre{j',\mathrm{H}}(y)}|\mathbf{O}\rangle}{\langle\mathbf{O}|\mathbf{O}\rangle}.
\label{eq: normal Green's function definition}
\end{align}
Here, $|\mathbf{O}\rangle$ is the ground state of the interacting system, and $\mathcal{T}$ and $\mathrm{H}$ denote the time ordering operator and the Heisenberg representation, respectively. 

In the presence of the condensate, we must take into account the collision processes in which two noncondensed atoms get into or out of the condensate. For this purpose, in addition to the normal Green's functions $G_{jj'}(x,y)$ defined in Eq.~\eqref{eq: normal Green's function definition}, it is necessary to introduce the so-called anomalous Green's functions which are defined as
\begin{align}
iG^{12}_{jj'}(x,y)\equiv&\, \frac{\langle\mathbf{O}|\mathcal{T}{\cre{j,\mathrm{H}}(x)\cre{j',\mathrm{H}}(y)}|\mathbf{O}\rangle}{\langle\mathbf{O}|\mathbf{O}\rangle}, \\
iG^{21}_{jj'}(x,y)\equiv&\, \frac{\langle\mathbf{O}|\mathcal{T}{\ann{j,\mathrm{H}}(x)\ann{j',\mathrm{H}}(y)}|\mathbf{O}\rangle}{\langle\mathbf{O}|\mathbf{O}\rangle}. 
\end{align}  

In energy-momentum space, the Dyson's equations for the noncondensate Green's functions are given by
\begin{align}
G^{\alpha\beta}_{jj'}(p)=(G^0)^{\alpha\beta}_{jj'}(p)+(G^0)^{\alpha\gamma}_{jm}\Sigma^{\gamma\delta}_{mm'}(p)G^{\delta\beta}_{m'j'}(p),
\label{eq: Dyson equation 1}
\end{align}
where $\hbar p\equiv \hbar (p_0,\mathbf{p})$ is the four-momentum, and $\alpha, \beta, \gamma, \delta=1,2$ are used to label the normal and anomalous Green's functions as matrix elements of a $6\times6$ matrix:
\begin{equation}
\begin{bmatrix}
G^{11}_{1,1}(p)&G^{11}_{1,0}(p)&G^{11}_{1,-1}(p)&G^{12}_{1,1}(p)&G^{12}_{1,0}(p)&G^{12}_{1,-1}(p)\\
G^{11}_{0,1}(p)&G^{11}_{0,0}(p)&G^{11}_{0,-1}(p)&G^{12}_{0,1}(p)&G^{12}_{0,0}(p)&G^{12}_{0,-1}(p)\\
G^{11}_{-1,1}(p)&G^{11}_{-1,0}(p)&G^{11}_{-1,-1}(p)&G^{12}_{-1,1}(p)&G^{12}_{-1,0}(p)&G^{12}_{-1,-1}(p)\\
G^{21}_{1,1}(p)&G^{21}_{1,0}(p)&G^{21}_{1,-1}(p)&G^{22}_{1,1}(p)&G^{22}_{1,0}(p)&G^{22}_{1,-1}(p)\\
G^{21}_{0,1}(p)&G^{21}_{0,0}(p)&G^{21}_{0,-1}(p)&G^{22}_{0,1}(p)&G^{22}_{00}(p)&G^{22}_{0,-1}(p)\\
G^{21}_{-1,1}(p)&G^{21}_{-1,0}(p)&G^{21}_{-1,-1}(p)&G^{22}_{-1,1}(p)&G^{22}_{-1,0}(p)&G^{22}_{-1,-1}(p)
\end{bmatrix}
\end{equation}
where
\begin{align}
G^{11}_{jj'}(p)\equiv G_{jj'}(p),\, G^{22}_{jj'}(p)\equiv G_{jj'}(-p).
\end{align}

\begin{figure}[tbp] % float placement: (h)ere, page (t)op, page (b)ottom, other (p)age
  \centering
  % file name: E:/Spinor Beliaev (June 27, 2011)/Figures/interaction.eps
  \includegraphics[width=4in,keepaspectratio]{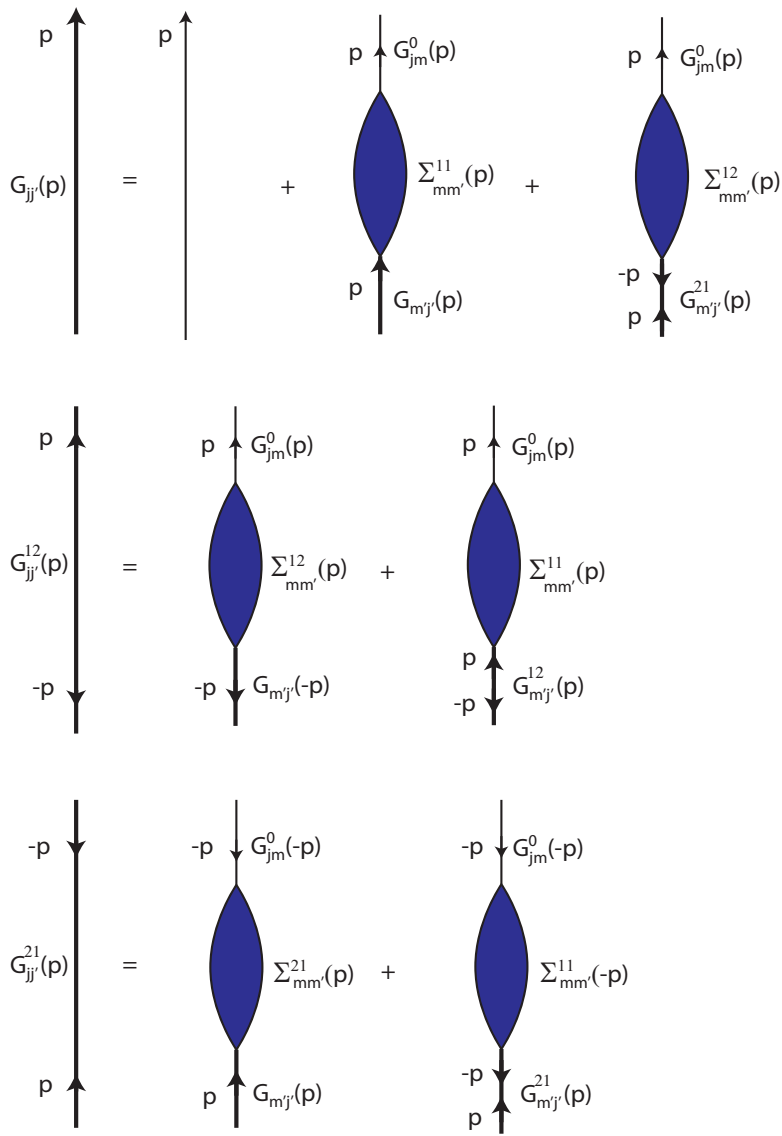}
  \caption{Dyson's equations for the normal and anomalous Green's functions. The thick line, thin line, and oval represent the interacting, non-interacting Green's functions, and the proper self-energies, respectively.}
  \label{fig: Dyson's equation}
\end{figure}

Equation~\eqref{eq: Dyson equation 1}, which is illustrated in Fig.~\ref{fig: Dyson's equation}, can be written in terms of $6\times6$ matrices as a matrix equation:
\begin{align}
\hat{G}(p)=\hat{G}^0(p)+\hat{G}^0(p)\hat{\Sigma}(p)\hat{G}(p),
\label{eq: Dyson equation}
\end{align}
where $\hat{G},\hat{G}^0$, and $\hat{\Sigma}$ denote the $6\times6$ matrices of Green's functions, non-interacting Green's functions, and proper self-energies, respectively. The normal and anomalous self-energies are labeled in the same way as the Green's functions. The solution to Eq.~\eqref{eq: Dyson equation} can be written formally as
\begin{align}
\hat{G}(p)=\Big[1-\hat{G}^0(p)\hat{\Sigma}(p)\Big]^{-1}\hat{G}^0(p).
\label{eq: Dyson eq's solution}
\end{align}

The non-interacting Green's function is defined as
\begin{align}
iG^0_{jj'}(x-y)\equiv \frac{\langle 0|\mathcal{T}\ann{j,\mathrm{H}_0}(x)\cre{j',\mathrm{H}_0}(y)|0 \rangle}{\langle 0|0 \rangle},
\label{eq: non-interacting Green's function 1}
\end{align}
where $|0\rangle$ is the non-interacting ground state, and $\mathrm{H}_0$ indicates the free time evolution in the Heisenberg representation under the non-interacting Hamiltonian $\hat{\mathcal{K}}_0$ given by Eq.~\eqref{eq: operator K0}. Here, $|0\rangle$ is the vacuum state with respect to noncondensate operators; that is, $\ahat_{\K,j}|0\rangle=0$ for all $\K\not=0$ and $j=\pm 1,0$. Substituting Eq.~\eqref{eq: operator K0} into Eq.~\eqref{eq: non-interacting Green's function 1}, we obtain the Fourier transform of $G^0_{jj'}(x-y)$ as
\begin{align}
G^0_{jj'}(p)=&\int \text{d}^4x \, e^{-ipx} G^0_{jj'}(x)\nonumber\\
=&\,\delta_{jj'}\frac{1}{p_0-\eps{\p}/\hbar+\mu/\hbar-q_Bj^2/\hbar+i\eta}\nonumber\\
\equiv& \,\delta_{jj'}G^0_j(p),
\label{eq: non-interacting Green's function}
\end{align}
where $\eta$ is an infinitesimal positive number. Note that the anomalous Green's functions in a non-interacting system are always zero, and thus, the matrix $\hat{G}^0(p)$ is diagonal with matrix elements given by Eq.~\eqref{eq: non-interacting Green's function}.

Now we consider two cases in which the mean-field ground state is in the ferromagnetic phase and in the polar phase.

%****************************
\subsubsection{Ferromagnetic phase}
\label{subsubsection: Ferromagnetic phase, section: Green's function formalism for a spinor Bose-Einstein condensate}
\begin{figure}[tbp] % float placement: (h)ere, page (t)op, page (b)ottom, other (p)age
  \centering
  % file name: E:/Spinor Beliaev (June 27, 2011)/Figures/interaction.eps
  \includegraphics[width=3in,keepaspectratio]{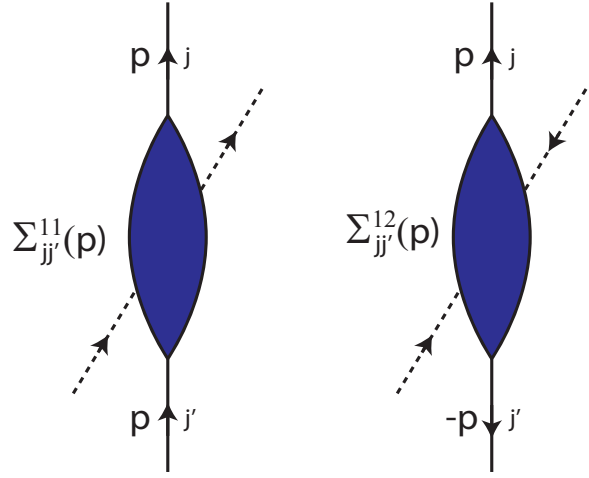}
  \caption{Normal and anomalous proper self-energies of noncondensed particles. $\hbar p$ is the four-momentum, while $j,j'$ label the spin components. The dashed lines represent condensed particles. }
  \label{fig: self-energies}
\end{figure}

If the system's ground state is in the ferromagnetic phase, the condensate's spinor is given by
\begin{align}
(\xi_1,\xi_0,\xi_{-1})=(1,0,0);
\end{align}
i.e., all condensed atoms reside in the $j=1$ magnetic sublevel. Then, the only nonzero matrix elements of $\hat{\Sigma}(p)$ are
\begin{equation}
\begin{bmatrix}
\Sigma^{11}_{1,1}(p)&0&0&\Sigma^{12}_{1,1}(p)&0&0\\
0&\Sigma^{11}_{0,0}(p)&0&0&0&0\\
0&0&\Sigma^{11}_{-1,-1}(p)&0&0&0\\
\Sigma^{21}_{1,1}(p)&0&0&\Sigma^{11}_{1,1}(-p)&0&0\\
0&0&0&0&\Sigma^{11}_{0,0}(-p)&0\\
0&0&0&0&0&\Sigma^{11}_{-1,-1}(-p)
\end{bmatrix}.
\label{eq: ferro, Sigma matrix}
\end{equation}
This can be understood by considering the spin conservation in normal and anomalous self-energies, which are illustrated in Fig.~\ref{fig: self-energies}. For normal self-energies $\Sigma^{11}_{jj'}(p)$, the conservation of the total projected spin allows only $j=j'$, i.e., diagonal elements. In contrast, for anomalous self-energies $\Sigma^{12}_{jj'}(p)$, only the $j=j'=1$ element is nonvanishing because the condensed atoms are all in the $m_F=1$ magnetic sublevel.

By substituting Eq.~\eqref{eq: ferro, Sigma matrix} into Eq.~\eqref{eq: Dyson eq's solution} and using the fact that $\hat{G}^0(p)$ is a diagonal matrix (see Eq.~\eqref{eq: non-interacting Green's function}), we find that the matrix $\hat{G}(p)$ of interacting Green's functions has the same form as $\hat{\Sigma}(p)$:
\begin{equation}
\begin{bmatrix}
G_{1,1}(p)&0&0&G^{12}_{1,1}(p)&0&0\\
0&G_{0,0}(p)&0&0&0&0\\
0&0&G_{-1,-1}(p)&0&0&0\\
G^{21}_{1,1}(p)&0&0&G_{1,1}(-p)&0&0\\
0&0&0&0&G_{0,0}(-p)&0\\
0&0&0&0&0&G_{-1,-1}(-p)
\end{bmatrix}.
\end{equation}
Both $\hat{G}(p)$ and $\hat{\Sigma}(p)$ are block-diagonal matrices composed of one $2\times2$ and four $1\times1$ sub-matrices.

The normal and anomalous Green's functions given by Eq.~\eqref{eq: Dyson eq's solution} then can be expressed in terms of the self-energies as
\begin{subequations}
\label{eq: ferro, solution to the Dyson equation}
\begin{align}
G_{1,1}(p)=&\frac{-[G^0_{1}(-p)]^{-1}+\Sigma^{11}_{1,1}(-p)}{D_1} =\frac{p_0+\eps{\p}/\hbar+q_B/\hbar+\Sigma^{11}_{1,1}(-p) -\mu/\hbar}{D_1}, \label{eq: G11(p) ferro} \\
G_{0,0}(p)=&\frac{1}{[G^0_{0}(p)]^{-1}-\Sigma^{11}_{0,0}(p)+i\eta}, \, G_{-1,-1}(p)=\frac{1}{[G^0_{-1}(p)]^{-1}-\Sigma^{11}_{-1,-1}(p)+i\eta},\\
G^{12}_{1,1}(p)=&\frac{-\Sigma^{12}_{1,1}(p)}{D_1},\, G^{21}_{1,1}(p)=\frac{-\Sigma^{21}_{1,1}(p)}{D_1},
\end{align}
\end{subequations}
where 
\begin{align}
D_1=&-[G^0_1(p)]^{-1}[G^0_{1}(-p)]^{-1}+\Sigma^{11}_{1,1}(p)[G^0_{1}(-p)]^{-1}+\Sigma^{11}_{1,1}(-p)[G^0_1(p)]^{-1}\nonumber\\
&-\Sigma^{11}_{1,1}(p)\Sigma^{11}_{1,1}(-p)+\Sigma^{21}_{1,1}(p)\Sigma^{12}_{1,1}(p)+i\eta \nonumber\\
=&\,p_0^2-\left[\Sigma^{11}_{1,1}(p)-\Sigma^{11}_{1,1}(-p)\right]p_0+\Sigma^{21}_{1,1}(p)\Sigma^{12}_{1,1}(p)\nonumber\\
&-\Big[\eps{\p}/\hbar-\mu/\hbar+q_B/\hbar+\frac{\Sigma^{11}_{1,1}(p)+\Sigma^{11}_{1,1}(-p)}{2}\Big]^2+\Big(\frac{\Sigma^{11}_{1,1}(p)-\Sigma^{11}_{1,1}(-p)}{2}\Big)^2+i\eta.
\label{eq: ferro, denominator of Green's functions}
\end{align}

From Eqs.~\eqref{eq: ferro, solution to the Dyson equation} and \eqref{eq: ferro, denominator of Green's functions}, we obtain the modified version of the Hugenholtz-Pines condition~\cite{Hugenholtz59} for an $F=1$ spinor BEC in the ferromagnetic phase, that is, for the three elementary excitations to be gapless, the following condition must be met:
\begin{align}
\Sigma^{11}_{j,j}(p_0=0,\mathbf{p}={\bf 0})-\Sigma^{12}_{j,j}(p_0=0,\mathbf{p}={\bf 0})=&\,(\mu-q_Bj^2)/\hbar.
\label{eq:Ferro, Hugenholz}
\end{align}
Here, the excitation modes with spin $j=0,-1$ are single-particle like, and thus, the corresponding anomalous self-energies and Green's functions vanish. The energy shift of $-q_B$ from the chemical potential on the right-hand side of Eq.~\eqref{eq:Ferro, Hugenholz} results from the difference in quadratic Zeeman energy between magnetic sublevels $j=\pm1$ and $j=0$~\cite{Szabo07}. For the ferromagnetic phase, the Hugenholtz-Pines condition~\eqref{eq:Ferro, Hugenholz} holds only for $j=1$ in the presence of the quadratic Zeeman effect; therefore, only the corresponding phonon mode ($j=1$) is gapless. When $q_B=0$, the spin-wave mode ($j=0$) also becomes gapless with a quadratic dispersion relation.  

%****************************
\subsubsection{Polar phase}
\label{subsubsection: Polar phase, section: Green's function formalism for a spinor Bose-Einstein condensate}
If the system's ground state is in the polar phase, the condensate's spinor is given by
\begin{align}
(\xi_1,\xi_0,\xi_{-1})=(0,1,0);
\end{align}
that is, all condensed atoms occupy the $j=0$ magnetic sublevel. With an argument similar to the ferromagnetic phase , the only nonzero matrix elements of $\hat{\Sigma}(p)$ and $\hat{G}(p)$ are the following:
\begin{equation}
\begin{bmatrix}
\Sigma^{11}_{1,1}(p)&0&0&0&0&\Sigma^{12}_{1,-1}(p)\\
0&\Sigma^{11}_{0,0}(p)&0&0&\Sigma^{12}_{0,0}(p)&0\\
0&0&\Sigma^{11}_{-1,-1}(p)&\Sigma^{12}_{-1,1}(p)&0&0\\
0&0&\Sigma^{21}_{1,-1}(p)&\Sigma^{11}_{1,1}(-p)&0&0\\
0&\Sigma^{21}_{0,0}(p)&0&0&\Sigma^{11}_{0,0}(-p)&0\\
\Sigma^{21}_{-1,1}(p)&0&0&0&0&\Sigma^{11}_{-1,-1}(-p)
\end{bmatrix},
\label{eq: polar, matrix form of Sigma}
\end{equation}
\begin{equation}
\begin{bmatrix}
G_{1,1}(p)&0&0&0&0&G^{12}_{1,-1}(p)\\
0&G_{0,0}(p)&0&0&G^{12}_{0,0}(p)&0\\
0&0&G_{-1,-1}(p)&G^{12}_{-1,1}(p)&0&0\\
0&0&G^{21}_{1,-1}(p)&G_{11}(-p)&0&0\\
0&G^{21}_{0,0}(p)&0&0&G_{0,0}(-p)&0\\
G^{21}_{-1,1}(p)&0&0&0&0&G_{-1,-1}(-p)
\end{bmatrix}.
\end{equation}
Both of these matrices are block-diagonal matrices composed of three $2\times2$ sub-matrices. Here, $\Sigma^{12}_{1,-1}(p)$ and $G^{12}_{1,-1}(p)$ are nonzero due to the projected-spin-conserved scattering process in which two condensed atoms both in the spin state $j=0$ collide with each other to produce two noncondensed atoms with spin components $j=\pm 1$ (see Fig.~\ref{fig: self-energies}).

The normal and anomalous Green's functions given by Eq.~\eqref{eq: Dyson eq's solution} can then be expressed in terms of the self-energies as
\begin{subequations}
\label{eq: solution to the Dyson equation}
\begin{align}
G_{1,1}(p)=&\frac{-[G^0_{-1}(-p)]^{-1}+\Sigma^{11}_{-1,-1}(-p)}{D_1}=\frac{p_0+\eps{\p}/\hbar+q_B/\hbar+\Sigma^{11}_{-1,-1}(-p) -\mu/\hbar}{D_1}, \label{eq: G11(p)} \\
G_{0,0}(p)=&\frac{-[G^0_0(-p)]^{-1}+\Sigma^{11}_{0,0}(-p)}{D_0}=\frac{p_0+\eps{\p}/\hbar+\Sigma^{11}_{0,0}(-p)-\mu/\hbar}{D_0}, \label{eq: G00(p)} \\
G_{-1,-1}(p)=&\frac{-[G^0_{1}(-p)]^{-1}+\Sigma^{11}_{1,1}(-p)}{D_{-1}}=\frac{p_0+\eps{\p}/\hbar+q_B/\hbar+\Sigma^{11}_{1,1}(-p)-\mu/\hbar}{D_{-1}},\\
G^{12}_{1,-1}(p)=&\frac{-\Sigma^{12}_{1,-1}(p)}{D_1},\, G^{21}_{1,-1}(p)=\frac{-\Sigma^{21}_{1,-1}(p)}{D_{-1}},\\
G^{12}_{0,0}(p)=&\frac{-\Sigma^{12}_{0,0}(p)}{D_0},\, G^{21}_{0,0}(p)=\frac{-\Sigma^{21}_{0,0}(p)}{D_0},\\
G^{12}_{-1,1}(p)=&\frac{-\Sigma^{12}_{-1,1}(p)}{D_{-1}},\, G^{21}_{-1,1}(p)=\frac{-\Sigma^{21}_{-1,1}(p)}{D_1},
\end{align}
\end{subequations}
where 
\begin{subequations}
\label{eq: denominator of Green's functions}
\begin{align}
D_1=&-[G^0_1(p)]^{-1}[G^0_{-1}(-p)]^{-1}+\Sigma^{11}_{1,1}(p)[G^0_{-1}(-p)]^{-1}+\Sigma^{11}_{-1,-1}(-p)[G^0_1(p)]^{-1}\nonumber\\
&-\Sigma^{11}_{1,1}(p)\Sigma^{11}_{-1,-1}(-p)+\Sigma^{21}_{-1,1}(p)\Sigma^{12}_{1,-1}(p)+i\eta \nonumber\\
=&\,p_0^2-\left[\Sigma^{11}_{1,1}(p)-\Sigma^{11}_{-1,-1}(-p)\right]p_0+\Sigma^{21}_{-1,1}(p)\Sigma^{12}_{1,-1}(p)\nonumber\\
&-\Big[\frac{\eps{\p}-\mu+q_B}{\hbar}+\frac{\Sigma^{11}_{1,1}(p)+\Sigma^{11}_{-1,-1}(-p)}{2}\Big]^2+\Big(\frac{\Sigma^{11}_{1,1}(p)-\Sigma^{11}_{-1,-1}(-p)}{2}\Big)^2+i\eta,\\
D_0=&-[G^0_0(p)]^{-1}[G^0_{0}(-p)]^{-1}+\Sigma^{11}_{0,0}(p)[G^0_{0}(-p)]^{-1}+\Sigma^{11}_{0,0}(-p)[G^0_0(p)]^{-1}\nonumber\\
&-\Sigma^{11}_{0,0}(p)\Sigma^{11}_{0,0}(-p)+\Sigma^{21}_{0,0}(p)\Sigma^{12}_{0,0}(p)+i\eta \nonumber\\
=&\,p_0^2-\left[\Sigma^{11}_{0,0}(p)-\Sigma^{11}_{0,0}(-p)\right]p_0+\Sigma^{21}_{0,0}(p)\Sigma^{12}_{0,0}(p)\nonumber\\
&-\Big[\frac{\eps{\p}-\mu}{\hbar}+\frac{\Sigma^{11}_{0,0}(p)+\Sigma^{11}_{0,0}(-p)}{2}\Big]^2+\Big(\frac{\Sigma^{11}_{0,0}(p)-\Sigma^{11}_{0,0}(-p)}{2}\Big)^2+i\eta,\\
D_{-1}=&-[G^0_{-1}(p)]^{-1}[G^0_{1}(-p)]^{-1}+\Sigma^{-1,-1}_{1,1}(p)[G^0_{1}(-p)]^{-1}+\Sigma^{11}_{1,1}(-p)[G^0_{-1}(p)]^{-1}\nonumber\\
&-\Sigma^{11}_{-1,-1}(p)\Sigma^{11}_{1,1}(-p)+\Sigma^{21}_{1,-1}(p)\Sigma^{12}_{-1,1}(p)+i\eta \nonumber\\
=&\,p_0^2-\left[\Sigma^{11}_{-1,-1}(p)-\Sigma^{11}_{1,1}(-p)\right]p_0+\Sigma^{21}_{1,-1}(p)\Sigma^{12}_{-1,1}(p)\nonumber\\
&-\Big[\frac{\eps{\p}-\mu+q_B}{\hbar}+\frac{\Sigma^{11}_{-1,-1}(p)+\Sigma^{11}_{1,1}(-p)}{2}\Big]^2+\Big(\frac{\Sigma^{11}_{-1,-1}(p)-\Sigma^{11}_{1,1}(-p)}{2}\Big)^2+i\eta.
\end{align}
\end{subequations}

From Eqs.~\eqref{eq: solution to the Dyson equation} and \eqref{eq: denominator of Green's functions}, we obtain the modified version of the Hugenholtz-Pines condition for an $F=1$ spinor BEC in the polar phase, that is, for the three elementary excitations to be gapless, the following condition must be met:
\begin{align}
\Sigma^{11}_{j,j}(p_0=0,\mathbf{p}={\bf 0})-\Sigma^{12}_{j,-j}(p_0=0,\mathbf{p}={\bf 0})=&(\mu-q_Bj^2)/\hbar.
\label{eq: polar, Hugenholtz}
\end{align}
For the polar phase, the Hugenholtz-Pines condition~\eqref{eq: polar, Hugenholtz} holds only for $j=0$ in the presence of the quadratic Zeeman effect ($q_B\not=0$); therefore, only the corresponding phonon mode ($j=0$) is gapless.

%&&&&&&&&&&&&&&&&&&&&&&&&&&&&
\subsection{T-matrix}
\label{subsection: T-matrix}
\begin{figure}[tbp] % float placement: (h)ere, page (t)op, page (b)ottom, other (p)age
  \centering
  % file name: E:/Spinor Beliaev (June 27, 2011)/Figures/effective potential.eps
  \includegraphics[width=5in,keepaspectratio]{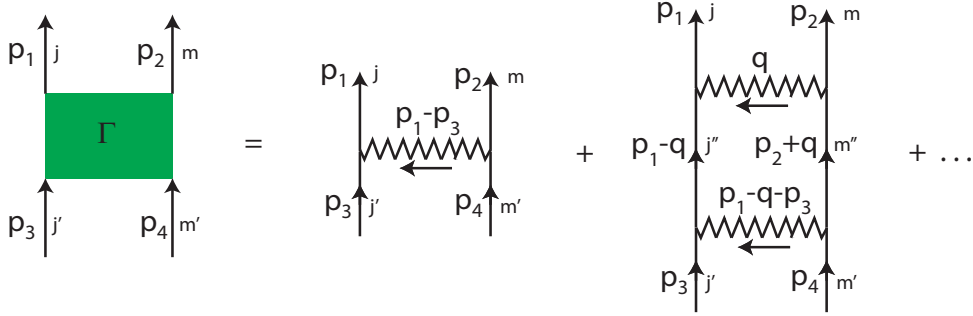}
  \caption{T-matrix of a two-body scattering. Two atoms with momenta $\hbar\mathbf{p}_3, \hbar \mathbf{p}_4$ and magnetic quantum numbers $j',m'$ collide to form two atoms with momenta $\hbar\mathbf{p}_1, \hbar\mathbf{p}_2$ and magnetic quantum numbers $j,m$. The T-matrix is defined as the sum of an infinite number of ladder-type diagrams which describe virtual multiple-scattering processes [see Eq.~\eqref{eq: effective potential}].}
  \label{fig:effective-potential}
\end{figure}

For a weakly interacting dilute Bose gas, the contributions from all ladder-type diagrams to the self-energies are shown to be of the same order of magnitude \cite{Beliaev1, Fetterbook}, and, therefore, all of these contributions must be taken into account. The T-matrix is defined as the sum of an infinite number of ladder-type diagrams as illustrated in Fig.~\ref{fig:effective-potential}. It is written as
\begin{align}
\Gamma_{jm,j'm'}(p_1,p_2;p_3,p_4)=&\,V_{jm,j'm'}(\p_1-\p_3)\nonumber\\
&+\frac{i}{\hbar}\sum_{j'',m''}\integralq G^0_{j''}(p_1-q)G^0_{m''}(p_2+q)\nonumber\\ 
&\times V_{jm,j''m''}(\Q)V_{j''m'',j'm'}(\p_1-\Q-\p_3)\nonumber\\
&+\dots \nonumber\\
=&\,V_{jm,j'm'}(\p_1-\p_3)+\sum_{j'',m''}\integralQ \nonumber\\
&\frac{1}{\hbar(p_1)_0+\hbar(p_2)_0-\eps{\p_1-\Q}-\eps{\p_2+\Q}+2\mu-q_B(j''^2+m''^2)+i\eta}\nonumber\\
&\times V_{jm,j''m''}(\Q)V_{j''m'',j'm'}(\p_1-\Q-\p_3)\nonumber\\
&+\dots \nonumber\\
=&\,V_{jm,j'm'}(\p_1-\p_3)+\sum_{j'',m''}\integralQ \nonumber\\
&\frac{1}{\hbar(p_1)_0+\hbar(p_2)_0-\eps{\p_1-\Q}-\eps{\p_2+\Q}+2\mu-q_B(j''^2+m''^2)+i\eta}\nonumber\\
&\times \Big[\langle jm|\mathcal{F}=0\rangle\langle \mathcal{F}=0|j''m''\rangle\langle j''m''|\mathcal{F}=0\rangle\langle \mathcal{F}=0|j'm'\rangle\nonumber\\
&\times V_0(\Q)V_0(\p_1-\Q-\p_3) \nonumber\\
&+\langle jm|\mathcal{F}=0\rangle\langle \mathcal{F}=0|j''m''\rangle\langle j''m''|\mathcal{F}=2\rangle\langle \mathcal{F}=2|j'm'\rangle \nonumber\\
&\times V_0(\Q)V_2(\p_1-\Q-\p_3) \nonumber\\
&+\langle jm|\mathcal{F}=2\rangle\langle \mathcal{F}=2|j''m''\rangle\langle j''m''|\mathcal{F}=0\rangle\langle \mathcal{F}=0|j'm'\rangle \nonumber\\
&\times V_2(\Q)V_0(\p_1-\Q-\p_3) \nonumber\\
&+\langle jm|\mathcal{F}=2\rangle\langle \mathcal{F}=2|j''m''\rangle\langle j''m''|\mathcal{F}=2\rangle\langle \mathcal{F}=2|j'm'\rangle \nonumber\\
&\times V_2(\Q)V_2(\p_1-\Q-\p_3)\Big] \nonumber\\
&+\dots
\label{eq: effective potential}
\end{align}
Here, the second identity in Eq.~\eqref{eq: effective potential} is obtained by using Eq.~\eqref{eq: non-interacting Green's function} for $G^0_{j}(p)$ in the integration of $G^0_{j''}(p_1-q)G^0_{m''}(p_2+q)$ with respect to $q_0$. 

In spinor BECs, the stable quantum phase of the ground state is determined as the competition between the spin-exchange interatomic interaction and the coupling of atoms to an external magnetic field via the quadratic Zeeman energy, and thus, the quadratic Zeeman energy usually has the same order of magnitude as the spin-exchange interaction: $q_B\sim |c_1|n\ll c_0n$. It can be shown that for such an external magnetic field, the spin dependence of intermediate states via the quadratic Zeeman energies $q_B(j''+m'')$ in the denominator of the right-hand side of Eq.~\eqref{eq: effective potential} only gives a small difference that is negligible up to the order of magnitude we are considering in this paper (see \ref{appendix: Relation between T-matrix and vacuum scattering amplitude}). Consequently, as a good approximation we can take the summation
\begin{align}
\sum_{g,h}|gh\rangle\langle gh|=1
\end{align}
out of the integral. Inside the integral, by using the fact that the $\mathcal{F}=0$ and $\mathcal{F}=2$ spin channels are orthogonal to each other: $\langle \mathcal{F}=0|\mathcal{F}=2\rangle=0$, the T-matrix can be rewritten as
\begin{align}
\Gamma_{jm,j'm'}(p_1,p_2;p_3,p_4)=&\langle j,m|\mathcal{F}=0\rangle\langle \mathcal{F}=0|j'm'\rangle \Gamma_0(p_1,p_2;p_3,p_4)\nonumber\\
&+\langle jm|\mathcal{F}=2\rangle\langle \mathcal{F}=2|j'm'\rangle \Gamma_2(p_1,p_2;p_3,p_4),
\label{eq: T-matrix written in terms of Gamma0,2}
\end{align}
where $\Gamma_\mathcal{F}(p_1,p_2;p_3,p_4)$ is the T-matrix in the $\mathcal{F}$ spin channel given by
\begin{align}
\Gamma_\mathcal{F}(p_1,p_2;p_3,p_4)=&V_\mathcal{F}(\p_1-\p_3)+\frac{i}{\hbar}\integralq G^0(p_1-q)G^0(p_2+q)V_\mathcal{F}(\Q)V_\mathcal{F}(\p_1-\Q-\p_3)\nonumber\\
&+\dots. 
\label{eq: T-matrices in spin channels F=0;2}
\end{align}
Here, $G^0(p)=1/(p_0-\eps{\p}+\mu+i\eta)$ is the spinless non-interacting Green's function.

The T-matrix $\Gamma_\mathcal{F}(p_1,p_2;p_3,p_4)$ can be expressed in terms of the vacuum scattering amplitude for the spin channel $\mathcal{F}=0$ and 2 as follows (see~\ref{appendix: Relation between T-matrix and vacuum scattering amplitude}) \cite{Beliaev1, Fetterbook}:
\begin{align}
\Gamma_\mathcal{F}(p_1,p_2;p_3,p_4)=&\Gamma_\mathcal{F}(\p,\p',P)\nonumber\\
=&\tilde{f}_\mathcal{F}(\p,\p')+\integralQ \tilde{f}_\mathcal{F}(\p,\Q)\Bigg(\frac{1}{\hbar P_0-\frac{\hbar^2\mathbf{P}^2}{4M}+2\mu-\frac{\hbar^2\Q^2}{M}+i\eta}\nonumber\\
&+\frac{1}{\frac{\hbar^2\Q^2}{M}-\frac{\hbar^2\p'^2}{M}-i\eta}\Bigg)\tilde{f}_\mathcal{F}^*(\p',\Q),
\label{eq: Gamma0, Gamma2}
\end{align}
where $-M\tilde{f}_\mathcal{F}(\p,\p')/(4\pi\hbar^2)$ is the vacuum scattering amplitude of the two-body collision in which the relative momentum changes from $\hbar\p'$ to $\hbar\p$. As seen in Eq.~\eqref{eq: Gamma0, Gamma2}, it can be shown that $\Gamma_\mathcal{F}(p_1,p_2;p_3,p_4)$ depends only on the four-vector total momentum $\hbar P\equiv \hbar p_1+\hbar p_2=\hbar p_3+\hbar p_4$ and the relative momenta $\hbar \p\equiv(\hbar \p_1-\hbar \p_2)/2$, $\hbar \p'\equiv(\hbar \p_3-\hbar \p_4)/2$, and depends on neither $p_0\equiv \left[(p_1)_0-(p_2)_0\right]/2$ nor $p'_0\equiv \left[(p_3)_0-(p_4)_0\right]/2$ (see~\ref{appendix: Relation between T-matrix and vacuum scattering amplitude}). 

%############################
\section{First-order approximation--Bogoliubov theory}
\label{section: First-order approximation}
In the approximation to first order in the interatomic interaction, we can neglect the $\Q$-integral in Eq.~\eqref{eq: Gamma0, Gamma2} because it give a contribution to second order. Indeed, its contribution is smaller in magnitude than the first-order contribution by a factor of the diluteness parameter $\sqrt{na_\mathcal{F}^3}\ll 1$, where $a_\mathcal{F}$ is the s-wave scattering length in spin channel $\mathcal{F}$ ($=0$, 2) (see Sec.~\ref{section: Second-order approximation}). On the other hand, in the low-energy regime $|\p|\ll 1/a_\mathcal{F}$, the momentum dependence of the vacuum scattering amplitudes is negligible, and $\tilde{f}_\mathcal{F}(\p,\Q)$ reduces to $f_\mathcal{F}\equiv 4\pi\hbar^2 a_\mathcal{F}/M$ in the limit of zero momenta: $\p,\Q\to 0$. The T-matrix then becomes
\begin{align}
\Gamma_{jm,j'm'}(\p,\p',P)\simeq\langle j,m|\mathcal{F}=0\rangle\langle \mathcal{F}=0|j',m'\rangle f_0+\langle j,m|\mathcal{F}=2\rangle\langle \mathcal{F}=2|j',m'\rangle f_2.
\end{align}
By using the following relations
\begin{align}
|\mathcal{F}=0\rangle\langle \mathcal{F}=0|+|\mathcal{F}=2\rangle\langle \mathcal{F}=2|=&1,\\
-2|\mathcal{F}=0\rangle\langle \mathcal{F}=0|+|\mathcal{F}=2\rangle\langle \mathcal{F}=2|=&\mathbf{F}\cdot \mathbf{F},
\end{align}
the T-matrix can be rewritten in the following form:
\begin{align} 
\Gamma_{jm,j'm'}(\p,\p',P)\simeq c_0\, \delta_{jj'}\delta_{mm'}+c_1\, \sum_\alpha (F_\alpha)_{jj'}(F_\alpha)_{mm'},
\label{eq: first-order T-matrix}
\end{align}
where $(F_\alpha)$ ($\alpha=x,y,z$) are the components of the spin-1 matrix vector
\begin{align}
F_x=\frac{1}{\sqrt{2}}
   \begin{pmatrix}
   0&1&0\\
   1&0&1\\
   0&1&0\\
   \end{pmatrix},
F_y=\frac{i}{\sqrt{2}}
   \begin{pmatrix}
   0&-1&0\\
   1&0&-1\\
   0&1&0\\
   \end{pmatrix},
F_z=
   \begin{pmatrix}
   1&0&0\\
   0&0&0\\
   0&0&-1\\
   \end{pmatrix},
\end{align}
and $c_0$ and $c_1$ are the coefficients of the spin-conserving and spin-exchange interactions, respectively. They are related to the s-wave scattering lengths as follows:
\begin{align}
c_0\equiv&\frac{f_0+2f_2}{3}=\frac{4\pi\hbar^2}{M}\frac{a_0+2a_2}{3}, \label{eq: definition of c0 in terms of a0 and a2}\\
c_1\equiv&\frac{f_2-f_0}{3}=\frac{4\pi\hbar^2}{M}\frac{a_2-a_0}{3}.
\end{align}
For a convenience, we define a characteristic length scale $a$ ($\tilde{a}$) that corresponds to the spin-conserving interaction in the T-matrix given by Eq.~\eqref{eq: first-order T-matrix}:
\begin{align}
a \equiv \frac{\tilde{a}}{4\pi}\equiv \frac{a_0+2a_2}{3},
\label{eq: define scattering length a}
\end{align}
from which we have $c_0=4\pi\hbar^2 a/M=\hbar^2 \tilde{a}/M$.

Now, we consider two cases in which the ground state is in the ferromagnetic and the polar phase.

%&&&&&&&&&&&&&&&&&&&&&&&&&&&&
\subsection{Ferromagnetic phase}
\label{subsection: Ferromagnetic phase, section: First-order approximation}
\begin{figure}[tbp] % float placement: (h)ere, page (t)op, page (b)ottom, other (p)age
  \centering
  % file name: E:/Spinor Beliaev (July 4, 2011)/Figures/first_order_approximation.eps
  \includegraphics[width=4in,keepaspectratio]{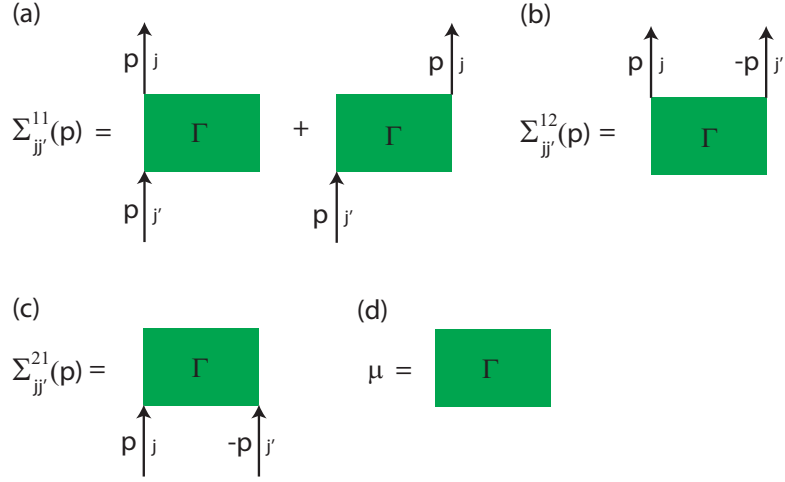}
  \caption{First-order contributions to the proper self-energies (a)-(c) and the chemical potential (d). Here, the squares represent the T-matrices while particles belonging to the condensate are not explicitly shown. In fact, in (a), there are one condensed atom moving in and one condensed atom moving out; in (b) and (c), there are two condensed atoms moving in and two condensed atoms moving out, respectively; in (d), all four atoms are condensed atoms. This convention helps simplify the second-order diagrams in Sec.~\ref{section: Second-order approximation}.}
  \label{fig:first_order_approximation}
\end{figure}

In the ferromagnetic phase, all condensed particles occupy the $j=1$ magnetic sub-level, and the condensate's spinor is 
\begin{align}
(\xi_1,\xi_0,\xi_{-1})=(1,0,0).
\end{align}
The proper self-energies and chemical potential in the first-order approximation, which are illustrated by diagrams in Fig. \ref{fig:first_order_approximation}, are then given by
\begin{subequations}
\label{eq: ferro, first-order proper self energies 1}
\begin{align}
\hbar\Sigma^{11}_{jj'}(p)=&\,\Gamma_{j1,j'1}(\p/2,\p/2,p)+\Gamma_{1j,j'1}(\p/2,-\p/2,p) \nonumber\\
\simeq&\,c_0n_0(\delta_{jj'}+\delta_{j,1}\delta_{j',1})+c_1n_0\sum_\alpha\left[(F_\alpha)_{jj'}(F_\alpha)_{11}+(F_\alpha)_{j,1}(f_\alpha)_{1,j'}\right]\nonumber\\
=&\,c_0n_0(\delta_{jj'}+\delta_{j,1}\delta_{j',1})+c_1n_0(j\delta_{jj'}+\delta_{j,1}\delta_{j',1}+\delta_{j,0}\delta_{j',0}), \label{eq: ferro, Sigma11 first-order approximation}\\
\hbar\Sigma^{12}_{jj'}(p)=\hbar\Sigma^{21}_{jj'}(p)=&\Gamma_{jj',11}(\p,{\bf 0},0) \nonumber\\
\simeq&\,c_0n_0\delta_{j,1}\delta_{j',1}+c_1n_0\sum_\alpha(F_\alpha)_{j,1}(F_\alpha)_{j',1}\nonumber\\
=&\,c_0n_0\delta_{j,1}\delta_{j',1}+c_1n_0\delta_{j,1}\delta_{j',1}, \label{eq: ferro, Sigma12 first-order approximation}\\
\mu=&\,\Gamma_{11,11}({\bf 0},{\bf 0},0)+q_B \nonumber\\
\simeq&\,c_0n_0+c_1n_0\sum_\alpha(F_\alpha)_{11}(F_\alpha)_{11}+q_B\nonumber\\
=&\,(c_0+c_1)n_0+q_B. \label{eq: ferro, mu first-order apprx}
\end{align}
\end{subequations}
Here, the quadratic Zeeman energy $q_B$ is added to the right-hand side of Eq.~\eqref{eq: ferro, mu first-order apprx} for the chemical potential to account for the fact that the condensate is in the magnetic sublevel $j=1$, whose energy is raised by $q_B$ due to the quadratic Zeeman effect. The matrix elements of $\hat{\Sigma}(p)$ in Eq.~\eqref{eq: ferro, Sigma matrix} are then given by 
\begin{subequations}
\label{eq: ferro, first-order proper self energies}
\begin{align}
\hbar\Sigma^{11}_{11}(p)=&\,2(c_0+c_1)n_0,\\
\hbar\Sigma^{11}_{00}(p)=&\,(c_0+c_1)n_0,\\
\hbar\Sigma^{11}_{-1,-1}(p)=&\,(c_0-c_1)n_0,\\
\hbar\Sigma^{12}_{11}(p)=\,\hbar\Sigma^{21}_{11}(p)=&\,(c_0+c_1)n_0,\\
\mathrm{others}=&\,0.
\end{align}
\end{subequations}

By substituting Eqs~\eqref{eq: ferro, mu first-order apprx} and \eqref{eq: ferro, first-order proper self energies} into Eq.~\eqref{eq: ferro, solution to the Dyson equation}, we obtain the first-order Green's functions:
\begin{subequations}
\label{eq: ferro, first-order Green's functions 1}
\begin{align}
G_{11}(p)=&\,\frac{p_0+\eps{\p}/\hbar+(c_0+c_1)n_0/\hbar}{p_0^2-\omega_{1,\p}^2+i\eta}, \\
G_{00}(p)=&\,\frac{1}{p_0-\omega_{0,\p}+i\eta},\\
G_{-1,-1}(p)=&\,\frac{1}{p_0-\omega_{-1,\p}+i\eta}, \\
G^{12}_{11}(p)=\,G^{21}_{11}=&\,-\frac{(c_0+c_1)n_0/\hbar}{p_0^2-\omega_{1,\p}^2+i\eta}, \\
\mathrm{others}=&\,0.
\end{align}
\end{subequations}

The energy spectra of the elementary excitations, which are given by the poles of the Green's functions, are
\begin{subequations}
\label{eq: ferro, first-order energy spectra}
\begin{align}
\hbar\omega_{1,\p}=&\,\sqrt{\eps{\p}[\eps{\p}+2(c_0+c_1)n_0]}, \label{eq: ferro, first-order energy spectra, E1p} \\
\hbar\omega_{0,\p}=&\,\eps{\p}-q_B, \label{eq: ferro, first-order energy spectra, E0p} \\
\hbar\omega_{-1,\p}=&\,\eps{\p}-2c_1n_0.
\end{align}
\end{subequations}
Thus, the Green's function approach gives the Bogoliubov energy spectra of elementary excitations as the first-order results~\cite{Uchino10}.

It will be useful for the second-order calculation in Sec.~\ref{section: Second-order approximation} to rewrite the first-order Green's functions in Eq.~\eqref{eq: ferro, first-order Green's functions 1} as follows:
\begin{subequations}
\label{eq: ferro, first-order Green's functions in convenient form}
\begin{align}
G_{11}(p)=&\frac{A_{1,\p}}{p_0-\omega_{1,\p}+i\eta}-\frac{B_{1,\p}}{p_0+\omega_{1,\p}-i\eta}, \label{eq: ferro, first-order G11}\\
G^{12}_{11}(p)=G^{21}_{11}=&-C_{1,\p}\left(\frac{1}{p_0-\omega_{1,\p}+i\eta}-\frac{1}{p_0+\omega_{1,\p}-i\eta}\right), \label{eq: ferro, first-order G12}
\end{align}
\end{subequations}
where 
\begin{align}
A_{1,\p}=&\frac{\hbar\omega_{1,\p}+\eps{\p}+(c_0+c_1)n_0}{2\hbar\omega_{1,\p}},\, B_{1,\p}=\frac{-\hbar\omega_{1,\p}+\eps{\p}+(c_0+c_1)n_0}{2\hbar\omega_{1,\p}}, \\
C_{1,\p}=&\frac{(c_0+c_1)n_0}{2\hbar\omega_{1,\p}}.
\end{align}

%&&&&&&&&&&&&&&&&&&&&&&&&&&&&
\subsection{Polar phase}
\label{subsection: Polar phase, section: First-order approximation}
In the polar phase, all condensed particles occupy the $j=0$ magnetic sublevel, and the condensate's spinor is  
\begin{align}
(\xi_1,\xi_0,\xi_{-1})=(0,1,0).
\end{align}
From Fig.~\ref{fig:first_order_approximation}, the first-order proper self-energies and chemical potential are given by
\begin{subequations}
\label{eq: polar, first-order self energies 1}
\begin{align}
\hbar\Sigma^{11}_{jj'}(p)=&\,\Gamma_{j0,j'0}(\p/2,\p/2,p)+\Gamma_{0j,j'0}(\p/2,-\p/2,p) \nonumber\\
\simeq&\,c_0n_0(\delta_{j,j'}+\delta_{j,0}\delta_{j',0})+c_1n_0\sum_\alpha\left[(F_\alpha)_{j,j'}(F_\alpha)_{0,0}+(F_\alpha)_{j,0}(F_\alpha)_{0,j'}\right]\nonumber\\
=&\,c_0n_0(\delta_{j,j'}+\delta_{j,0}\delta_{j',0})+c_1n_0(\delta_{j,1}\delta_{j',1}+\delta_{j,-1}\delta_{j',-1}), \label{eq: Sigma11 first-order approximation}\\
\hbar\Sigma^{12}_{jj'}(p)=\hbar\Sigma^{21}_{jj'}(p)=&\,\Gamma_{jj',00}(\p,{\bf 0},0) \nonumber\\
\simeq&\,c_0n_0\delta_{j,0}\delta_{j',0}+c_1n_0\sum_\alpha(F_\alpha)_{j,0}(F_\alpha)_{j',0}\nonumber\\
=&\,c_0n_0\delta_{j,0}\delta_{j',0}+c_1n_0(\delta_{j,1}\delta_{j',-1}+\delta_{j,-1}\delta_{j',1}), \label{eq: Sigma12 first-order approximation}\\
\mu=&\,\Gamma_{00,00}({\bf 0},{\bf 0},0) \nonumber\\
\simeq&\,c_0n_0, \label{eq: polar, first-order chemical potential}
\end{align}
\end{subequations}
The matrix elements of $\hat{\Sigma}(p)$ in Eq.~\eqref{eq: polar, matrix form of Sigma} are then given by
\begin{subequations}
\label{eq: polar, first-order self energies 2}
\begin{align}
\hbar\Sigma^{11}_{11}(p)=\hbar\Sigma^{11}_{-1,-1}(p)=&\,(c_0+c_1)n_0,\\
\hbar\Sigma^{11}_{00}(p)=&\,2c_0n_0,\\
\hbar\Sigma^{12}_{1,-1}(p)=\hbar\Sigma^{12}_{-1,1}(p)=\hbar\Sigma^{21}_{1,-1}(p)=\hbar\Sigma^{21}_{-1,1}(p)=&\,c_1n_0,\\
\hbar\Sigma^{12}_{00}(p)=\hbar\Sigma^{21}_{00}(p)=&\,c_0n_0,\\
\mathrm{others}=&\,0.
\end{align}
\end{subequations}

Substituting Eqs.~\eqref{eq: polar, first-order chemical potential} and \eqref{eq: polar, first-order self energies 2} into Eqs.~\eqref{eq: solution to the Dyson equation}, we obtain the first-order Green's functions as follows:
\begin{subequations}
\label{eq: polar, first-order Green's functions 1}
\begin{align}
G_{11}(p)=G_{-1,-1}(p)=&\,\frac{p_0+\left(\eps{\p}+c_1n_0+q_B\right)/\hbar}{p_0^2-\omega_{1,\p}^2+i\eta}, \\
G_{00}(p)=&\,\frac{p_0+\left(\eps{\p}+c_0n_0\right)/\hbar}{p_0^2-\omega_{0,\p}^2+i\eta}, \\
G^{12}_{1,-1}(p)=G^{12}_{-1,1}(p)=G^{21}_{1,-1}(p)=G^{21}_{1,-1}(p)=&\,-\frac{c_1n_0/\hbar}{p_0^2-\omega_{1,\p}^2+i\eta}, \\
G^{12}_{00}(p)=G^{21}_{00}(p)=&\,-\frac{c_0n_0/\hbar}{p_0^2-\omega_{0,\p}^2+i\eta},\\
\mathrm{others}=&\,0.
\end{align}
\end{subequations}
The energy spectra of the elementary excitations, which are given by the poles of the Green's functions, are
\begin{subequations}
\label{eq: polar, first-order energy spectra}
\begin{align}
\hbar\omega_{1,\p}=\hbar\omega_{-1,\p}=&\sqrt{(\eps{\p}+q_B)(\eps{\p}+q_B+2c_1n_0)}, \label{eq: first-order energy spectrum E1p}\\
\hbar\omega_{0,\p}=&\sqrt{\eps{\p}(\eps{\p}+2c_0n_0)}\label{eq: first-order energy spectrum E0p}.
\end{align}
\end{subequations}
Here, there is a two-fold degeneracy in the energy spectra: $\omega_{1,\p}=\omega_{-1,\p}$ for the polar phase due to the symmetry between two magnetic sublevels $j=\pm 1$. Similarly to the ferromagnetic phase, by using the Green's function approach, we have obtained the Bogoliubov energy spectra of elementary excitations for the polar phase as the first-order results~\cite{Uchino10}.

The first-order Green's functions given by Eq.~\eqref{eq: polar, first-order Green's functions 1} can be rewritten in the following form:
\begin{subequations}
\label{eq: polar, first-order Green's functions in convenient form}
\begin{align}
G_{11}(p)=&\,G_{-1,-1}(p)=\,\frac{A_{1,\p}}{p_0-\omega_{1,\p}+i\eta}-\frac{B_{1,\p}}{p_0+\omega_{1,\p}-i\eta}, \label{eq: first-order G11}\\
G_{00}(p)=&\,\frac{A_{1,\p}}{p_0-\omega_{0,\p}+i\eta}-\frac{B_{0,\p}}{p_0+\omega_{0,\p}-i\eta}, \label{eq: first-order G00}\\
G^{12}_{1,-1}(p)=&\,G^{12}_{-1,1}(p)=\,G^{21}_{1,-1}(p)=\,G^{21}_{1,-1}(p)\nonumber\\
=&\,-C_{1,\p}\Bigg(\frac{1}{p_0-\omega_{1,\p}+i\eta}-\frac{1}{p_0+\omega_{1,\p}-i\eta}\Bigg), \label{eq: first-order G1-1}\\
G^{12}_{00}(p)=&\,G^{21}_{00}(p)=\,-C_{0,\p}\Bigg(\frac{1}{p_0-\omega_{0,\p}+i\eta}-\frac{1}{p_0+\omega_{0,\p}-i\eta}\Bigg),\label{eq: first-order G00}
\end{align}
\end{subequations}
where 
\begin{align}
A_{1,\p}=&\frac{\hbar\omega_{1,\p}+\eps{\p}+c_1n_0+q_B}{2\hbar\omega_{1,\p}},\, B_{1,\p}=\frac{-\hbar\omega_{1,\p}+\eps{\p}+c_1n_0+q_B}{2\hbar\omega_{1,\p}}, \\
A_{0,\p}=&\frac{\hbar\omega_{0,\p}+\eps{\p}+c_0n_0}{2\hbar\omega_{0,\p}},\, B_{0,\p}=\frac{-\hbar\omega_{0,\p}+\eps{\p}+c_0n_0}{2\hbar\omega_{0,\p}},\\
C_{1,\p}=&\frac{c_1n_0}{2\hbar\omega_{1,\p}},\, C_{0,\p}=\frac{c_0n_0}{2\hbar\omega_{0,\p}}.
\end{align}
These expressions will be used in the following sections.

%############################
\section{Second-order approximation--Beliaev theory}
\label{section: Second-order approximation}
We now investigate how the effect of quantum depletion at absolute zero alters the energy spectra of elementary excitations in an $F=1$ spinor condensate of $\Rb$ by calculating the energy spectra to the second-order in interaction. The spin-exchange interaction for $\Rb$ atoms is known to be ferromagnetic ($c_1<0$). Here, we only consider the case of $q_B<0$ and $q_B>2|c_1|n$ for the respective ferromagnetic and polar phases, where the corresponding first-order energy spectra of elementary excitations show that the system is dynamically stable [see Eqs.~\eqref{eq: ferro, first-order energy spectra} and \eqref{eq: polar, first-order energy spectra}]. On the other hand, when considering the second-order corrections to the first-order results, we only need to take into account the spin-conserving interaction since the spin-exchange interaction would make a much smaller contribution to the already very small second-order quantities. This is due to the large ratio of spin-conserving to spin-exchange interactions of $\Rb$ atoms: $c_0/|c_1|\simeq 200$. However, for a usual atomic density in experiments of ultracold atoms, the second-order contribution to the proper self-energies from the spin-conserving interaction is of the order of $c_0n\sqrt{na^3}\sim 0.01c_0n$, which is of the same order of magnitude as the first-order contribution from the spin-exchange interaction~$\sim |c_1|n$. We may thus expect an interplay between quantum depletion and spinor physics.

%&&&&&&&&&&&&&&&&&&&&&&&&&&&&
\subsection{Second-order proper self-energies and chemical potential}
\label{subsection: Second-order self energies}
\begin{figure}[tbp] % float placement: (h)ere, page (t)op, page (b)ottom, other (p)age
  \centering
  % file name: E:/Spinor Beliaev (July 4, 2011)/Figures/second_order_Sigma11.eps
  \includegraphics[width=5in,keepaspectratio]{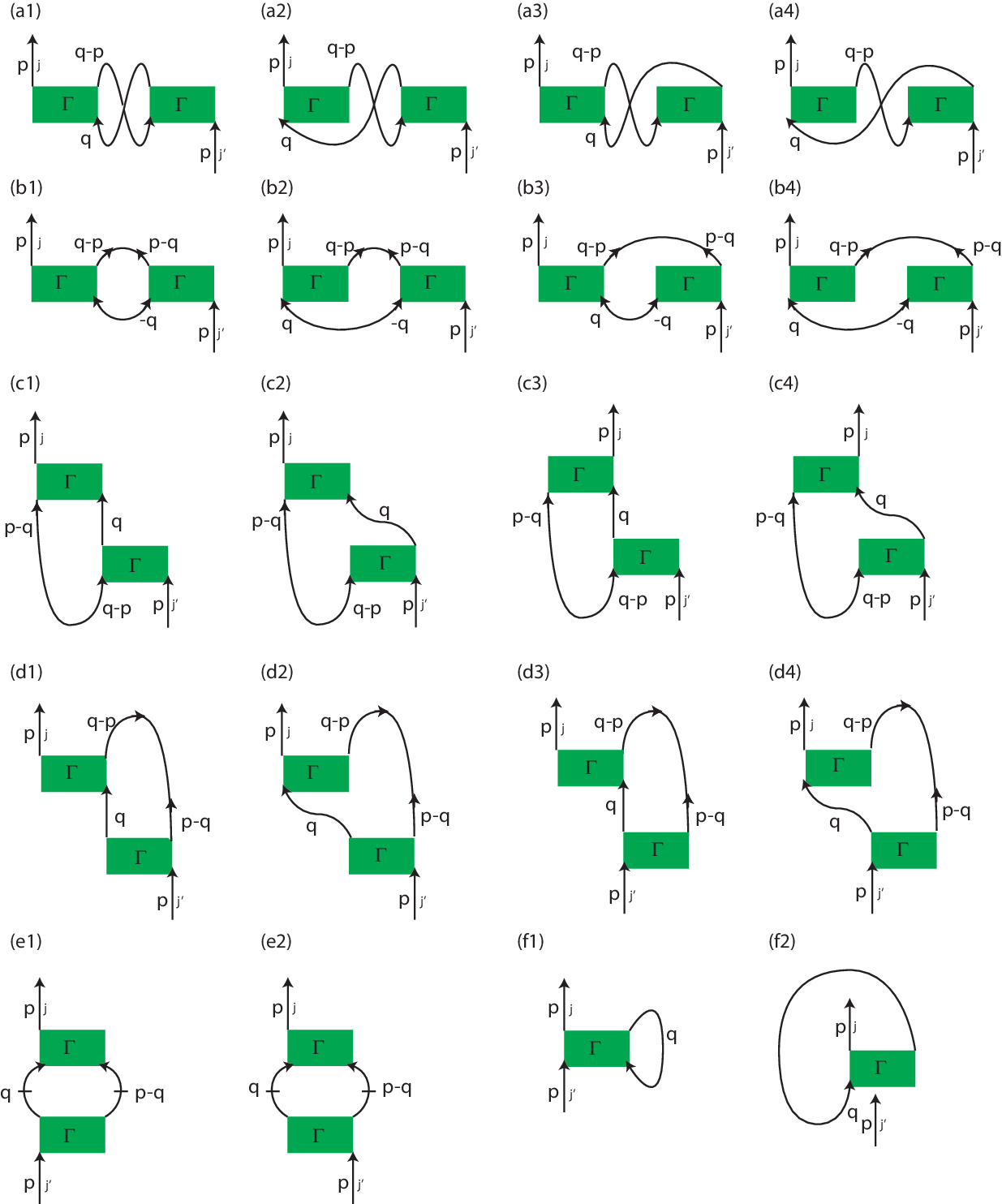}
  \caption{Second-order diagrams for the proper self-energy $\Sigma^{11}_{jj'}(p)$. The intermediate propagators are classified into three different groups, depending on the number of noncondensed atoms moving into and out of the condensate. They are represented by curves with one arrow ($\longrightarrow$), two out-arrows ($\longleftrightarrow$), and two in-arrows ($\rightarrow\leftarrow$), and are described respectively by the first-order normal Green's function $G_{jj'}(p)$ and two anomalous Green's functions $G^{12}_{jj'}(p)$ and $G^{21}_{jj'}(p)$, respectively. Here, the two horizontal dashes in diagrams (e1) and (e2) represent the fact that we need to subtract from these diagrams terms containing non-interacting Green's functions to avoid double counting of the contributions that have already been taken into account by the definition of the T-matrix and the first-order diagrams [see Eqs.~\eqref{eq: diagram c1} and \eqref{eq: diagram c2}]. As in Fig.~\ref{fig:first_order_approximation}, here, we use the convention that the condensed particles in diagrams (a1)-(e2) are not explicitly shown.}
  \label{fig:second_order_Sigma11}
\end{figure}

\begin{figure}[tbp] % float placement: (h)ere, page (t)op, page (b)ottom, other (p)age
  \centering
  % file name: E:/Spinor Beliaev (July 4, 2011)/Figures/second_order_Sigma12.eps
  \includegraphics[width=5in,keepaspectratio]{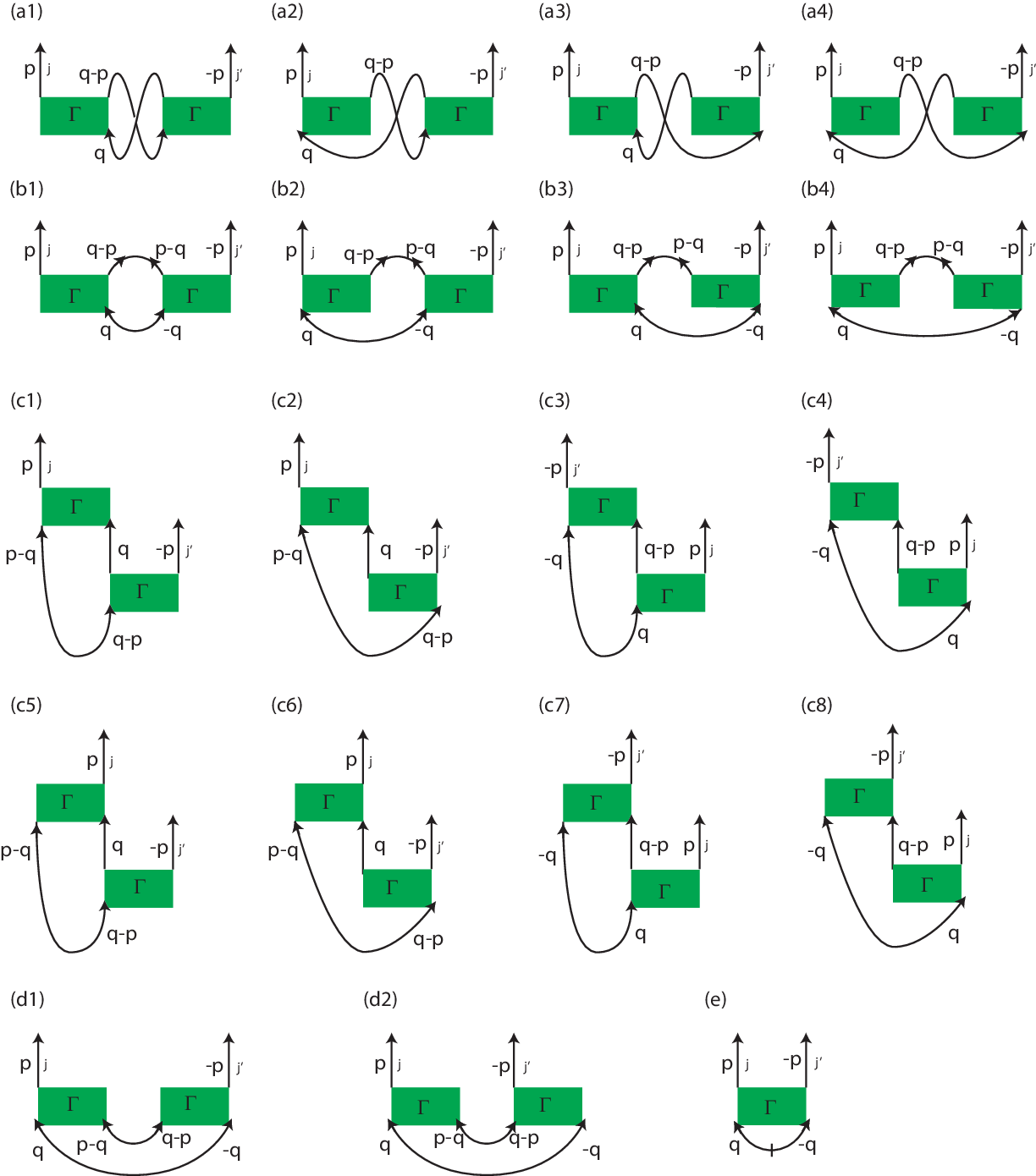}
  \caption{Second-order diagrams for the proper self-energy $\Sigma^{12}_{jj'}(p)$. Similar to the horizontal dashes in diagrams (e1) and (e2) of Fig.~\ref{fig:second_order_Sigma11}, the vertical dash in diagram (e) represent the fact that we need to subtract from this diagram a term containing non-interacting Green's functions to avoid double counting of the contribution that has already been taken into account by the definition of the T-matrix and the first-order diagrams [see Eq.~\eqref{eq: ferro, diagram (i) for Sigma12}].}
  \label{fig:second_order_Sigma12}
\end{figure}

\begin{figure}[tbp] % float placement: (h)ere, page (t)op, page (b)ottom, other (p)age
  \centering
  % file name: E:/Spinor Beliaev (July 4, 2011)/Figures/second_order_Sigma21.eps
  \includegraphics[width=5in,keepaspectratio]{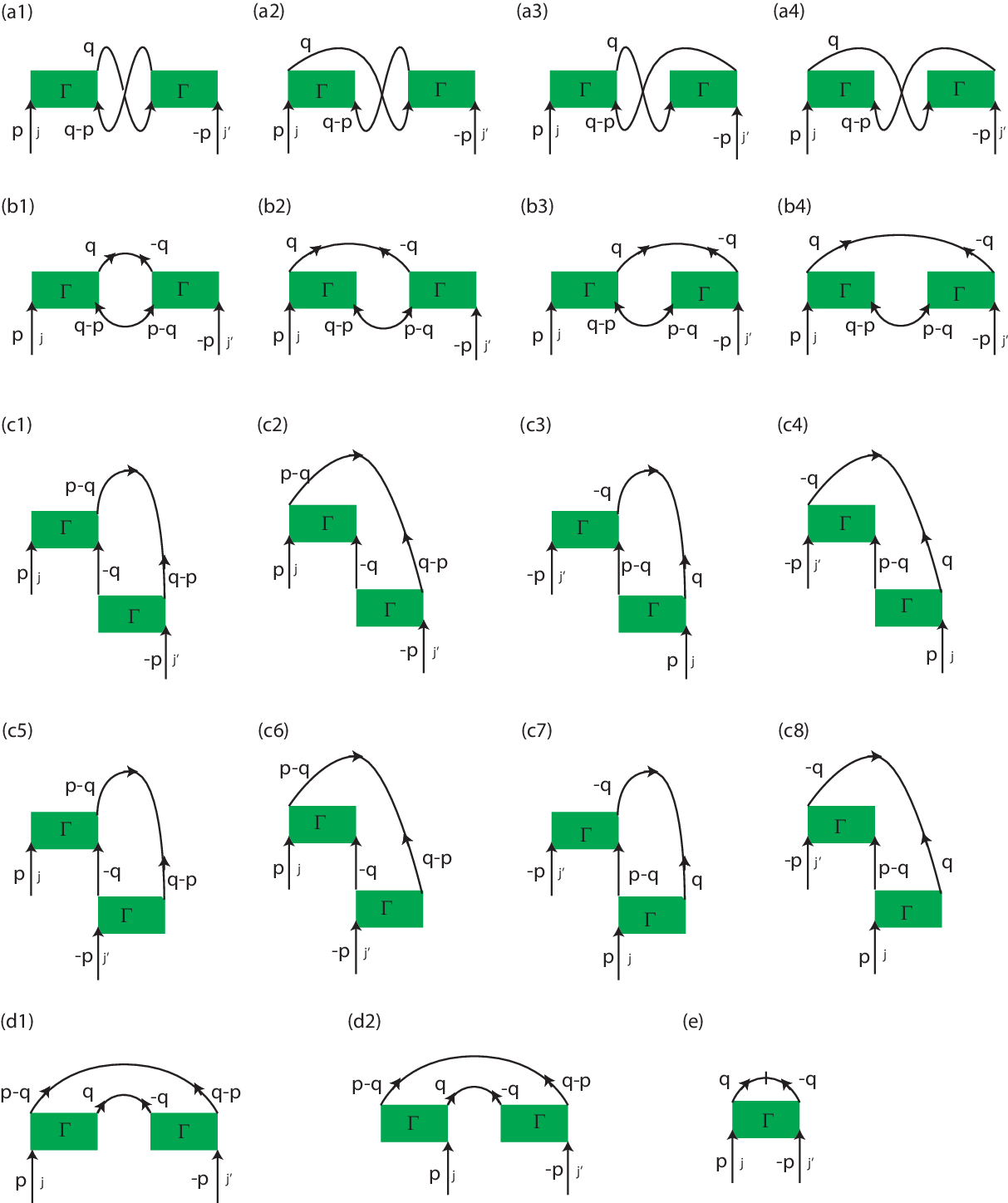}
  \caption{Second-order diagrams for the proper self-energy $\Sigma^{21}_{jj'}(p)$.}
  \label{fig:second_order_Sigma21}
\end{figure}

\begin{figure}[tbp] % float placement: (h)ere, page (t)op, page (b)ottom, other (p)age
  \centering
  % file name: E:/Spinor Beliaev (July 4, 2011)/Figures/second_order_mu.eps
  \includegraphics[width=3in,keepaspectratio]{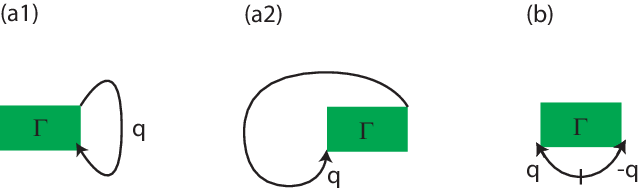}
  \caption{Second-order diagrams for the chemical potential $\mu$.}
  \label{fig:second_order_mu}
\end{figure}

The second-order correction of the proper self-energies and chemical potential involves two terms. One is the second-order correction to $\Gamma_{\mathcal{F}=0,2}(p_1,p_2;p_3,p_4)$ in the first-order diagrams (see Fig.~\ref{fig:first_order_approximation}), that is, the $\Q$-integrals and the imaginary part of $f_{\mathcal{F}=0,2}(\p,\p')$ in Eq.~\eqref{eq: Gamma0, Gamma2}. The other is the contribution from the second-order diagrams given in Figs.~\ref{fig:second_order_Sigma11}-\ref{fig:second_order_mu}.

%****************************
\subsubsection{Ferromagnetic phase}
\label{subsubsection: Ferromagnetic phase, subsection: Second-order self energies}
First, we consider the second-order corrections to the self-energies and chemical potential that result from the correction to the T-matrix in the first-order diagrams. They are obtained by substituting the $\Q$-integrals and the imaginary part of $f_\mathcal{F}(\p,\p')$ in Eq.~\eqref{eq: Gamma0, Gamma2} into the first lines of Eqs.~\eqref{eq: ferro, Sigma11 first-order approximation}-\eqref{eq: ferro, mu first-order apprx} (for more details, see \ref{appendix: Derivation of second-order correction from q-integral for both ferro and polar phases}):
\begin{subequations}
\label{eq: ferro, second-order correction from q-integral}
\begin{align}
\hbar\Sigma^{11}_{jj'}(p):\,&i\,\mathrm{Im}\{c_0(\p/2,\p/2)\}n_0\delta_{jj'}+i\,\mathrm{Im} \{c_0(\p/2,-\p/2)\}n_0\delta_{j,1}\delta_{j',1}\nonumber\\
&+n_0\left(\frac{f_0^2+2f_2^2}{3}\right)\integralQ \Bigg(\frac{1}{\hbar p_0+2[(c_0+c_1)n_0+q_B]-\eps{\Q}-\eps{\K}+i\eta}\nonumber\\
&-\frac{1}{\eps{\p}-\eps{\Q}-\eps{\K}+i\eta}\Bigg)\times(\delta_{jj'}+\delta_{j,1}\delta_{j',1}), \label{eq: ferro, first contribution to second-order Sigma11}\\
\hbar\Sigma^{12,21}_{jj'}(p):\,&\,n_0\left(\frac{f_0^2+2f_2^2}{3}\right)\integralQ \Bigg(\frac{1}{2[(c_0+c_1)n_0+q_B]-2\eps{\Q}+i\eta}+\frac{1}{2\eps{\Q}}\Bigg)\delta_{j,1}\delta_{j',1}, \label{eq: ferro, first contribution to second-order Sigma12}\\
\mu:\,&\,n_0\left(\frac{f_0^2+2f_2^2}{3}\right)\integralQ \Bigg(\frac{1}{2[(c_0+c_1)n_0+q_B]-2\eps{\Q}+i\eta}+\frac{1}{2\eps{\Q}}\Bigg), \label{eq: ferro, first contribution to mu}
\end{align}
\end{subequations}
where $\mathrm{Im}$ denotes the imaginary part of a complex number, $\K\equiv \Q-\p$, and  
\begin{align}
c_0(\p/2,\pm \p/2) \equiv& \frac{\tilde{f}_0(\p/2,\pm \p/2)+2\tilde{f}_2(\p/2,\pm \p/2)}{3}.
\label{eq: c0(p/2,+-p/2)}
\end{align}

Using the optical theorem for scattering, the imaginary part of an on-shell vacuum scattering amplitude $\tilde{f}_\mathcal{F}(\p,\p')$ with $|\p|=|\p'|$ is given by \cite{Fetterbook}
\begin{align}
\mathrm{Im}\{\tilde{f}_\mathcal{F}(\p,\p')\}=&-\frac{\pi M}{\hbar^2}\integralQ \tilde{f}_\mathcal{F}(\p,\Q)\tilde{f}_\mathcal{F}^*(\p',\Q)\delta(\p^2-\Q^2)\nonumber\\
=&\frac{-|\p|M}{16\pi^2\hbar^2}\int d\Omega_\Q\tilde{f}_\mathcal{F}(\p,\Q)\tilde{f}_\mathcal{F}^*(\p',\Q),
\label{eq: imaginary part of on-shell scattering amplitude}
\end{align}
where $\Omega_\Q$ denotes the solid angle of the on-shell momentum $\Q$: $|\Q|=|\p|=|\p'|$. Consequently, the imaginary parts of $\tilde{f}_\mathcal{F}(\p/2,\pm \p/2)$ and $c_0(\p/2,\pm \p/2)$ in Eqs.~\eqref{eq: ferro, second-order correction from q-integral} and \eqref{eq: c0(p/2,+-p/2)} are given in the second-order approximation as follows:
\begin{align}
\mathrm{Im}\{\tilde{f}_\mathcal{F}(\p/2,\pm \p/2)\}=&\frac{-|\p|M}{8\pi\hbar^2}f_\mathcal{F},\\
\mathrm{Im}\{c_0(\p/2,\pm \p/2)\}=&\frac{-|\p|M}{8\pi\hbar^2}\left(\frac{f_0^2+2f_2^2}{3}\right),
\end{align}
where we have replaced $\tilde{f}_\mathcal{F}(\p,\p')$ on the right-hand side of Eq.~\eqref{eq: imaginary part of on-shell scattering amplitude} by its zero-momentum limit $f_\mathcal{F}$.

Next, we calculate the second-order contributions to the proper self-energies and chemical potential from the second-order diagrams illustrated in Figs.~\ref{fig:second_order_Sigma11}-\ref{fig:second_order_mu} by using the first-order Green's functions given in Eq.~\eqref{eq: ferro, first-order Green's functions in convenient form} (for more details, see \ref{appendix: Contribution of second-order diagrams, subsection: Ferrromagnetic phase}). By summing up the second-order corrections that arise from the correction to the T-matrix [Eq.~\eqref{eq: ferro, second-order correction from q-integral}] and the contributions from the second-order diagrams [Eqs.~\eqref{eq: ferro, diagram (a1) for Sigma11}-\eqref{eq: ferro, diagram (i) for Sigma12}], we obtain the second-order self-energies and chemical potential as follows:
\begin{align}
\hbar\Sigma^{11(2)}_{11}(p)=&\,\frac{-i|\p|Mn_0}{4\pi\hbar^2}\left(\frac{f_0^2+2f_2^2}{3}\right)+2n_0\left(\frac{f_0^2+2f_2^2}{3}\right)\integralQ \nonumber\\
&\times\Bigg(\frac{1}{\hbar p_0+2[(c_0+c_1)n_0+q_B]-\eps{\Q}-\eps{\K}+i\eta}-\frac{1}{\eps{\p}-\eps{\Q}-\eps{\K}+i\eta}\Bigg)\nonumber\\
&+n_0c_0^2 \integralQ \Bigg(\frac{2\left\{A_{1,\Q}, B_{1,\K}\right\}+4C_{1,\Q} C_{1,\K}-4\left\{A_{1,\Q},C_{1,\K}\right\}+2A_{1,\Q} A_{1,\K}}{\hbar\left(p_0-\omega_{1,\Q}-\omega_{1,\K}\right)+i\eta}\nonumber\\
&-\frac{2\left\{A_{1,\Q}, B_{1,\K}\right\}+4C_{1,\Q} C_{1,\K}-4\left\{B_{1,\Q},C_{1,\K}\right\}+2B_{1,\Q} B_{1,\K}}{\hbar\left(p_0+\omega_{1,\Q}+\omega_{1,\K}\right)-i\eta}\nonumber\\
&-\frac{2}{\hbar p_0-\eps{\Q}-\eps{\K}+2(c_0+c_1)n_0+i\eta}\Bigg)+2c_0\integralQ B_{1,\Q}, 
\label{eq: ferro, second-order diagrams' contribution, Sigma 11-11}
\end{align}
\begin{align}
\hbar\Sigma^{11(2)}_{00}(p)=&\, \frac{-i|\p|Mn_0}{8\pi\hbar^2}\left(\frac{f_0^2+2f_2^2}{3}\right)+n_0\left(\frac{f_0^2+2f_2^2}{3}\right)\integralQ \nonumber\\
&\times\Bigg(\frac{1}{\hbar p_0+2[(c_0+c_1)n_0+q_B]-\eps{\Q}-\eps{\K}+i\eta}-\frac{1}{\eps{\p}-\eps{\Q}-\eps{\K}+i\eta}\Bigg)\nonumber\\
&+n_0c_0^2 \integralQ \Bigg(\frac{A_{1,\K}+B_{1,\K}-2C_{1,\K}}{\hbar\left(p_0-\omega_{0,\Q}-\omega_{1,\K}\right)+i\eta}\nonumber\\
&-\frac{1}{\hbar p_0-\eps{\Q}-\eps{\K}+2(c_0+c_1)n_0+q_B+i\eta}\Bigg)+c_0\integralQ B_{1,\Q},
\label{eq: ferro, second-order diagrams' contribution, Sigma 11-00}
\end{align}
\begin{align}
\hbar\Sigma^{11(2)}_{-1,-1}(p)=&\,\frac{-i|\p|Mn_0}{8\pi\hbar^2}\left(\frac{f_0^2+2f_2^2}{3}\right)+n_0\left(\frac{f_0^2+2f_2^2}{3}\right)\integralQ \nonumber\\
&\Bigg(\frac{1}{\hbar p_0+2[(c_0+c_1)n_0+q_B]-\eps{\Q}-\eps{\K}+i\eta}-\frac{1}{\eps{\p}-\eps{\Q}-\eps{\K}+i\eta}\Bigg)\nonumber\\
&+n_0c_0^2 \integralQ \Bigg(\frac{A_{1,\K}+B_{1,\K}-2C_{1,\K}}{\hbar \left(p_0-\omega_{-1,\Q}-\omega_{1,\K}\right)+i\eta}\nonumber\\
&-\frac{1}{\hbar p_0-\eps{\Q}-\eps{\K}+2(c_0+c_1)n_0+i\eta}\Bigg)+c_0\integralQ B_{1,\Q},
\label{eq: ferro, second-order diagrams' contribution, Sigma 11,-1-1}
\end{align}
\begin{align}
\hbar\Sigma^{12(2)}_{11}(p)=&\,\hbar\Sigma^{21(2)}_{11}(p)\nonumber\\
=&\,n_0\left(\frac{f_0^2+2f_2^2}{3}\right) \integralQ \Bigg(\frac{1}{2[(c_0+c_1)n_0+q_B]-2\eps{\Q}+i\eta}+\frac{1}{2\eps{\Q}}\Bigg)\nonumber\\
&+n_0c_0^2 \integralQ \Bigg[\Big(2\left\{A_{1,\Q},B_{1,\K}\right\}+6C_{1,\Q} C_{1,\K}-2\left\{A_{1,\Q}+B_{1,\Q},C_{1,\K}\right\}\Big) \nonumber\\
&\times \Bigg(\frac{1}{\hbar\left(p_0-\omega_{1,\Q}-\omega_{1,\K}\right)+i\eta}-\frac{1}{\hbar\left(p_0+\omega_{1,\Q}+\omega_{1,\K}\right)-i\eta}\Bigg)\Bigg]\nonumber\\
&+c_0\integralQ \Bigg( -C_{1,\Q}+\frac{c_0n_0}{2\eps{\Q}-2(c_0+c_1)n_0-i\eta} \Bigg), 
\label{eq: ferro, second-order diagrams' contribution, Sigma 12,11}
\end{align}
\begin{align}
\mu^{(2)}=&\,n_0\left(\frac{f_0^2+2f_2^2}{3}\right)\integralQ \left(\frac{1}{2[(c_0+c_1)n_0+q_B]-2\eps{\Q}+i\eta}+\frac{1}{2\eps{\Q}}\right) \nonumber\\
&+2c_0\integralQ B_{1,\Q}+c_0\integralQ \left( -C_{1,\Q}+\frac{c_0n_0}{2\eps{\Q}-2(c_0+c_1)n_0-i\eta} \right),
\label{eq: ferro, second-order diagrams' contribution, mu}
\end{align}
where $\K\equiv \Q-\p$ and $\left\{A_{1,\Q},B_{1,\K}\right\}\equiv A_{1,\Q} B_{1,\K}+A_{1,\K} B_{1,\Q}$.

Here we consider only the case in which the external magnetic field satisfies $q_B\sim |c_1|n\ll c_0n$ (see Sec.~\ref{subsection: T-matrix}), and ignore terms smaller than $c_0n\sqrt{na^3}$, which is the order of magnitude of the second-order approximation under consideration. Then Eqs.~\eqref{eq: ferro, second-order diagrams' contribution, Sigma 11-11}, \eqref{eq: ferro, second-order diagrams' contribution, Sigma 12,11}, and \eqref{eq: ferro, second-order diagrams' contribution, mu} for $\Sigma^{11(2)}_{11}(p)$, $\Sigma^{12(2)}_{11}(p)$, and $\mu^{(2)}$, respectively, are the same as those for a spinless Bose-Einstein condensate \cite{Beliaev1}. It is because the condensate is in the $j=1$ sublevel, and the elementary excitation given by $\Sigma^{11;12(2)}_{11}(p)$ is the density-wave excitation as in a spinless system. Consequently, it has a phonon-like second-order energy spectrum in the low-momentum regime ($\eps{\p}\ll c_0n$):
\begin{align}
\hbar p_0=&\left(1+\frac{7}{6\pi^2}\sqrt{n_0\tilde{a}^3}\right)\sqrt{2n_0(c_0+c_1)}\sqrt{\eps{\p}}-i\frac{3}{640\pi}n_0c_0\sqrt{n_0\tilde{a}^3}\frac{\left(\eps{\p}\right)^{5/2}}{\left(n_0c_0\right)^{5/2}}\nonumber\\
=&\left(1+\frac{28}{3\sqrt{\pi}}\sqrt{n_0a^3}\right)\sqrt{2n_0(c_0+c_1)}\sqrt{\eps{\p}}-i\frac{3\sqrt{\pi}}{80}n_0c_0\sqrt{n_0a^3}\frac{\left(\eps{\p}\right)^{5/2}}{\left(n_0c_0\right)^{5/2}}\nonumber\\
=&\left(1+\frac{8}{\sqrt{\pi}}\sqrt{na^3}\right)\sqrt{2n(c_0+c_1)}\sqrt{\eps{\p}}-i\frac{3\sqrt{\pi}}{80}nc_0\sqrt{na^3}\frac{\left(\eps{\p}\right)^{5/2}}{\left(nc_0\right)^{5/2}},
\label{eq: ferro, second-order energy spectrum of phonon mode}
\end{align}
where $a$ and $\tilde{a}$ are defined in Eq.~\eqref{eq: define scattering length a}. Here, in the derivation of the last line of Eq.~\eqref{eq: ferro, second-order energy spectrum of phonon mode}, we used the expression for the condensate fraction in a homogeneous system \cite{Pethickbook, Fetterbook}:
\begin{align}
 n_0=n\left(1-\frac{8}{3\sqrt{\pi}}\sqrt{na^3}\right).
\label{eq: condensate fraction}
\end{align}
The first term (real part) on the right-hand side of Eq.~\eqref{eq: ferro, second-order energy spectrum of phonon mode} shows an increase in the sound velocity of a density-wave excitation due to quantum depletion, while the second term (imaginary part) is the so-called Beliaev damping, which shows a finite lifetime of phonons due to their collisions with the condensate. The second-order contribution to the chemical potential is given by~\cite{Beliaev1}
\begin{align}
\mu^{(2)}=\frac{5}{3\pi^2}n_0c_0\sqrt{n_0\tilde{a}^3}.
\label{eq: ferro, mu(2)}
\end{align}

Now, to evaluate $\Sigma^{11(2)}_{00}(p)$ we take a Taylor expansion of it around $p_0=\omega_{0,\p}$, where $\hbar\omega_{0,\p}$ is the first-order energy spectrum given by Eq.~\eqref{eq: ferro, first-order energy spectra, E0p}: 
\begin{align}
\Sigma^{11(2)}_{00}(p)=\Sigma^{11(2)}_{00}(p_0=\omega_{0,\p})+\frac{\partial \Sigma^{11(2)}_{00}(p)}{\partial p_0}\Bigg|_{p_0=\omega_{0,\p}}\left(p_0-\omega_{0,\p}\right)+\mathcal{O}\left[\left(p_0-\omega_{0,\p}\right)^2\right]+\cdots.
\label{eq: a Taylor expansion around p0=omega0p}
\end{align}
We can stop at the linear term in this Taylor expansion, provided that the difference between the second-order energy spectrum and the first-order one is small:
\begin{align}
|p_0-\omega_{0,\p}| \ll \frac{\Sigma^{11(2)}_{00}(p_0=\omega_{0,\p})}{\left[\partial \Sigma^{11(2)}_{00}/\partial p_0\right](p_0=\omega_{0,\p})} \sim \frac{c_0n_0}{\hbar},
\label{eq: condition for a Taylor expansion around p0=omega0p}
\end{align}
which is justified by the fact that the system is a dilute weakly interacting Bose gas, and will be confirmed later by the second-order energy spectrum obtained below in Eq.~\eqref{eq: ferro, second-order energy spectrum G00 (2)}. This will be discussed in more detail at the end of Sec.~\ref{subsubsection: Ferromagnetic phase, subsection: Second-order energy spectra of elementary excitations}.

It can be shown that the imaginary parts of $\Sigma^{11(2)}_{00}(p_0=\omega_{0,\p})$ and $\left[\partial \Sigma^{11(2)}_{00}/\partial p_0\right](p_0=\omega_{0,\p})$ vanish for any value of $\p$ (see~\ref{appendix: Imaginary part of self energies, subsection: Ferrromagnetic phase}), which results in 
\begin{align}
\mathrm{Im}\Sigma^{11(2)}_{00}(p)=0+\mathcal{O}\left[(p_0-\omega_{0,\p})^2\right].
\label{eq: imaginary part, second order Sigma11-00(p)}
\end{align}
This result implies that there is no damping for the elementary excitation given by $\Sigma^{11}_{00}(p)$ up to the order of magnitude under consideration.

For the real parts of $\Sigma^{11(2)}_{00}(p_0=\omega_{0,\p})$ and $\left[\partial \Sigma^{11(2)}_{00}/\partial p_0\right](p_0=\omega_{0,\p})$, we can make their Taylor expansions around $\p=0$ in the low-momentum regime $\eps{\p}\ll c_0n_0$:
\begin{subequations}
\label{eq: Taylor expansion in low momentum region}
\begin{align}
\mathrm{Re}\Sigma^{11(2)}_{00}(p_0=\omega_{0,\p})=&\mathrm{Re}\Sigma^{11(2)}_{00}(p_0=\omega_{0,\p})\Big|_{\p=0}+\frac{\partial \mathrm{Re}\Sigma^{11(2)}_{00}(p_0=\omega_{0,\p})}{\partial \omega^1_\p}\Bigg|_{\p=0}\times \omega_{1,\p}\nonumber\\
&+\frac{1}{2}\frac{\partial^2 \mathrm{Re}\Sigma^{11(2)}_{00}(p_0=\omega_{0,\p})}{\partial (\omega_{1,\p})^2}\Bigg|_{\p=0}\times\omega_{1,\p}^2+\cdots,\\
\frac{\partial \mathrm{Re}\Sigma^{11(2)}_{00}(p)}{\partial p_0}\Bigg|_{p_0=\omega_{0,\p}}=&\frac{\partial \mathrm{Re}\Sigma^{11(2)}_{00}(p)}{\partial p_0}\Bigg|_{p_0=\omega_{0,\p},\p=0}+\frac{\partial}{\partial \omega_{1,\p}}\Bigg(\frac{\partial \mathrm{Re}\Sigma^{11(2)}_{00}(p)}{\partial p_0}\Bigg|_{p_0=\omega_{0,\p}}\Bigg)\Bigg|_{\p=0}\nonumber\\
&\times E_{1,\p} +\frac{1}{2} \frac{\partial^2}{\partial (\omega_{1,\p})^2}\Bigg(\frac{\partial \mathrm{Re}\Sigma^{11(2)}_{00}(p)}{\partial p_0}\Bigg|_{p_0=\omega_{0,\p}}\Bigg)\Bigg|_{\p=0} \times \omega_{1,\p}^2\nonumber\\
&+\cdots,
\end{align}
\end{subequations}
where $\omega_{1,\p}$ is given in Eq.~\eqref{eq: ferro, first-order energy spectra, E1p}. Note that $\p={\bf 0}$ is equivalent to $\omega_{1,\p}=0$, and $\eps{\p}\ll c_0n_0$ is equivalent to $\hbar\omega_{1,\p} \ll c_0n_0$.

With straightforward calculations, we obtain (see~\ref{appendix: Real part of self energies, subsection: Ferrromagnetic phase} for details):
\begin{align}
\hbar \mathrm{Re}\Sigma^{11(2)}_{00}(p_0=\omega_{0,\p})\Big|_{\p=0}=&\,\frac{5}{3\pi^2}n_0c_0\sqrt{n_0\tilde{a}^3},\\
\frac{\partial \mathrm{Re}\Sigma^{11(2)}_{00}(p_0=\omega_{0,\p})}{\partial \omega_{1,\p}}\Bigg|_{\p=0}=&\,0,\\
\frac{1}{\hbar}\frac{\partial^2 \mathrm{Re}\Sigma^{11(2)}_{00}(p_0=\omega_{0,\p})}{\partial (\omega_{1,\p})^2}\Bigg|_{\p=0}=&\,- \,\frac{49}{360\pi^2}\sqrt{n_0\tilde{a}^3}\frac{1}{n_0c_0},\\
\frac{\partial \mathrm{Re}\Sigma^{11(2)}_{00}(p)}{\partial p_0}\Bigg|_{p_0=\omega_{0,\p},\p=0}=&\,-\,\frac{1}{3\pi^2}\sqrt{n_0\tilde{a}^3},\\
\frac{\partial}{\partial \omega_{1,\p}}\Bigg(\frac{\partial \mathrm{Re}\Sigma^{11(2)}_{00}(p)}{\partial p_0}\Bigg|_{p_0=\omega_{0,\p}}\Bigg)\Bigg|_{\p=0}=&\,0, \\
\frac{1}{\hbar^2}\frac{\partial^2}{\partial (\omega_{1,\p})^2}\Bigg(\frac{\partial \mathrm{Re}\Sigma^{11(2)}_{00}(p)}{\partial p_0}\Bigg|_{p_0=\omega_{0,\p}}\Bigg)\Bigg|_{\p=0}=&\,-\,\frac{13}{60\pi^2}\sqrt{n_0\tilde{a}^3}\frac{1}{(n_0c_0)^2}.
\end{align}
Hence, we obtain:
\begin{align}
\hbar \mathrm{Re}\Sigma^{11(2)}_{00}(p)=&\frac{5}{3\pi^2}\sqrt{n_0\tilde{a}^3}\Bigg[1-\frac{49}{1200}\left(\frac{\hbar\omega_{1,\p}}{n_0c_0}\right)^2\Bigg]n_0c_0\nonumber\\
&-\,\frac{1}{3\pi^2}\sqrt{n_0\tilde{a}^3}\Bigg[1+\frac{13}{40}\left(\frac{\hbar\omega_{1,\p}}{n_0c_0}\right)^2\Bigg]\hbar\left(p_0-\omega_{0,\p}\right)\nonumber\\
&+\mathcal{O}\left[(p_0-\omega_{0,\p})^2\right].
\label{eq: real part, second order Sigma11-00}
\end{align}
Equation~\eqref{eq: real part, second order Sigma11-00} shows the modification of the self-energy $\Sigma^{11}_{00}(p)$ due to the effect of quantum depletion. The first term in the first line is the value for $\p={\bf 0}$, $p_0=0$, the second term in the first line is the correction for a nonzero momentum, while the second line is the correction for a nonzero energy. It can be seen that the self-energy $\Sigma^{11}_{00}(p)$, which describe the effect of interaction with other particles on the propagation of a quasiparticle, decreases with increasing momentum or frequency. 

%****************************
\subsubsection{Polar phase}
\label{subsubsection: Polar phase, subsection: Second-order self energies}
Following a procedure similar to the ferromagnetic case, the second-order corrections to the self-energies and chemical potential that result from the correction to the T-matrix in the first-order diagrams are given by
\begin{subequations}
\label{eq: polar, second-order correction from q-integral}
\begin{align}
\hbar\Sigma^{11}_{jj'}(p):\,&i\,\mathrm{Im}\{c_0(\p/2,\p/2)\}n_0\delta_{jj'}+i\,\mathrm{Im} \{c_0(\p/2,-\p/2)\}n_0\delta_{j,0}\delta_{j',0}\nonumber\\
&+n_0\left(\frac{f_0^2+2f_2^2}{3}\right)\integralQ \Bigg(\frac{1}{\hbar p_0+2c_0n_0-\eps{\Q}-\eps{\K}+i\eta}\nonumber\\
&-\frac{1}{\eps{\p}-\eps{\Q}-\eps{\K}+i\eta}\Bigg)\times(\delta_{jj'}+\delta_{j,0}\delta_{j',0}), \label{eq: polar, first contribution to second-order Sigma11}\\
\hbar\Sigma^{12,21}_{jj'}(p):\,&\,n_0\left(\frac{f_0^2+2f_2^2}{3}\right)\integralQ \Bigg(\frac{1}{2c_0n_0-2\eps{\Q}+i\eta}+\frac{1}{2\eps{\Q}}\Bigg)\delta_{j,0}\delta_{j',0}, \label{eq: polar, first contribution to second-order Sigma12}\\
\mu:\,&\,n_0\left(\frac{f_0^2+2f_2^2}{3}\right)\integralQ \Bigg(\frac{1}{2c_0n_0-2\eps{\Q}+i\eta}+\frac{1}{2\eps{\Q}}\Bigg). \label{eq: polar, first contribution to mu}
\end{align}
\end{subequations}
By summing up the second-order corrections that arise from the correction to the T-matrix [Eq.~\eqref{eq: polar, second-order correction from q-integral}] and the contributions from the second-order diagrams (see~\ref{appendix: Contribution of second-order diagrams, subsection: Polar phase}), we obtain the second-order self-energies and chemical potential as follows:
\begin{align}
\hbar\Sigma^{11(2)}_{11}(p)=&\,\hbar\Sigma^{11(2)}_{-1,-1}(p)\nonumber\\
=&\, \frac{-i|\p|Mn_0}{8\pi\hbar^2}\left(\frac{f_0^2+2f_2^2}{3}\right)+n_0\left(\frac{f_0^2+2f_2^2}{3}\right)\integralQ \nonumber\\
&\times\Bigg(\frac{1}{\hbar p_0+2c_0n_0-\eps{\Q}-\eps{\K}+i\eta}-\frac{1}{\eps{\p}-\eps{\Q}-\eps{\K}+i\eta}\Bigg)+n_0c_0^2 \integralQ\nonumber\\
&\times  \Bigg[\left(A_{0,\K}+ B_{0,\K}-2C_{0,\K}\right)\Bigg(\frac{A_{1,\Q}}{\hbar\left(p_0-\omega_{1,\Q}-\omega_{0,\K}\right)+i\eta}\nonumber\\
&-\frac{B_{1,\Q}}{\hbar\left(p_0+\omega_{1,\Q}+\omega_{0,\K}\right)-i\eta}\Bigg)-\frac{1}{\hbar p_0-\eps{\Q}-\eps{\K}+2c_0n_0-q_B+i\eta}\Bigg]\nonumber\\
&+c_0\integralQ \left( 3B_{1,\Q}+B_{0,\Q} \right),
\label{eq: polar, Sigma11(2) 11}
\end{align}
\begin{align}
\hbar\Sigma^{11(2)}_{00}(p)=&\, \frac{-i|\p|Mn_0}{4\pi\hbar^2}\left(\frac{f_0^2+2f_2^2}{3}\right)+2n_0\left(\frac{f_0^2+2f_2^2}{3}\right)\integralQ \nonumber\\
&\times\Bigg(\frac{1}{\hbar p_0+2c_0n_0-\eps{\Q}-\eps{\K}+i\eta}-\frac{1}{\eps{\p}-\eps{\Q}-\eps{\K}+i\eta}\Bigg)\nonumber\\
&+n_0c_0^2 \integralQ \Bigg[\frac{2\left\{A_{0,\Q}, B_{0,\K}\right\}+4C_{0,\Q} C_{0,\K}-4\left\{A_{0,\Q},C_{0,\K}\right\}+2A_{0,\Q} A_{0,\K}}{\hbar\left(p_0-\omega_{0,\Q}-\omega_{0,\K}\right)+i\eta}\nonumber\\
&-\frac{2\left\{A_{0,\Q}, B_{0,\K}\right\}+4C_{0,\Q} C_{0,\K}-4\left\{B_{0,\Q},C_{0,\K}\right\}+2B_{0,\Q} B_{0,\K}}{\hbar\left(p_0+\omega_{0,\Q}+\omega_{0,\K}\right)-i\eta}\nonumber\\
&-\frac{2}{\hbar p_0-\eps{\Q}-\eps{\K}+2c_0n_0+i\eta}+\Big(\left\{A_{1,\Q}, B_{1,\K}\right\}+2C_{1,\Q} C_{1,\K}\Big)\nonumber\\
&\times\Bigg(\frac{1}{\hbar\left(p_0-\omega_{1,\Q}-\omega_{1,\K}\right)+i\eta}-\frac{1}{\hbar\left(p_0+\omega_{1,\Q}+\omega_{1,\K}\right)-i\eta}\Bigg)\Bigg]\nonumber\\
&+c_0\integralQ \left( 2B_{0,\Q}+2B_{1,\Q} \right),
\label{eq: Sigma11(2) 00}
\end{align}
\begin{align}
\hbar\Sigma^{12(2)}_{1,-1}(p)=&\,\hbar\Sigma^{12(2)}_{-1,1}(p)=\hbar\Sigma^{21(2)}_{1,-1}(p)=\hbar\Sigma^{21(2)}_{-1,1}(p)\nonumber\\
=&\,n_0c_0^2 \integralQ C_{1,\Q} \left(2C_{0,\K}-A_{0,\K}-B_{0,\K}\right) \nonumber\\
&\times \Bigg(\frac{1}{\hbar\left(p_0-\omega_{1,\Q}-\omega_{0,\K}\right)+i\eta}-\frac{1}{\hbar\left(p_0+\omega_{1,\Q}+\omega_{0,\K}\right)-i\eta}\Bigg),
\label{eq: Sigma12,1,-1(2)}
\end{align}
\begin{align}
\hbar\Sigma^{12(2)}_{00}(p)=&\,n_0\left(\frac{f_0^2+2f_2^2}{3}\right) \integralQ \Bigg(\frac{1}{2c_0n_0-2\eps{\Q}+i\eta}+\frac{1}{2\eps{\Q}}\Bigg)\nonumber\\
&+n_0c_0^2 \integralQ \Bigg[\Big(2\left\{A_{0,\Q},B_{0,\K}\right\}+6C_{0,\Q} C_{0,\K}-2\left\{A_{0,\Q}+B_{0,\Q},C_{0,\K}\right\}\Big) \nonumber\\
&\times\Bigg(\frac{1}{\hbar\left(p_0-\omega_{0,\Q}-\omega_{0,\K}\right)+i\eta}-\frac{1}{\hbar\left(p_0+\omega_{0,\Q}+\omega_{0,\K}\right)-i\eta}\Bigg)\nonumber\\
&+\Big(\left\{A_{1,\Q},B_{1,\K}\right\}+2C_{1,\Q} C_{1,\K} \Big)\Bigg(\frac{1}{\hbar\left(p_0-\omega_{1,\Q}-\omega_{1,\K}\right)+i\eta}\nonumber\\
&-\frac{1}{\hbar\left(p_0+\omega_{1,\Q}+\omega_{1,\K}\right)-i\eta}\Bigg)\Bigg]+c_0\integralQ \Bigg( -C_{0,\Q}+\frac{c_0n_0}{2\eps{\Q}-2c_0n_0-i\eta} \Bigg),
\label{eq: Sigma12,00(2)}
\end{align}
\begin{align}
\mu^{(2)}=&\,n_0\left(\frac{f_0^2+2f_2^2}{3}\right)\integralQ \Bigg(\frac{1}{2c_0n_0-2\eps{\Q}+i\eta}+\frac{1}{2\eps{\Q}}\Bigg)+2c_0\integralQ \left( B_{0,\Q}+B_{1,\Q} \right)\nonumber\\
&+c_0\integralQ \left( -C_{0,\Q}-\frac{c_0n_0}{2c_0n_0-2\eps{\Q}+i\eta} \right),
\label{eq: mu(2)}
\end{align} 
It can be seen that the integrand of the $\Q$-integral in each of Eqs.~\eqref{eq: Sigma11(2) 00}, \eqref{eq: Sigma12,00(2)}, and \eqref{eq: mu(2)} is a sum of a term that contain only $A_{0,\Q;\K}$, $B_{0,\Q;\K}$, $C_{0,\Q;\K}$, $\omega_{0,\Q;\K}$, $c_0n_0$ and a term that contain only $A_{1,\Q;\K}$, $B_{1,\Q;\K}$, $C_{1,\Q;\K}$, $\omega_{1,\Q;\K}, c_1n$. By rewritting the corresponding $\Q$-integrals using dimensionless variables $\eps{\Q}/(c_0n_0)$ and $\eps{\Q}/(|c_1|n_0)$, we find that the value of the latter integral is smaller than that of the former one by a factor of $\sqrt{|c_1|/c_0}\ll 1$, and thus, the latter integral can be ignored. Here, we used $\hbar p_0\simeq \hbar\omega_{0,\p} \ll \sqrt{|c_1|n_0c_0n_0} \ll c_0n_0$ for the low-momentum region $\eps{\p}\ll |c_1|n$ under consideration for the case of the polar phase. Consequently, $\Sigma^{11,12(2)}_{00}(p)$ and $\mu^{(2)}$ are the same as the second-order self-energies and chemical potential of a spinless Bose-Einstein condensate~\cite{Beliaev1}. Namely, the second-order contribution to the chemical potential is given by
\begin{align}
\mu^{(2)}=\frac{5}{3\pi^2}n_0c_0\sqrt{n_0\tilde{a}^3}.
\label{eq: polar, mu(2)}
\end{align}
Here, the elementary excitation given by $\Sigma^{11;12(2)}_{00}(p)$ is a density-wave excitation as in a spinless system. It, therefore, has a phonon-like second-order energy spectrum in the low-momentum regime:
\begin{align}
\hbar p_0=\left(1+\frac{8}{\sqrt{\pi}}\sqrt{na^3}\right)\sqrt{2nc_0}\sqrt{\eps{\p}}-i\frac{3\sqrt{\pi}}{80}nc_0\sqrt{na^3}\frac{\left(\eps{\p}\right)^{5/2}}{\left(nc_0\right)^{5/2}}.
\label{eq: polar, second-order energy spectrum of phonon mode}
\end{align}

On the other hand, in Eq.~\eqref{eq: Sigma12,1,-1(2)} for $\Sigma^{12(2)}_{1,-1}$, the factor $c_1n_0$, which arises from $C_{1,\Q}$, can be taken out of the integral, and thus, $\Sigma^{12(2)}_{1,-1}$ is negligibly small compared to the order of magnitude under consideration:
\begin{align}
\Sigma^{12(2)}_{1,-1}(p)=&\,\Sigma^{12(2)}_{-1,1}(p)=\Sigma^{21(2)}_{1,-1}(p)=\Sigma^{21(2)}_{-1,1}(p)\nonumber\\
=&\,0+\mathcal{O}\left[|c_1|n_0\sqrt{n_0c_0^3}\right].
\label{eq: polar, second-order Sigma12 1,-1=0}
\end{align}

Now, to calculate $\Sigma^{11(2)}_{11}(p)$ we make its Taylor expansion around $p_0=\omega_{1,\p}$, where $\hbar\omega_{1,\p}$ is the first-order energy spectrum given by Eq.~\eqref{eq: first-order energy spectrum E1p}: 
\begin{align}
\Sigma^{11(2)}_{11}(p)=\Sigma^{11(2)}_{11}(p_0=\omega_{1,\p})+\frac{\partial \Sigma^{11(2)}_{11}(p)}{\partial p_0}\Bigg|_{p_0=\omega_{1,\p}}\left(p_0-\omega_{1,\p}\right)+\mathcal{O}\left[\left(p_0-\omega_{1,\p}\right)^2\right]+\cdots.
\label{eq: polar, Taylor expansion around p0=omega 1,p}
\end{align}
We can stop at the linear term in this Taylor expansion, provided that the difference between the second-order energy spectrum and the first-order one is small:
\begin{align}
p_0-\omega_{1,\p} \ll \frac{\Sigma^{11(2)}_{11}(p_0=\omega_{1,\p})}{\left[\partial \Sigma^{11(2)}_{11}/\partial p_0\right](p_0=\omega_{1,\p})} \sim \frac{c_0n_0}{\hbar},
\end{align}
which is justified by the fact that the system is a dilute weakly interacting Bose gas, and will be confirmed later by the second-order energy spectrum obtained below in Eq.~\eqref{eq: energy spectrum at second-order approximation}.

It can be shown that the imaginary parts of $\Sigma^{11(2)}_{11}(p_0=\omega_{1,\p})$ and $\left[\partial \Sigma^{11(2)}_{11}/\partial p_0\right](p_0=\omega_{1,\p})$ vanish for any value of $\p$ (see~\ref{appendix: Imaginary part of self energies, subsection: Polar phase}), resulting in 
\begin{align}
\mathrm{Im}\Sigma^{11(2)}_{11}(p)=0+\mathcal{O}\left[(p_0-\omega_{1,\p})^2\right].
\label{eq: polar, imaginary part of second order Sigma11-11(p)}
\end{align}
Physically, this implies that there is no damping for this elementary excitation up to the order of magnitude under consideration.

For the real parts of $\Sigma^{11(2)}_{11}(p_0=\omega_{1,\p})$ and $\left[\partial \Sigma^{11(2)}_{11}/\partial p_0\right](p_0=\omega_{1,\p})$, we can make their Taylor expansions around $\p={\bf 0}$ in the low-momentum regime $\eps{\p}\ll |c_1|n_0 \ll c_0n_0$:
\begin{subequations}
\label{eq: polar, Taylor expansion around p=0}
\begin{align}
\mathrm{Re}\Sigma^{11(2)}_{11}(p)\Big|_{p_0=\omega_{1,\p}}=&\mathrm{Re}\Sigma^{11(2)}_{11}(p_0=\omega_{1,\p})\Big|_{\p=0}+\frac{\partial \mathrm{Re}\Sigma^{11(2)}_{11}(p_0=\omega_{1,\p})}{\partial \omega_{0,\p}}\Bigg|_{\p=0}\times \omega_{0,\p}\nonumber\\
&+\frac{1}{2}\frac{\partial^2 \mathrm{Re}\Sigma^{11(2)}_{11}(p_0=\omega_{1,\p})}{\partial (\omega_{0,\p})^2}\Bigg|_{\p=0}\times\omega_{0,\p}^2+\cdots,\\
\frac{\partial \mathrm{Re}\Sigma^{11(2)}_{11}(p)}{\partial p_0}\Bigg|_{p_0=\omega_{1,\p}}=&\frac{\partial \mathrm{Re}\Sigma^{11(2)}_{11}(p)}{\partial p_0}\Bigg|_{p_0=\omega_{1,\p},\p=0}+\frac{\partial}{\partial \omega_{0,\p}}\Bigg(\frac{\partial \mathrm{Re}\Sigma^{11(2)}_{11}(p)}{\partial p_0}\Bigg|_{p_0=\omega_{1,\p}}\Bigg)\Bigg|_{\p=0}\nonumber\\
& \times \omega_{0,\p}+\frac{1}{2} \frac{\partial^2}{\partial (\omega_{0,\p})^2}\Bigg(\frac{\partial \mathrm{Re}\Sigma^{11(2)}_{11}(p)}{\partial p_0}\Bigg|_{p_0=\omega_{1,\p}}\Bigg)\Bigg|_{\p=0} \times \omega_{0,\p}^2\nonumber\\
&+\cdots,
\end{align}
\end{subequations}
where $\omega_{0,\p}$ is given in Eq.~\eqref{eq: first-order energy spectrum E0p}. Note that $\p={\bf 0}$ is equivalent to $\omega_{0,\p}=0$, and $\eps{\p}\ll c_0n_0$ is equivalent to $\hbar\omega_{0,\p} \ll c_0n_0$.

With straightforward calculations, we obtain (see~\ref{appendix: Real part of self energies, subsection: Polar phase} for details):
\begin{align}
\hbar\mathrm{Re}\Sigma^{11(2)}_{11}(p_0=\omega_{1,\p})\Big|_{\p=0}=&\,\frac{5}{3\pi^2}n_0c_0\sqrt{n_0\tilde{a}^3},\\
\frac{\partial \mathrm{Re}\Sigma^{11(2)}_{11}(p_0=\omega_{1,\p})}{\partial \omega_{0,\p}}\Bigg|_{\p=0}=&\,0,\\
\frac{1}{\hbar}\frac{\partial^2 \mathrm{Re}\Sigma^{11(2)}_{11}(p_0=\omega_{1,\p})}{\partial (\omega_{0,\p})^2}\Bigg|_{\p=0}=&\,\left(- \,\frac{1}{3\pi^2}\frac{q_B+c_1n_0}{\sqrt{q_B(q_B+2c_1n_0)}}+\frac{71}{360\pi^2}\right)\nonumber\\
&\times \sqrt{n_0\tilde{a}^3}\frac{1}{n_0c_0},\\
\frac{\partial \mathrm{Re}\Sigma^{11(2)}_{11}(p)}{\partial p_0}\Bigg|_{p_0=\omega_{1,\p},\p=0}=&\,-\,\frac{1}{3\pi^2}\sqrt{n_0\tilde{a}^3},\\
\frac{\partial}{\partial \omega_{0,\p}}\Bigg(\frac{\partial \mathrm{Re}\Sigma^{11(2)}_{11}(p)}{\partial p_0}\Bigg|_{p_0=\omega_{1,\p}}\Bigg)\Bigg|_{\p=0}=&\,0, \\
\frac{1}{\hbar^2}\frac{\partial^2}{\partial (\omega_{0,\p})^2}\Bigg(\frac{\partial \mathrm{Re}\Sigma^{11(2)}_{11}(p)}{\partial p_0}\Bigg|_{p_0=\omega_{1,\p}}\Bigg)\Bigg|_{\p=0}=&\,\left(- \,\frac{1}{3\pi^2}\frac{q_B+c_1n_0}{\sqrt{q_B(q_B+2c_1n_0)}}+\frac{7}{60\pi^2}\right)\nonumber\\
&\times \sqrt{n_0\tilde{a}^3}\frac{1}{(n_0c_0)^2}.
\end{align}
Therefore, we have
\begin{align}
\hbar\mathrm{Re}\Sigma^{11(2)}_{11}(p)=&\,\frac{5}{3\pi^2}n_0c_0\sqrt{n_0\tilde{a}^3}\Bigg[1+\left(-\,\frac{1}{10}\frac{q_B+c_1n_0}{\sqrt{q_B(q_B+2c_1n_0)}}+\frac{71}{1200}\right) \left(\frac{\hbar\omega_{0,\p}}{n_0c_0}\right)^2\Bigg]\nonumber\\
&-\frac{1}{3\pi^2}\sqrt{n_0\tilde{a}^3}\Bigg[1+\left(\frac{1}{2}\frac{q_B+c_1n_0}{\sqrt{q_B(q_B+2c_1n_0)}}-\frac{7}{40}\right) \left(\frac{\hbar\omega_{0,\p}}{n_0c_0}\right)^2\Bigg]\hbar\left(p_0-\omega_{1,\p}\right)\nonumber\\
&+\mathcal{O}\left[(p_0-\omega_{1,\p})^2\right].
\label{eq: real part, second order Sigma11-11, polar}
\end{align}
Equation~\eqref{eq: real part, second order Sigma11-11, polar} shows the modification of the self-energy $\Sigma^{11}_{11}(p)$ due to the effect of quantum depletion. The first term in the first line is the value for $\p={\bf 0}$, $p_0=0$, the second term in the first line is the correction for a nonzero momentum, while the second line is the correction for a nonzero frequency. It can be seen that because $(q_B+c_1n_0)/\sqrt{q_B(q_B+2c_1n_0)}\geqslant 1$ for any $q_B \geqslant 2|c_1|n_0$, the self-energy $\Sigma^{11}_{11}(p)$, which describe the effect of interaction with other particles on the propagation of a quasiparticle, decreases for increasing momentum or frequency, regardless of the strength of the external magnetic field. Similarly, we have
\begin{align}
\hbar\mathrm{Re}\Sigma^{22(2)}_{11}(p)\equiv&\,\hbar\mathrm{Re}\Sigma^{11(2)}_{11}(-p)\nonumber\\
=&\,\frac{5}{3\pi^2}n_0c_0\sqrt{n_0\tilde{a}^3}\Bigg[1+\left(\frac{1}{10}\frac{q_B+c_1n_0}{\sqrt{q_B(q_B+2c_1n_0)}}+\frac{71}{1200}\right) \left(\frac{\hbar\omega_{0,\p}}{n_0c_0}\right)^2\Bigg]\nonumber\\
&+\frac{1}{3\pi^2}\sqrt{n_0\tilde{a}^3}\Bigg[1+\left(\frac{1}{2}\frac{q_B+c_1n_0}{\sqrt{q_B(q_B+2c_1n_0)}}-\frac{7}{40}\right) \left(\frac{\hbar\omega_{0,\p}}{n_0c_0}\right)^2\Bigg]\hbar\left(p_0-\omega_{1,\p}\right)\nonumber\\
&+\mathcal{O}\left[(p_0-\omega_{1,\p})^2\right].
\label{eq: real part, second order Sigma22-11, polar}
\end{align}

%&&&&&&&&&&&&&&&&&&&&&&&&&&&&
\subsection{Second-order energy spectra of elementary excitations}
\label{subsection: Second-order energy spectra of elementary excitations}
With the second-order self energies and chemical potential obtained in Sec.~\ref{subsection: Second-order self energies}, we are now in a position to evaluate the second-order energy spectra of elementary excitations, which can be obtained from the poles of the second-order Green's functions. As shown in Sec.~\ref{subsection: Second-order self energies}, there is always one density-wave elementary excitation, which is given by $G^{11;12}_{11}$ and $G^{11;12}_{00}$ for the ferromagnetic and polar phases, respectively. It has a linear dispersion relation as the phonon mode in spinless BECs [see Eqs.~\eqref{eq: ferro, second-order energy spectrum of phonon mode} and \eqref{eq: polar, second-order energy spectrum of phonon mode}]. As a consequence of quantum depletion, the sound velocity increases by a universal factor of $1+(8/\sqrt{\pi})\sqrt{na^3}$, while there appears the so-called Beliaev damping due to the collisons between quasiparticles in the elementary excitation and the condensate. The second-order energy spectra of the other elementary excitations will be discussed in the following.

%****************************
\subsubsection{Ferromagnetic phase}
\label{subsubsection: Ferromagnetic phase, subsection: Second-order energy spectra of elementary excitations}
The poles of the Green's functions $G_{00}(p)$ and $G_{-1,-1}(p)$ given by Eqs.~\eqref{eq: ferro, solution to the Dyson equation} and \eqref{eq: ferro, denominator of Green's functions} are the solutions of the following equations:
\begin{subequations}
\label{eq: ferro, poles of G00 and G-1,-1}
\begin{align}
G_{00}(p):\, \hbar p_0=&\eps{\p}-\mu+\hbar\Sigma^{11}_{00}(p)\nonumber\\
=&\eps{\p}-q_B+\left[\hbar\Sigma^{11(2)}_{00}(p)-\mu^{(2)}\right]\nonumber\\
=&\hbar\omega_{0,\p}+\left[\hbar\Sigma^{11(2)}_{00}(p)-\mu^{(2)}\right], \label{eq: ferro, pole of G00}\\
G_{-1,-1}(p)\,: \hbar p_0=&\eps{\p}-\mu+q_B+\hbar\Sigma^{11}_{-1,-1}(p)\nonumber\\
=&\eps{\p}-2c_1n_0+\left[\hbar\Sigma^{11(2)}_{-1,-1}(p)-\mu^{(2)}\right]\nonumber\\
=&\hbar\omega_{-1,\p}+\left[\hbar\Sigma^{11(2)}_{-1,-1}(p)-\mu^{(2)}\right].\label{eq: ferro, pole of G-1,-1}
\end{align}
\end{subequations}
Here, we used the first-order self-energies and chemical potential, which are given in Eq.~\eqref{eq: ferro, first-order proper self energies}, and the first-order energy spectra $\hbar\omega_{0;-1,\p}$ given in Eq.~\eqref{eq: ferro, first-order energy spectra}. Note that on the right-hand sides of Eqs.~\eqref{eq: ferro, pole of G00} and \eqref{eq: ferro, pole of G-1,-1}, the self energies are functions of both $p_0$ and $\p$.

By substituting Eqs.~\eqref{eq: imaginary part, second order Sigma11-00(p)} and \eqref{eq: real part, second order Sigma11-00} for $\Sigma^{11(2)}_{00}(p)$ and Eq.~\eqref{eq: ferro, mu(2)} for the chemical potential $\mu^{(2)}$ into Eq.~\eqref{eq: ferro, pole of G00}, the equation for the pole of $G_{00}(p)$ becomes
\begin{align}
p_0=\omega_{0,\p}+\alpha_\p+\beta_\p\left(p_0-\omega_{0,\p}\right),
\label{eq: ferro, equation for the pole of G00}
\end{align}
where $\alpha_\p$, $\beta_\p$ are the lowest-order coefficients in the Taylor expansion around $p_0=\omega_{0,\p}$ and given by 
\begin{align}
\hbar\alpha_\p=&\,-\frac{49}{720\pi^2}n_0c_0\sqrt{n_0\tilde{a}^3}\left(\frac{\hbar\omega_{1,\p}}{n_0c_0}\right)^2,\\
\beta_\p=&\,-\frac{1}{3\pi^2}\sqrt{n_0\tilde{a}^3}-\frac{13}{120\pi^2}\sqrt{n_0\tilde{a}^3}\left(\frac{\hbar\omega_{1,\p}}{n_0c_0}\right)^2.
\end{align}
Using the fact that $\beta_\p \sim \sqrt{n_0a^3} \ll 1$, the solution to Eq.~\eqref{eq: ferro, equation for the pole of G00} is given by
\begin{align}
p_0=&\,\omega_{0,\p}+\frac{\alpha_\p}{1-\beta_\p}\nonumber\\
\simeq&\,\omega_{0,\p}+\alpha_\p\nonumber\\
=&\,\omega_{0,\p}-\frac{1}{\hbar}\frac{49}{720\pi^2}n_0c_0\sqrt{n_0\tilde{a}^3}\left(\frac{\hbar\omega_{1,\p}}{n_0c_0}\right)^2.
\label{eq: ferro, second-order energy spectrum G00 (1)}
\end{align}
In the low-momentum region $\eps{\p} \ll c_0n_0$,  $\hbar\omega_{1,\p}$ given by Eq.~\eqref{eq: ferro, first-order energy spectra, E1p} can be approximated by $\hbar\omega_{1,\p} \simeq \sqrt{2(c_0+c_1)n_0\eps{\p}} $. Substituting this and Eq.~\eqref{eq: ferro, first-order energy spectra, E0p} into Eq.~\eqref{eq: ferro, second-order energy spectrum G00 (1)}, and neglecting all the terms that are smaller than the order of magnitude under consideration, we obtain the energy spectrum:
\begin{align}
\hbar p_0=&\eps{\p}-q_B-\frac{49}{720\pi^2}\frac{\sqrt{n_0\tilde{a}^3}}{n_0c_0}\times 2c_0n_0\eps{\p}\nonumber\\
=&\left(1-\frac{49}{360\pi^2}\sqrt{n_0\tilde{a}^3}\right)\eps{\p}-q_B \nonumber\\
=&\left(1-\frac{49}{45\sqrt{\pi}}\sqrt{n_0a^3}\right)\eps{\p}-q_B \nonumber\\
\simeq&\left(1-\frac{49}{45\sqrt{\pi}}\sqrt{na^3}\right)\eps{\p}-q_B.
\label{eq: ferro, second-order energy spectrum G00 (2)}
\end{align}
Here, we used Eqs.~\eqref{eq: define scattering length a} and \eqref{eq: condensate fraction} in deriving the last two equalities. Equation~\eqref{eq: ferro, second-order energy spectrum G00 (2)} shows that the energy spectrum of the elementary excitation with spin state $j=0$ has a quadratic dispersion relation at low momenta, which can be expressed using an effective mass $M_\mathrm{eff}$ as 
\begin{align}
\hbar p_0=\frac{\hbar^2\p^2}{2M_\mathrm{eff}}-q_B,
\label{eq: ferro, energy spectrum with effective mass}
\end{align}
where
\begin{align}
M_\mathrm{eff}=\frac{M}{1-\frac{49}{45\sqrt{\pi}}\sqrt{na^3}}.
\label{eq: ferro, effective mass}
\end{align}
Compared with the first-order energy spectrum $\hbar\omega_{0,\p}$ [see Eq.~\eqref{eq: ferro, first-order energy spectra, E0p}], the energy gap remains unchanged while the effective mass $M_\mathrm{eff}$ of the corresponding quasiparticles increases by a factor of $1/[1-49/(45\sqrt{\pi})\sqrt{na^3}]$. From Eq.~\eqref{eq: condensate fraction}, it can be seen that enhancement factor of the effective mass is proportional to the number of quantum depleted atoms, both of which are proportional to $\sqrt{na^3}$. This can be understood as the effect of the interaction between a quasiparticle and the quantum depleted atoms, which hinders the motion of the quasiparticle.

\begin{figure}[tbp] % float placement: (h)ere, page (t)op, page (b)ottom, other (p)age
  \centering
  % file name: E:/Spinor Beliaev (July 4, 2011)/Figures/second_order_Sigma21.eps
  \includegraphics[width=3in,keepaspectratio]{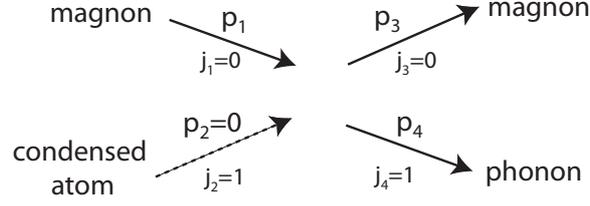}
  \caption{A collision between a magnon in spin state $j=0$ and a condensed atom in spin state $j=1$. Because the dominant interaction is spin-conserving, the collision produces another magnon in spin state $j=0$ and a phonon in spin state $j=1$. The condensed atom is represented by a dashed line.}
  \label{fig: magnon's scattering}
\end{figure}

Furthermore, because the imaginary part of $\Sigma^{11(2)}_{00}$ vanishes up to the order of $n_0c_0\sqrt{n_0a^3}$ [see Eq.~\eqref{eq: imaginary part, second order Sigma11-00(p)}], the damping of this spin-wave elementary excitation is much smaller than that of the density-wave excitation mode (the mode with spin state $j=1$). In other words, the lifetime of the corresponding magnons is much longer than that of phonons. (The fact that the excitation with spin state $j=0$ and its quasiparticles can be identified as a spin wave and magnons will be discussed in Sec.~\ref{section: Spin-density waves} below.) This agrees with the mechanism of the Beliaev damping via collisions between quasiparticles and condensed atoms. Physically, the Beliaev damping can be understood by considering the conservation of momentum and energy in the collisional process between a magnon and a condensed atom. Because $c_0/|c_1|\simeq 200 \gg 1$, the interaction between $\Rb$ atoms in a scattering process is dominated by the spin-conserving interaction. Consequently, the collision between a magnon (spin state $j=0$) and a condensed atom (spin state $j=1$) would produce another magnon (spin state $j=0$) and a phonon (spin state $j=1$). This is illustrated in Fig. \ref{fig: magnon's scattering}.

The conservation of the total momentum and energy in the collision requires that the following simultaneous equations be satisfied:
\begin{align}
\p_1=&\,\p_3+\p_4, \label{eq: momentum conservation}\\
-q_B+\frac{\hbar^2\p_1^2}{2M_\mathrm{eff}}=&\,-q_B+\frac{\hbar^2\p_3^2}{2M_\mathrm{eff}}+\hbar v_\mathrm{s}|\p_4| \label{eq: energy conservation}. 
\end{align}
Here, we used the following energy spectra of magnons and phonons:
\begin{align}
E^\mathrm{mag}_\p=&\,-q_B+\frac{\hbar^2\p^2}{2M_\mathrm{eff}},\\
E^\mathrm{pho}_\p=&\,\hbar v_\mathrm{s}|\p|,
\end{align}
where the effective mass $M_\mathrm{eff}$ and the sound velocity $v_\mathrm{s}$ are given by [see Eqs.~\eqref{eq: ferro, second-order energy spectrum G00 (2)} and \eqref{eq: ferro, second-order energy spectrum of phonon mode}]
\begin{align}
M_\mathrm{eff}=&\,\frac{M}{1-\frac{49}{45\sqrt{\pi}}\sqrt{na^3}},\label{eq: effective mass}\\
v_\mathrm{s}=&\,\left(1+\frac{8}{\sqrt{\pi}}\sqrt{na^3}\right)\sqrt{\frac{(c_0+c_1)n}{M}}\label{eq: sound velocity}.
\end{align}
From Eqs.~\eqref{eq: momentum conservation} and \eqref{eq: energy conservation}, we obtain an equation
\begin{align}
\frac{\hbar|\p_1+\p_3|\cos\theta}{2M_\mathrm{eff}}=v_\mathrm{s},
\label{eq: equation for energy conservation}
\end{align}
where $\theta$ is the angle between $\p_1+\p_3$ and $\p_1-\p_3=\p_4$. By using Eqs.~\eqref{eq: effective mass} and \eqref{eq: sound velocity} for the effective mass and sound velocity, we can evaluate the ratio of the left-hand side to the right-hand side of Eq.~\eqref{eq: equation for energy conservation}:
\begin{align}
\frac{\hbar|\p_1+\p_3||\cos\theta|}{2M_\mathrm{eff}v_\mathrm{s}}<&\,\frac{\hbar|\p_1|}{M_\mathrm{eff}v_\mathrm{s}}\nonumber\\
\ll&\, \frac{\sqrt{2Mc_0n}}{M_\mathrm{eff}}\nonumber\\
=&\,\frac{1-\frac{49}{45\sqrt{\pi}}\sqrt{na^3}}{1+\frac{8}{\sqrt{\pi}}\sqrt{na^3}}\sqrt{\frac{2c_0}{c_0+c_1}}\nonumber\\
\sim&\, \mathcal{O}(1).
\label{eq: necessary condition} 
\end{align}
Here, we used $|\p_3|<|\p_1|$, which results from Eq.~\eqref{eq: energy conservation}, and $|\p_1|\ll \sqrt{2Mc_0n}$ for the low-momentum region $\eps{\p_1}\ll c_0n$ where the obtained second-order energy spectrum [Eq.~\eqref{eq: ferro, second-order energy spectrum G00 (2)}] is valid. From Eq.~\eqref{eq: necessary condition}, it is obvious that there is no collision between a magnon and a condensed atom in which the momentum and energy conservations [Eqs.~\eqref{eq: momentum conservation} and \eqref{eq: energy conservation}, or equivalently, Eqs.~\eqref{eq: momentum conservation} and \eqref{eq: equation for energy conservation}] are satisfied. Consequently, the damping of magnons vanishes up to the order of magnitude under consideration in this paper as shown by Eq.~\eqref{eq: imaginary part, second order Sigma11-00(p)}. 

Similarly, the energy spectrum of the elementary excitation given by $G_{-1,-1}(p)$ at low momenta $\eps{\p}\ll c_0n_0$ is given by
\begin{align}\
\hbar p_0\simeq&\,\hbar\omega_{-1,\p}-\frac{49}{720\pi^2}n_0c_0\sqrt{n_0\tilde{a}^3}\left(\frac{\hbar\omega_{1,\p}}{n_0c_0}\right)^2\nonumber\\
\simeq&\,\eps{\p}-2c_1n_0-\frac{49}{720\pi^2}\frac{\sqrt{n_0\tilde{a}^3}}{n_0c_0}\times 2(c_0+c_1)n_0\eps{\p}\nonumber\\
=&\,\left(1-\frac{49}{360\pi^2}\sqrt{n_0\tilde{a}^3}\right)\eps{\p}-2c_1n_0 \nonumber\\
=&\,\left(1-\frac{49}{45\sqrt{\pi}}\sqrt{na^3}\right)\eps{\p}-2c_1n
\label{eq: ferro, second-order energy spectrum G-1,-1 (2)}
\end{align}
It can be seen from Eq.~\eqref{eq: ferro, second-order energy spectrum G-1,-1 (2)} that the energy spectrum of the excitation with spin state $j=-1$ also has a quadratic dispersion relation at low momenta, and compared with the first-order energy spectrum, the energy gap remains unchanged while the effective mass increases by the same factor of $1/[1-49/(45\sqrt{\pi})\sqrt{na^3}]$ as the excitation with spin state $j=0$.

Now we are in a position to evaluate the validity of the \textit{a priori} assumption that the difference between the second-order and first-order energy spectra is small [see Eq.~\eqref{eq: condition for a Taylor expansion around p0=omega0p}]. This assumption has been used in Sec.~\ref{subsection: Second-order self energies} as the self-energies were Taylor expanded around $p_0=\omega_{0,\p}$ [Eq.~\eqref{eq: a Taylor expansion around p0=omega0p}], and the expansions were stopped at the linear term. The condition for the validity of the Taylor expansion can be obtained from Eq.~\eqref{eq: ferro, equation for the pole of G00}, that is, 
\begin{align}
\beta_\p \left|p_0-\omega_{0,\p}\right| \ll \alpha_\p.
\label{eq: condition for validity of expansion}
\end{align}
By using Eq.~\eqref{eq: ferro, second-order energy spectrum G00 (1)} for the second-order energy spectrum, we find that the left-hand side of Eq.~\eqref{eq: condition for validity of expansion} is almost equal to $\alpha_\p\beta_\p$, and thus, Eq.~\eqref{eq: condition for validity of expansion} is satisfied, provided that
\begin{align}
\beta_\p \sim \sqrt{n_0a^3} \ll 1.
\label{eq: condition for validity of expansion (2)}
\end{align}
Equation~\eqref{eq: condition for validity of expansion (2)} is nothing but the diluteness condition, and it is usually satisfied in conventional experiments of ultracold Bose gases.

%****************************
\subsubsection{Polar phase}
\label{subsubsection: Polar phase, subsection: Second-order energy spectra of elementary excitations}
For the polar phase, there is a two-fold degeneracy in the energy spectra of elementary excitations due to the symmetry between the $j=\pm1$ sublevels. The poles of the Green's function $G_{11}(p)$ (or equivalently, $G_{-1,-1}(p)$) given in Eqs.~\eqref{eq: solution to the Dyson equation} and \eqref{eq: denominator of Green's functions} are
\begin{align}
p_0=&\,\frac{\Sigma_{+}-\Sigma_{-}}{2} \pm \sqrt{\Bigg[\frac{\eps{\p}-\mu+q_B}{\hbar}+\left(\frac{\Sigma_{+}+\Sigma_{-}}{2}\right)\Bigg]^2-\Sigma^{12}_{1,-1}\Sigma^{21}_{-1,1}} \nonumber\\
=&\,\frac{\Sigma^{(2)}_{+}-\Sigma^{(2)}_{-}}{2} \pm \Bigg\{\Bigg[\frac{\eps{\p}+c_1n_0+q_B}{\hbar}+\left(\frac{\Sigma^{(2)}_{+}+\Sigma^{(2)}_{-}}{2}\right)-\frac{\mu^{(2)}}{\hbar}\Bigg]^2-\left(\frac{c_1n_0}{\hbar}+\Sigma^{12(2)}_{1,-1}\right)^2 \Bigg\}^{1/2} \nonumber\\
\simeq &\,\frac{\Sigma^{(2)}_{+}-\Sigma^{(2)}_{-}}{2} \pm \Bigg\{ \omega_{1,\p}^2+ \frac{\left(\Sigma^{(2)}_{+}+\Sigma^{(2)}_{-}-2\mu^{(2)}/\hbar\right)\left(\eps{\p}+c_1n_0+q_B\right)}{\hbar}-2\frac{c_1n_0}{\hbar}\Sigma^{12(2)}_{1,-1} \Bigg\}^{1/2} \nonumber\\
\simeq&\, \frac{\Sigma^{(2)}_{+}-\Sigma^{(2)}_{-}}{2} \pm \Bigg[ \omega_{1,\p}+\frac{\left(\Sigma^{(2)}_{+}+\Sigma^{(2)}_{-}-2\mu^{(2)}/\hbar\right)(\eps{\p}+c_1n_0+q_B)}{2\hbar\omega_{1,\p}}-\frac{c_1n_0}{\hbar\omega_{1,\p}}\Sigma^{12(2)}_{1,-1} \Bigg] \nonumber\\
=&\, \pm \left( \omega_{1,\p}+ \Lambda^{\mp}_{1,p} \right),
\label{eq: second-order pole of G11}
\end{align}
where $\Sigma_{\pm}$ denote $\Sigma^{11}_{11}(\pm p)$, and 
\begin{align}
\Lambda^{\mp}_{1,p}\equiv&\, \frac{\eps{\p}+c_1n_0+q_B}{2\hbar\omega_{1,\p}}\left(\Sigma^{(2)}_{+}+\Sigma^{(2)}_{-}-2\mu^{(2)}/\hbar\right)-\frac{c_1n_0}{\hbar\omega_{1,\p}}\Sigma^{12(2)}_{1,-1}\pm \frac{\Sigma^{(2)}_{+}-\Sigma^{(2)}_{-}}{2}.
\label{eq: lambda1mp}
\end{align}
Here, we used the first-order self-energies and chemical potential given in Eq.~\eqref{eq: polar, first-order self energies 2}, the first-order energy spectrum given in Eq.~\eqref{eq: first-order energy spectrum E1p}, and the fact that $\Sigma^{12(2)}_{1,-1}$ and $\Sigma^{(2)}_{+}+\Sigma^{(2)}_{-}-2\mu^{(2)}/\hbar$ are much smaller than $|c_1|n_0/\hbar \sim q_B/\hbar \sim \omega_{1,\p}$ in the low-momentum region $\eps{\p}\ll |c_1|n_0$ [see Eqs.~ \eqref{eq: polar, mu(2)}, \eqref{eq: polar, second-order Sigma12 1,-1=0}, \eqref{eq: polar, imaginary part of second order Sigma11-11(p)}, \eqref{eq: real part, second order Sigma11-11, polar}, and \eqref{eq: real part, second order Sigma22-11, polar}].

By substituting Eqs.~ \eqref{eq: polar, mu(2)}, \eqref{eq: polar, second-order Sigma12 1,-1=0}, \eqref{eq: polar, imaginary part of second order Sigma11-11(p)}, \eqref{eq: real part, second order Sigma11-11, polar}, and \eqref{eq: real part, second order Sigma22-11, polar} into Eqs.~\eqref{eq: second-order pole of G11} and \eqref{eq: lambda1mp}, we obtain the equation that determines the pole of $G_{11}(p)$:
\begin{align}
p_0=\,\omega_{1,\p}+\alpha_\p+\beta_\p \left(p_0-\omega_{1,\p}\right),
\label{eq: polar, equation for the pole of G11}
\end{align}
where
\begin{align}
\hbar\alpha_\p=&\Bigg[\frac{71}{720\pi^2}\frac{\eps{\p}+q_B+c_1n_0}{\hbar\omega_{1,\p}}-\frac{1}{6\pi^2}\frac{(q_B+c_1n_0)}{\sqrt{q_B(q_B+2c_1n_0)}}\Bigg] n_0c_0\sqrt{n_0\tilde{a}^3}\left(\frac{\hbar\omega_{0,\p}}{n_0c_0}\right)^2,
\end{align}
\begin{align}
\beta_\p=&\,-\frac{1}{3\pi^2}\sqrt{n_0\tilde{a}^3}+\Bigg[\frac{7}{120\pi^2}-\frac{1}{6\pi^2}\frac{(q_B+c_1n_0)}{\sqrt{q_B(q_B+2c_1n_0)}}\Bigg]\sqrt{n_0\tilde{a}^3}\left(\frac{\hbar\omega_{0,\p}}{n_0c_0}\right)^2.
\label{eq: beta p}
\end{align}
Using the fact that $\beta_\p\sim \sqrt{n_0a^3}\ll 1$, the solution to Eq.~\eqref{eq: polar, equation for the pole of G11} is given by
\begin{align}
p_0=&\,\omega_{1,\p}+\frac{\alpha_\p}{1-\beta_\p} \nonumber\\
\simeq &\, \omega_{1,\p}+ \alpha_\p \nonumber\\
=&\, \omega_{1,\p}+\Bigg[\frac{71}{720\pi^2}\frac{\eps{\p}+q_B+c_1n_0}{\hbar\omega_{1,\p}}-\frac{1}{6\pi^2}\frac{(q_B+c_1n_0)}{\sqrt{q_B(q_B+2c_1n_0)}}\Bigg] n_0c_0\sqrt{n_0\tilde{a}^3}\left(\frac{\hbar\omega_{0,\p}}{n_0c_0}\right)^2.
\label{eq: polar, pole of G11 (2)}
\end{align}
In the low-momentum region $\eps{\p} \ll |c_1|n_0\sim q_B \ll c_0n_0$, $\omega_{1,\p}$ and $\omega_{0,\p}$ can be approximated as
\begin{subequations}
\label{eq: polar, E1p, E0p at low momenta}
\begin{align}
\hbar\omega_{1,\p}=&\sqrt{(\eps{\p}+q_B)(\eps{\p}+q_B+2c_1n_0)} \simeq \sqrt{q_B(q_B+2c_1n_0)}+\frac{(q_B+c_1n_0)}{\sqrt{q_B(q_B+2c_1n_0)}} \eps{\p},\label{eq: polar, low-momentum E1p}\\
\hbar\omega_{0,\p}=&\sqrt{\eps{\p}(\eps{\p}+2c_0n_0)} \simeq \sqrt{2c_0n_0 \eps{\p}}.
\end{align}
\end{subequations}
Substituting Eq.~\eqref{eq: polar, E1p, E0p at low momenta} into Eq.~\eqref{eq: polar, pole of G11 (2)}, we obtain the energy spectrum which is correct up to the second order: 
\begin{align}
\hbar p_0=&\sqrt{q_B(q_B+2c_1n_0)}+\frac{(q_B+c_1n_0)}{\sqrt{q_B(q_B+2c_1n_0)}} \eps{\p}-\frac{49}{360\pi^2}\sqrt{n_0\tilde{a}^3}\frac{(q_B+c_1n_0)}{\sqrt{q_B(q_B+2c_1n_0)}} \eps{\p} \nonumber\\
=&\sqrt{q_B(q_B+2c_1n_0)}+ \left(1-\frac{49}{360\pi^2}\sqrt{n_0\tilde{a}^3}\right)\frac{(q_B+c_1n_0)}{\sqrt{q_B(q_B+2c_1n_0)}} \eps{\p}\nonumber\\
=&\sqrt{q_B(q_B+2c_1n_0)}+ \left(1-\frac{49}{45\sqrt{\pi}}\sqrt{n_0a^3}\right)\frac{(q_B+c_1n_0)}{\sqrt{q_B(q_B+2c_1n_0)}} \eps{\p}\nonumber\\
=&\sqrt{q_B(q_B+2c_1n)}+ \left(1-\frac{49}{45\sqrt{\pi}}\sqrt{na^3}\right)\frac{(q_B+c_1n)}{\sqrt{q_B(q_B+2c_1n)}} \eps{\p}.
\label{eq: energy spectrum at second-order approximation}
\end{align}
Here, in deriving the last equality, we used $na^3\ll 1$ and
\begin{align}
\frac{(q_B+c_1n_0)}{\sqrt{q_B(q_B+2c_1n_0)}}\simeq\frac{(q_B+c_1n)}{\sqrt{q_B(q_B+2c_1n)}}\left[1-\frac{(c_1n)^2}{(q_B+c_1n)^2}\left(\frac{8}{3\sqrt{\pi}}\right)^2 na^3\right].
\end{align}
From Eq.~\eqref{eq: energy spectrum at second-order approximation}, it can be seen that the energy spectrum of the elementary excitation given by $G_{11}(p)$ has a quadratic dispersion relation at low momenta, which can be expressed using the effective mass $M_\mathrm{eff}$ as
\begin{align}
\hbar p_0=\frac{\hbar^2\p^2}{2M_\mathrm{eff}}+\sqrt{q_B(q_B+2c_1n)}.
\end{align}
The effective mass depends on the quadratic Zeeman energy $q_B$ as:
\begin{align}
M_\mathrm{eff}=\frac{\sqrt{q_B(q_B+2c_1n)}}{(q_B+c_1n)}\frac{M}{1-\frac{49}{45\sqrt{\pi}}\sqrt{na^3}}.
\label{eq: polar, effective mass}
\end{align}
Compared with the first-order energy spectrum given by Eq.~\eqref{eq: polar, low-momentum E1p}, the energy gap remains unchanged while the effective mass increases by a factor of $1/[1-49/(45\sqrt{\pi})\sqrt{na^3}]$ as a consequence of quantum depletion. It can be seen that the effect of quantum depletion on the effective mass is characterized by the same enhancement factor, regardless of whether the system is ferromagnetic or polar, and independent of external parameters of the system. Furthermore, since the imaginary part of $\Sigma^{11(2)}_{11}$ vanishes up to the order of $n_0c_0\sqrt{n_0a^3}$, the damping of this spin-wave excitation (see Sec. \ref{section: Spin-density waves}) is much smaller than that of the density-wave excitation (with spin state $j=0$). In other words, the lifetime of the corresponding magnons is much longer than that of phonons. This agrees with the mechanism of Beliaev damping and can be understood by considering the conservation of momentum and energy in a collision between a quasiparticle and a condensed atom (see Sec.~\ref{subsubsection: Ferromagnetic phase, subsection: Second-order energy spectra of elementary excitations}).

%############################
\section{Transverse magnetization and spin wave}
\label{section: Spin-density waves}
As shown in Secs.~\ref{section: First-order approximation} and \ref{section: Second-order approximation}, the elementary excitations of a spinor BEC include both density-wave and spin-wave excitations with quasiparticles being phonons and magnons, respectively. The energy spectrum of phonons with a linear dispersion relation can be experimentally measured by using the neutron scattering or the Bragg spectroscopy. The former has been widely used in experiments of the superfluid  helium \cite{Palevsky57, Palevsky58, Henshaw58, Yarnell58, Yarnell59}, while the later has been applied to measurements of ultracold atoms \cite{Stenger99, Stamper-kurn99, Papp08, Veeravalli08}. Similarly, the neutron scattering has been used to measure the dispersion relation of magnons in solid crystals \cite{Shirane65, Nikotin69}, though its application to ultracold atomic systems is limited by a huge difference in energy scales between neutrons and atoms. The Bragg spectroscopy can also be generalized to measure the energy spectrum of magnons by using appropriately polarized laser beams to make spin-selective transitions. In this section, we propose an alternative experimental setup to indirectly get information of the effective mass of magnons in a spinor BEC of ultracold atoms. In ultracold atom experiments, atoms can be optically excited to a higher energy level to produce an appreciable number of quasiparticles. Furthermore, an in-situ, non-destructive measurement of magnetization can be made with a high resolution \cite{Higbie05}. It has been applied in a wide range of spinor BECs' experiments to investigate, for instance, the formation of spin textures and topological excitations as well as their dynamics \cite{Sadler06, Vengalattore08, Leslie09, Vengalattore10, Guzman11}. 

First, we show that in a spinor BEC, the elementary excitations with quadratic dispersion relations (see Secs.~\ref{section: First-order approximation} and \ref{section: Second-order approximation}) can be interpreted as waves of transverse magnetization. The transverse magnetization density operators $\hat{F}_x(\R,t)$ and $\hat{F}_y(\R,t)$ in the Heisenberg representation are defined by
\begin{subequations}
\label{eq: transverse spin density operator, definition}
\begin{align}
\hat{F}_{+}(\R,t)\equiv\,\hat{F}_x(\R,t)+i\hat{F}_y(\R,t)=\,\sqrt{2}\left[\creat{1}(\R,t)\annih{0}(\R,t)+\creat{0}(\R,t)\annih{-1}(\R,t)\right],\\
\hat{F}_{-}(\R,t)\equiv\,\hat{F}_x(\R,t)-i\hat{F}_y(\R,t)=\,\sqrt{2}\left[\creat{0}(\R,t)\annih{1}(\R,t)+\creat{-1}(\R,t)\annih{0}(\R,t)\right].
\end{align}
\end{subequations}
The squared modulus of the transverse magnetization is written in terms of $\hat{F}_{+}(\R,t)$ and $\hat{F}_{-}(\R,t)$ as
\begin{align}
\hat{F}_{\perp}^2(\R,t)\equiv &\,\hat{F}_{x}^2(\R,t)+\hat{F}_{y}^2(\R,t)\nonumber\\
=&\frac{1}{2}\left[\hat{F}_{+}(\R,t)\hat{F}_{-}(\R,t)+\hat{F}_{-}(\R,t)\hat{F}_{+}(\R,t)\right].
\end{align}

%&&&&&&&&&&&&&&&&&&&&&&&&&&&&
\subsection{Ferromagnetic phase}
\label{subsection: Ferromagnetic phase, section: Spin-density waves}
The condensate wavefunction is given by
\begin{align}
\bm{\phi}=\sqrt{n_0}(1,0,0).
\end{align}
The lowest-order transverse magnetization $\hat{F}_{+}(\R,t)$ and $\hat{F}_{-}(\R,t)$  are then given by
\begin{subequations}
\label{eq: ferro, lowest-order transverse spin density operators}
\begin{align}
\hat{F}_{+}(\R,t)=&\sqrt{2n_0}\annih{0}(\R,t)=\sqrt{2n_0}\sum_\K e^{i\K\cdot \R} \ahat_{0,\K}(t),\\
\hat{F}_{-}(\R,t)=&\sqrt{2n_0}\creat{0}(\R,t)=\sqrt{2n_0}\sum_\K e^{-i\K\cdot \R} \adagg_{0,\K}(t),
\end{align}
\end{subequations}
where $\ahat_{0,\K}$ is the Fourier transform of the field operator $\annih{0}(\R)=\ann{0}(\R)$ as defined in Eq.~\eqref{eq: Fourier transform of the field operator}. Here, we replaced operators $\annih{1}(\R,t)$ and $\creat{1}(\R,t)$ in Eqs.~\eqref{eq: transverse spin density operator, definition} by the condensate wavefunction $\phi_1=\sqrt{n_0}$.

At the lowest order (first-order) approximation, by using the Bogoliubov transformation
\begin{align}
\hat{b}_{1,\K}=u_\K \ahat_{1,\K}-v_\K \adagg_{1,-\K},
\end{align}
the Hamiltonian for the noncondensate part given by Eq.~\eqref{eq: operator K'} can be effectively written as
\begin{align}
\hat{\mathcal{H}}=\sum_{\K\not=0}\hbar \omega_{1,\K} \hat{b}^\dagger_{1,\K}\hat{b}_{1,\K}+\sum_{\K}\left(\hbar \omega_{0,\K} \hat{a}^\dagger_{0,\K}\hat{a}_{0,\K}+\hbar \omega_{-1,\K} \hat{a}^\dagger_{-1,\K}\hat{a}_{-1,\K}\right),
\label{eq: ferro, effective Hamiltonian}
\end{align}
where $\hbar\omega_{\pm 1;0,\K}$ are the lowest-order energy spectra of elementary excitations given by Eq.~\eqref{eq: ferro, first-order energy spectra}. The system's ground state is defined as the vacuum of annihilation operators $\hat{b}_{1,\K}, \hat{a}_{0,\K}$ and $\hat{a}_{-1,\K}$:
\begin{align}
\hat{b}_{1,\K}|\text{g}\rangle=0,\\
\hat{a}_{0,\K}|\text{g}\rangle=0,\\
\hat{a}_{-1,\K}|\text{g}\rangle=0.
\end{align}

If a particle with momentum $\hbar \K_0$ and spin $j=0$ is created above the ground state, the system is in an excited state given by
\begin{align}
|\text{e}\rangle=\hat{a}^\dagger_{0,\K_0}|\text{g}\rangle.
\label{eq: ferro, plane wave excited state}
\end{align}
We now calculate the equal-time spatial correlation of transverse magnetization in the system for this plane-wave excited state. Using Eq.~\eqref{eq: ferro, lowest-order transverse spin density operators}, we have
\begin{align}
\langle\text{e}|\hat{F}_{+}(\R,t)\hat{F}_{-}(\R',t)|\text{e}\rangle=&2n_0\sum_{\K,\K'}e^{i(\K\cdot\R-\K'\cdot\R')}\langle\text{e}| \ahat_{0,\K}(t)\adagg_{0,\K'}(t)|\text{e}\rangle\nonumber\\
=&2n_0\sum_{\K,\K'}e^{i(\K\cdot\R-\K'\cdot\R')}e^{-i(\omega_{0,\K}-\omega_{0,\K'})t}\left(\delta_{\K,\K'}+\delta_{\K,\K_0}\delta_{\K',\K_0}\right)\nonumber\\
=&2n_0\sum_\K e^{i\K\cdot(\R-\R')}+2n_0e^{i\K_0\cdot(\R-\R')},
\label{eq: ferro, spatial correlation transverse spin}
\end{align}
where 
\begin{align}
\hat{a}_{0,\K}(t)\equiv&\, e^{\frac{i}{\hbar}\hat{\mathcal{H}}t}\hat{a}_{0,\K}e^{-\frac{i}{\hbar}\hat{\mathcal{H}}t}=\,e^{-i\omega_{0,\K} t}\hat{a}_{0,\K},\\
\hat{a}^\dagger_{0,\K}(t)\equiv&\, e^{\frac{i}{\hbar}\hat{\mathcal{H}}t}\hat{a}^\dagger_{0,\K}e^{-\frac{i}{\hbar}\hat{\mathcal{H}}t}=\,e^{i\omega_{0,\K} t}\hat{a}^\dagger_{0,\K}.
\end{align}
Here, the first term in the last line of Eq.~\eqref{eq: ferro, spatial correlation transverse spin}, which is proportional to $\delta(\R-\R')$, describes the self-correlation, and it exists for all states including the ground state. Similarly, we have
\begin{align}
\langle\text{e}|\hat{F}_{-}(\R,t)\hat{F}_{+}(\R',t)|\text{e}\rangle=&2n_0\sum_{\K,\K'}e^{-i(\K\cdot\R-\K'\cdot\R')}\langle\text{e}| \adagg_{0,\K}(t)\ahat_{0,\K'}(t)|\text{e}\rangle\nonumber\\
=&2n_0e^{-i\K_0\cdot(\R-\R')}.
\label{eq: ferro, spatial correlation transverse spin 2}
\end{align}
Using Eqs.~\eqref{eq: ferro, spatial correlation transverse spin} and \eqref{eq: ferro, spatial correlation transverse spin 2}, we obtain the spatial correlation of transverse magnetization in the system: 
\begin{align}
&\langle\text{e}|\hat{\mathbf{F}}_{\perp}(\R,t)\cdot\hat{\mathbf{F}}_{\perp}(\R',t)|\text{e}\rangle
-\langle\text{g}|\hat{\mathbf{F}}_{\perp}(\R,t)\cdot\hat{\mathbf{F}}_{\perp}(\R',t)|\text{g}\rangle\nonumber\\
=&\frac{1}{2}\Bigg[\langle\text{e}|\hat{F}_{+}(\R,t)\hat{F}_{-}(\R',t)+\hat{F}_{-}(\R,t)\hat{F}_{+}(\R',t)|\text{e}\rangle-\langle\text{g}|\hat{F}_{+}(\R,t)\hat{F}_{-}(\R',t)+\hat{F}_{-}(\R,t)\hat{F}_{+}(\R',t)|\text{g}\rangle\Bigg]\nonumber\\
=&2n_0\cos\left[\K_0\cdot(\R-\R')\right].
\label{eq: ferro, spatial correlation of spin wave}
\end{align}
Here, we subtract the correlation in the ground state from that in the excited state to remove the self-correlation term in Eq.~\eqref{eq: ferro, spatial correlation transverse spin} which is not a physical observable. Equation~\eqref{eq: ferro, spatial correlation of spin wave} shows that the elementary excitation given by Eq.~\eqref{eq: ferro, plane wave excited state} can be interpreted as a spatial modulation of the transverse magnetization, or a transverse spin wave.

Now let us assume that one particle is excited to create a superposition state of different momenta:
\begin{align}
|\text{e}_\mathrm{sp}\rangle=\int \text{d}^3\K f(\K) \hat{a}^\dagger_{0,\K}|\text{g}\rangle,
\label{eq: ferro, superposition excited state}
\end{align}
where $f(\K)$ is the weight of the superposition. In a manner similar to the above calculation for the plane-wave excited state, the expectation value of the squared modulus of the transverse magnetization density $\hat{\mathbf{F}}_{\perp}^2(\R,t)$ with respect to this excited state is given by
\begin{align}
\langle\text{e}_\mathrm{sp}|\hat{\mathbf{F}}_{\perp}^2(\R,t)|\text{e}_\mathrm{sp}\rangle-\langle\text{g}|\hat{\mathbf{F}}_{\perp}^2(\R,t)|\text{g}\rangle=&\frac{1}{2}\Bigg[\langle\text{e}_\mathrm{sp}|\hat{F}_{+}(\R,t)\hat{F}_{-}(\R,t)+\hat{F}_{-}(\R,t)\hat{F}_{+}(\R,t)|\text{e}_\mathrm{sp}\rangle\nonumber\\
&-\langle\text{g}|\hat{F}_{+}(\R,t)\hat{F}_{-}(\R,t)+\hat{F}_{-}(\R,t)\hat{F}_{+}(\R,t)|\text{g}\rangle\Bigg]\nonumber\\
=&n_0\sum_{\K,\K'}\Bigg[e^{i(\K-\K')\cdot\R}e^{-i(\omega_{0,\K}-\omega_{0,\K'})t}f(\K')^*f(\K)+\text{c.c.}\Bigg]\nonumber\\
=&2n_0\Bigg|\sum_\K e^{i(\K\cdot\R-\omega_{0,\K} t)}f(\K)\Bigg|^2.
\label{eq: ferro, spin wave packet propagation}
\end{align}
The expression inside the vertical bars in the last line of Eq.~\eqref{eq: ferro, spin wave packet propagation} is nothing but the time evolution of a wave packet, which is initially constructed by a superposition of plane waves with a weight function $f(\K)$. Although the results in this section are derived at the lowest-order approximation, as we move to the second-order approximation, the physical properties of the elementary excitations do not change (see Sec.~\ref{section: First-order approximation} and \ref{section: Second-order approximation}). Namely, the elementary excitation given by Eq.~\eqref{eq: ferro, plane wave excited state} always has a quadratic dispersion relation and can be interpreted as a transverse spin wave. The only difference between the first-order and second-order results is the enhancement factor for the effective mass of magnons as a consequence of quantum depletion [see, for example, Eqs.~\eqref{eq: ferro, first-order energy spectra, E0p} and \eqref{eq: ferro, second-order energy spectrum G00 (2)}]. Therefore, we can apply the time evolution of the transverse magnetization density for a spinor wave packet, which is given by Eq.~\eqref{eq: ferro, spin wave packet propagation}, to the second-order approximation with just a replacement of the first-order energy spectrum $\hbar\omega_{0,\K}=\eps{\K}-q_B$ by the second-order one $\hbar\omega^{(2)}_{0,\K}=\left[1-49\sqrt{na^3}/(45\sqrt{\pi})\right]\eps{\K}-q_B$. 

As an example, let us consider a Gaussian wave packet in one dimension, which is a superposition state with spectral weight $f(k)$ in momentum space given by
\begin{align} 
f(k)=e^{-d^2(k-k_0)^2}.
\label{eq: ferro, Gaussian spectral weight}
\end{align}
This Gaussian wave packet has a width of the order of $d$ in the coordinate space and the center of mass moves with momentum $\hbar k_0$. Although a generalization to three dimensions is straightforward, a quasi-one-dimensional atomic system is relevant to the experiments of ultracold atoms which are tightly confined in the radial direction. Now, we can see how a quadratic dispersion relation $\hbar\omega^{(2)}_{0,k}=\hbar^2 k^2/(2M_\mathrm{eff})$ with $M_\mathrm{eff}$ given by Eq.~\eqref{eq: ferro, effective mass} affects the propagation of a spinor wave packet. Note that the energy gap $-q_B$ in the energy spectrum has no influence on the squared modulus of the transverse magnetization density because its contribution drops out upon taking the absolute value in Eq.~\eqref{eq: ferro, spin wave packet propagation}. We then have
\begin{align}
\sum_k e^{i(kx-\omega^{(2)}_{0,k} t)}f(k)=&\frac{V}{2\pi}\int\limits_{-\infty}^{\infty} \text{d}k \exp\left[i\left(kx-\frac{\hbar k^2}{2M_\mathrm{eff}}t\right)-d^2(k-k_0)^2\right]\nonumber\\
=&\frac{V}{2\pi}\times \frac{\sqrt{2\pi}}{\sqrt{2d^2+\frac{i\hbar t}{M_\mathrm{eff}}}}\exp\left[\frac{-M_\mathrm{eff}x^2-2id^2k_0(\hbar k_0t-2M_\mathrm{eff}x)}{4d^2M_\mathrm{eff}+2i\hbar t}\right],
\end{align}
and thus, 
\begin{align}
\Bigg|\sum_k e^{i(kx-\omega^{(2)}_{0,k} t)}f(k)\Bigg|^2\propto\frac{2\pi}{2d^2\sqrt{1+\frac{\hbar^2 t^2}{4d^4M_\mathrm{eff}^2}}}\exp\left[\frac{-(x-\hbar k_0t/M_\mathrm{eff})^2}{2d^2\left(1+\frac{\hbar^2 t^2}{4d^4M_\mathrm{eff}^2}\right)}\right].
\label{eq: ferro, time evolution of a Gaussian wave packet}
\end{align}
From Eq.~\eqref{eq: ferro, time evolution of a Gaussian wave packet}, it can be seen that due to the nonlinear dispersion relation, the transverse magnetization of a spinor wave packet expands in space during its propagation with a group velocity $v_\mathrm{g}=\hbar k_0/M_\mathrm{eff}$. The time dependence of the width of the wave packet is also governed by the effective mass $M_\mathrm{eff}$:
\begin{align}
d(t)=d\sqrt{1+\frac{\hbar^2 t^2}{4d^4M_\mathrm{eff}^2}}.
\label{eq: time dependence of wavepacket's width}
\end{align}
Consequently, by measuring either the group velocity or the rate of expansion of the transverse magnetization of a spinor wave packet, we can find the effective mass of magnons.

\begin{figure}[tbp] % float placement: (h)ere, page (t)op, page (b)ottom, other (p)age
  \centering
  % file name: E:/Spinor Beliaev (July 4, 2011)/Figures/second_order_Sigma12.eps
  \includegraphics[width=5in,keepaspectratio]{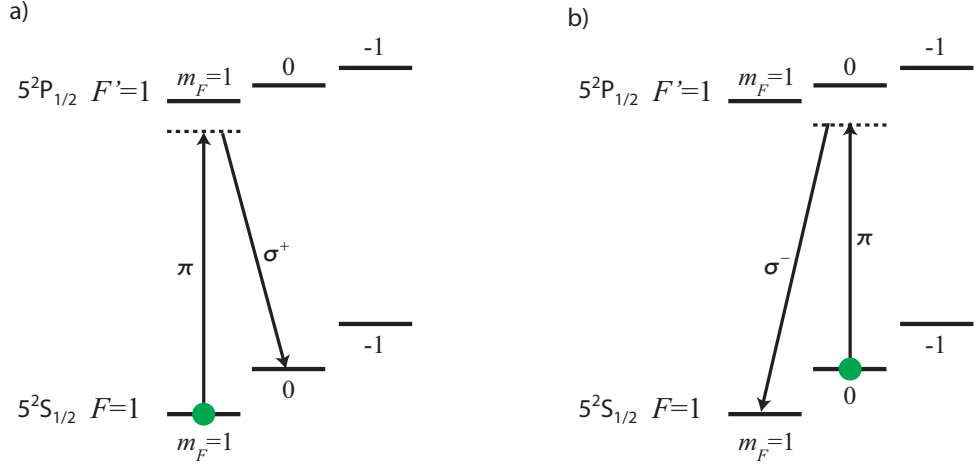}
  \caption{Two-photon Raman optical transition used to transfer $\Rb$ atoms between the $j=1$ and $j=0$ spin states in the ground-state manifold $5^{2}\text{S}_{1/2}$, $F=1$. To produce a localized wave packet of transverse magnetization, atoms in a small region of the atomic cloud need to be transferred from the $j=1$ to $j=0$ spin state for the ferromagnetic phase (a), and from the $j=0$ to $j=1$ spin state for the polar phase (b) (see Sec.~\ref{subsection: Polar phase, section: Spin-density waves}). Here, $\sigma^+$ ($\sigma^-$) denotes the right (left) circularly polarized light, and $\pi$ the linearly polarized light.}
  \label{Raman transition}
\end{figure}

\begin{figure}[tbp] % float placement: (h)ere, page (t)op, page (b)ottom, other (p)age
  \centering
  % file name: E:/Spinor Beliaev (July 4, 2011)/Figures/second_order_Sigma12.eps
  \includegraphics[width=4in,keepaspectratio]{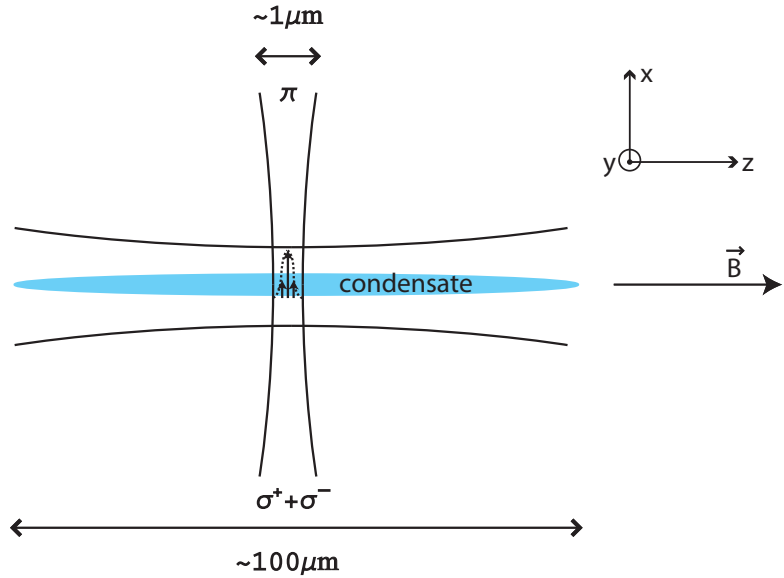}
  \caption{A schematic configuration of laser beams used to produce a wave packet of transverse magnetization. A pair of laser beams both along the $x$-axis are used to transfer atoms between the $j=1$ and $j=0$ spin states: one is linearly polarized ($\pi$) in a direction parallel to the external magnetic field, which determines the quantization axis ($z$-axis), while the other is linearly polarized in a perpendicular direction ($y$-axis), which can be regarded as a superposition of two circular polarizations ($\sigma^+ + \sigma^-$). An additional single laser beam along the $z$-axis is used as a trapping potential.}
  \label{laser beam's setup}
\end{figure}

To produce a spinor wave packet, which is a localized excited state as given in Eqs.~\eqref{eq: ferro, superposition excited state} and \eqref{eq: ferro, Gaussian spectral weight}, a small region of the atomic condensate can be exposed locally to a pair of laser beams which couple the states in the ground-state manifold to the electronically excited states. Via a Raman optical transition, which is a two-photon process, a fraction of the atoms in the $j=1$ sublevel are transferred locally into the $j=0$ state (see Fig.~\ref{Raman transition}). As can be seen from Eqs.~\eqref{eq: Taylor expansion in low momentum region} and \eqref{eq: real part, second order Sigma11-00}, which are Taylor expansions in powers of the momentum, the second-order energy spectra obtained in Sec. \ref{section: Second-order approximation} are valid only in the low-momentum region $\eps{\p}\ll c_0n_0$. Using the parameters of $\Rb$ \cite{Kempen02}, the maximum momentum $\hbar p_\mathrm{max}$ is given by
\begin{align}
p_\mathrm{max} \ll \sqrt{8\pi an} \sim 10^7 \, \mathrm{m}^{-1}.
\end{align}
Therefore, as a superposition of plane waves with momenta in the above region, the spinor wave packet should have a width of the order of
\begin{align}
\Delta x \sim \frac{1}{p_\mathrm{max}} \gg 10^{-7} \, \mathrm{m}.
\label{eq: width of the wave packet}
\end{align}
The condition \eqref{eq: width of the wave packet} turns out to be well satisfied with the parameters of laser beams used in the experimental setup. The pair of laser beams is set to be perpendicular to the single laser beam used as the trapping potential (see Fig.~\ref{laser beam's setup}). The wavelength of the pair of laser beams that couple the ground-state manifold to electronically excited states is of the order of $0.5\,\mu\mathrm{m}$ and their beam waist is a couple of the wavelength, i.e., of the order of  a micrometer. Finally, using Eq.~\eqref{eq: time dependence of wavepacket's width} and the parameters of $\Rb$ \cite{Kempen02}, we can estimate the time it takes for the spinor wave packet to expand to the entire atomic cloud. For an atomic cloud whose axial length is 100 $\mu$m, the evolution time of the spinor wave packet is about 40 ms. It is well within the lifetime of the condensate, which is of the order of a second. Furthermore, the enhancement of the effective mass of magnons as a consequence of quantum depletion is manifested by a difference in the width of the spinor wave packet, which is of the order of $1\, \mu\mathrm{m}$ after 40 ms of propagation.

%&&&&&&&&&&&&&&&&&&&&&&&&&&&&
\subsection{Polar phase}
\label{subsection: Polar phase, section: Spin-density waves}
The condensate wavefunction for the polar phase is
\begin{align}
\bm{\phi}=\sqrt{n_0}(0,1,0).
\end{align}
The lowest-order transverse magnetization $\hat{F}_{+}(\R,t)$ and $\hat{F}_{-}(\R,t)$ are then given by
\begin{subequations}
\label{eq: polar, lowest-order F+-(r,t)}
\begin{align}
\hat{F}_{+}(\R,t)=&\sqrt{2n_0}\left[\creat{1}(\R,t)+\annih{-1}(\R,t)\right]\nonumber\\
=&\sqrt{2n_0}\sum_\K \left[e^{-i\K\cdot \R} \adagg_{1,\K}(t)+e^{i\K\cdot \R} \ahat_{-1,\K}(t)\right]\nonumber\\
=&\sqrt{2n_0}\sum_\K e^{-i\K\cdot \R} \left[\adagg_{1,\K}(t)+\ahat_{-1,-\K}(t)\right],\\
\hat{F}_{-}(\R,t)=&\sqrt{2n_0}\left[\annih{1}(\R,t)+\creat{-1}(\R,t)\right]\nonumber\\
=&\sqrt{2n_0}\sum_\K \left[e^{i\K\cdot \R} \ahat_{1,\K}(t)+e^{-i\K\cdot \R} \adagg_{-1,\K}(t)\right]\nonumber\\
=&\sqrt{2n_0}\sum_\K e^{-i\K\cdot \R} \left[\adagg_{-1,\K}(t)+\ahat_{1,-\K}(t)\right].
\end{align}
\end{subequations}
At the lowest-order approximation, by using the Bogoliubov transformations
\begin{subequations}
\label{eq: polar, Bogoliubov transformation}
\begin{align}
\hat{a}_{0,\K}=&\,u_{0,\K} \hat{b}_{0,\K}+v_{0,\K} \hat{b}^\dagger_{0,-\K},\\
\hat{a}_{1,\K}=&\,u_{1,\K} \hat{b}_{1,\K}+v_{-1,\K} \hat{b}^\dagger_{-1,-\K},\\
\hat{a}_{-1,\K}=&\,u_{-1,\K} \hat{b}_{-1,\K}+v_{1,\K} \hat{b}^\dagger_{1,-\K},
\end{align}
\end{subequations}
the effective Hamiltonian for the noncondensate part can be written as
\begin{align}
\hat{\mathcal{H}}=\sum_{\K\not=0}\hbar\omega_{0,\K} \hat{b}^\dagger_{0,\K}\hat{b}_{0,\K}+\sum_{\K}\left(\hbar\omega_{1,\K} \hat{b}^\dagger_{1,\K}\hat{b}_{1,\K}+\hbar\omega_{-1,\K} \hat{b}^\dagger_{-1,\K}\hat{b}_{-1,\K}\right),
\end{align}
where $\hat{b}_{0;\pm 1,\K}$ are the annihilation operators of quasiparticles. As seen in Eq.~\eqref{eq: first-order energy spectrum E1p}, there is a two-fold degeneracy in energy $\hbar\omega_{1,\K}=\hbar\omega_{-1,\K}$ due to the symmetry between the $j=\pm 1$ sublevels. The ground state is defined as the vacuum of annihilation operators $\hat{b}_{0;\pm 1,\K}$:
\begin{align}
\hat{b}_{0,\K}|\text{g}\rangle=0,\\
\hat{b}_{\pm1,\K}|\text{g}\rangle=0.
\end{align}

From Eqs.~\eqref{eq: polar, lowest-order F+-(r,t)} and \eqref{eq: polar, Bogoliubov transformation}, the transverse magnetization density operators $\hat{F}_{+}(\R,t)$ and $\hat{F}_{-}(\R,t)$ can be written in terms of quasiparticle operators $\hat{b}_{0;\pm 1,\K}$ as
\begin{subequations}
\label{eq: polar, F+- written in terms of quasiparticle operators}
\begin{align}
\hat{F}_{+}(\R,t)=&\sqrt{2n_0}\sum_\K e^{-i\K\cdot \R} \left[\left(u_{1,\K}^*\hat{b}^\dagger_{1,\K}+v_{-1,\K}^*\hat{b}_{-1,-\K}\right)+\left(u_{-1,-\K}\hat{b}_{-1,-\K}+v_{1,-\K}\hat{b}^\dagger_{1,\K}\right)\right]\nonumber\\
=&\sqrt{2n_0}\sum_\K e^{-i\K\cdot \R} \left[\left(u_{1,\K}^*+v_{1,\K}\right)\hat{b}^\dagger_{1,\K}+\left(v_{-1,\K}^*+u_{-1,\K}\right)\hat{b}_{-1,-\K}\right],\\
\hat{F}_{-}(\R,t)=&\sqrt{2n_0}\sum_\K e^{-i\K\cdot \R} \left[\left(u_{-1,\K}^*\hat{b}^\dagger_{-1,\K}+v_{1,\K}^*\hat{b}_{1,-\K}\right)+\left(u_{1,-\K}\hat{b}_{1,-\K}+v_{-1,-\K}\hat{b}^\dagger_{-1,\K}\right)\right]\nonumber\\
=&\sqrt{2n_0}\sum_\K e^{-i\K\cdot \R} \left[\left(u_{-1,\K}^*+v_{-1,\K}\right)\hat{b}^\dagger_{-1,\K}+\left(v_{1,\K}^*+u_{1,\K}\right)\hat{b}_{1,-\K}\right].
\end{align}
\end{subequations}
Here, we used $u_{\pm1,\K}=u_{\pm1,-\K},v_{\pm1,\K}=v_{\pm1,-\K}$ for coefficients of the Bogoliubov transformations.

If a particle with momentum $\hbar \K_0$ and spin $j=1$ is created above the ground state, the system is in an excited state given by
\begin{align}
|\text{e}\rangle=&\hat{a}^\dagger_{1,\K_0}|\text{g}\rangle=\left[u^*_{1,\K_0} \hat{b}^\dagger_{1,\K_0}+v^*_{-1,\K_0} \hat{b}_{-1,-\K_0}\right]|\text{g}\rangle=u^*_{1,\K_0} \hat{b}^\dagger_{1,\K_0}|\text{g}\rangle.
\label{eq: polar, plain wave excited state}
\end{align}
Using Eqs.~\eqref{eq: polar, F+- written in terms of quasiparticle operators} and \eqref{eq: polar, plain wave excited state}, we can calculate the equal-time spatial correlation of transverse magnetization with respect to this excited state as follows:
\begin{align}
&\langle\text{e}|\hat{F}_{+}(\R,t)\hat{F}_{-}(\R',t)|\text{e}\rangle\nonumber\\
=&2n_0\sum_{\K,\K'}e^{-i(\K\cdot\R+\K'\cdot\R')}\langle\text{e}|\left[\left(u_{1,\K}^*+v_{1,\K}\right)\hat{b}^\dagger_{1,\K}(t)+\left(v_{-1,\K}^*+u_{-1,\K}\right)\hat{b}_{-1,-\K}(t)\right] \nonumber\\
&\times\left[\left(u_{-1,\K'}^*+v_{-1,\K'}\right)\hat{b}^\dagger_{-1,\K'}(t)+\left(v_{1,\K'}^*+u_{1,\K'}\right)\hat{b}_{1,-\K'}(t)\right] |\text{e}\rangle\nonumber\\
=&2n_0\sum_{\K,\K'}e^{-i(\K\cdot\R+\K'\cdot\R')}\Bigg[\left(u_{1,\K}^*+v_{1,\K}\right)\left(v_{1,\K'}^*+u_{1,\K'}\right)e^{-i(\omega_{1,-\K'}-\omega_{1,\K})t}|u_{1,\K_0}|^2\delta_{\K,\K_0}\delta_{-\K',\K_0}\nonumber\\
&+\left(v_{-1,\K}^*+u_{-1,\K}\right)\left(u_{-1,\K'}^*+v_{-1,\K'}\right)e^{-i(\omega_{-1,-\K}-\omega_{-1,\K'})t}\delta_{-\K,\K'}\nonumber\\
=&2n_0\left|u_{1,\K_0}^*+v_{1,\K_0}\right|^2|u_{1,\K_0}|^2e^{-i\K_0\cdot(\R-\R')}+2n_0\sum_\K \left|v_{-1,\K}^*+u_{-1,\K}\right|^2e^{-i\K\cdot(\R-\R')}.
\label{eq: polar, spatial correlation transverse spin}
\end{align}
Here, the last term in Eq.~\eqref{eq: polar, spatial correlation transverse spin} also exists in the spatial correlation of transverse magnetization for the ground state. Similarly, we have
\begin{align}
\langle\text{e}|\hat{F}_{-}(\R,t)\hat{F}_{+}(\R',t)|\text{e}\rangle=&2n_0\left|u_{1,\K_0}^*+v_{1,\K_0}\right|^2|u_{1,\K_0}|^2e^{i\K_0\cdot(\R-\R')}\nonumber\\
&+2n_0\sum_\K \left|v_{-1,\K}^*+u_{-1,\K}\right|^2 e^{i\K\cdot(\R-\R')}.
\end{align}
The equal-time spatial correlation of transverse magnetization with respect to the plane-wave excited state $|\text{e}\rangle$ then takes the following form:
\begin{align}
&\langle\text{e}|\hat{\mathbf{F}}_{\perp}(\R,t)\cdot\hat{\mathbf{F}}_{\perp}(\R',t)|\text{e}\rangle
-\langle\text{g}|\hat{\mathbf{F}}_{\perp}(\R,t)\cdot\hat{\mathbf{F}}_{\perp}(\R',t)|\text{g}\rangle\nonumber\\
=&\frac{1}{2}\Bigg[\langle\text{e}|\hat{F}_{+}(\R,t)\hat{F}_{-}(\R',t)+\hat{F}_{-}(\R,t)\hat{F}_{+}(\R',t)|\text{e}\rangle-\langle\text{g}|\hat{F}_{+}(\R,t)\hat{F}_{-}(\R',t)+\hat{F}_{-}(\R,t)\hat{F}_{+}(\R',t)|\text{g}\rangle\Bigg]\nonumber\\
=&2n_0\left|u_{1,\K_0}^*+v_{1,\K_0}\right|^2|u_{1,\K_0}|^2\cos\left[\K_0\cdot(\R-\R')\right].
\label{eq: polar, spatial correlation}
\end{align}
From Eq.~\eqref{eq: polar, spatial correlation}, it is clear that, similar to the elementary excitation given by Eq.~\eqref{eq: ferro, plane wave excited state} for the ferromagnetic phase, the elementary excitation given by Eq.~\eqref{eq: polar, plain wave excited state} for the polar phase also shows a periodic spatial modulation in the system's transverse magnetization, and thus, can be interpreted as a transverse spin wave. We can, therefore, produce a localized wave packet of transverse magnetization by locally exciting atoms initially in the $j=0$ to the $j=1$ state (or equivalently, to the $j=-1$ state). Due to the nonlinear dispersion relation [see Eqs.~\eqref{eq: polar, low-momentum E1p} and \eqref{eq: energy spectrum at second-order approximation}], the prepared spinor wave packet expands in space during its propagation. By measuring the group velocity or the rate of expansion of the wave packet, we can obtain information about the effective mass of the corresponding magnons. In the case of polar phase, the low-momentum region for which the dispersion relation of magnons is in a quadratic form is given by $\eps{\p} \ll |c_1|n_0$ [see, for example, Eq.~\eqref{eq: polar, low-momentum E1p}]. The width of the initially prepared spinor wave packet, therefore, should be of the order of $10\, \mu$m, which can be produced by using a pair of laser beams whose beam waist is larger than that used in the ferromagnetic phase. Furthermore, in contrast to the ferromagnetic phase, the effective mass of magnons for the polar phase, and in turn, the time-dependent width of the spinor wave packet depends on the external magnetic field via the quadratic Zeeman effect as given by Eq.~\eqref{eq: polar, effective mass}. A small variation of $q_B$ near $2|c_1|n$, which corresponds to a magnetic field of the order of hundreds milliGauss, can make a big difference in the dynamics of a spin wave, and thus, the time evolution of the spinor wave packet can also be exploited to perform a precise measurement on a magnetic field.

%############################
\section{Conclusion}
\label{section: Conclusion}
In this paper, we have studied the effect of quantum depletion at absolute zero on the energy spectra of elementary excitations for a Bose-Einstein condensate (BEC) of $\Rb$ atoms in the $F=1$ hyperfine spin manifold. We have generalized the Beliaev theory, which is a diagrammatic Green's function approach, to describe a system with internal degrees of freedom. The investigation was done on an atomic system whose ground state is in one of the two characteristic quantum phases of an $F=1$ spinor BEC: the fully polarized ferromagnetic phase and the unmagnetized polar phase. In contrast to a spinless BEC, there are spin-wave elementary excitations in a spinor BEC in addition to the conventional density-wave excitation. We showed that the corresponding magnons in a spinor BEC have quadratic dispersion relations as opposed to the linear dispersion relation of phonons. We also found that in both cases of ferromagnetic and polar phases, the quantum depletion leads to an increase in the effective mass of magnons, while it does not alter the energy gap at the leading order. Under an external magnetic field, the effective mass of magnons for the polar phase depends on the strength of the quadratic Zeeman energy relative to the spin-exchange interatomic interaction, as opposed to the ferromagnetic phase. This demonstrates the difference in the coupling of the motion of magnons to the external field for the ferromagnetic and polar phases. Nevertheless, we found that the enhancement factor of the effective mass of magnons due to the quantum depletion turns out to be the same for both phases, and also independent of the external parameters of the system. This implies a universal mechanism whereby the quantum depletion affects the motion of magnons in spinor Bose gases: the motion of magnons is hindered by the interaction with the quantum depleted atoms. Furthermore, in the case of $\Rb$ atoms where the spin-conserving interaction is much larger than the spin-exchange one, the lifetime of magnons becomes much larger than that of phonons. This agrees with the mechanism of Beliaev damping as due to collisions between quasiparticles and the condensate, and can be understood by considering the momentum and energy conservations in the collisions.

For a system of ultracold atoms with a particle density $n\sim10^{15}\, \text{cm}^{-3}$, the effective mass of magnons increases by a factor of $1/[1-49\sqrt{na^3}/(45\sqrt{\pi})]\sim 1.01$. The increase is about $1\%$, which is expected to be measurable with high-resolution experiments. Moreover, by using a technique to effectively increase the scattering length $a$ of the atoms, the quantum effect can become much larger, and easily measurable. To measure the effective mass of magnons in a dilute ultracold spinor Bose gas, we have proposed an experimental scheme which exploits the effect of a nonlinear dispersion relation on the spatial expansion of a spinor wave packet during its time evolution. By measuring either the group velocity or the rate of expansion of the wave packet of transverse magnetization, the information about the magnons' effective mass can be obtained, from which the quantum depletion effect can be probed. We also evaluated the time needed for the spinor wave packet to expand to the entire atomic cloud, and it is well within the lifetime of BECs in experiments of ultracold atoms. Using the fact that the effective mass of magnons for the polar phase is a function of the magnitude of external magnetic field, this kind of measurement can be used for numerous practical applications as, for example, to identify spinor quantum phases, or to be used for precision magnetometry.

%############################
\section*{Acknowledgments}
This work was supported by KAKENHI (22340114, 22740265, 22103005), a Global COE Program \textquotedblleft the Physical Sciences Frontier", and the Photon Frontier Network Program, from MEXT of Japan, and by JSPS and FRST under the Japan-New Zealand Research Cooperative Program.

\appendix
%***************************
\section{Relation between T-matrix and vacuum scattering amplitude}
\label{appendix: Relation between T-matrix and vacuum scattering amplitude}
\begin{figure}[tbp] % float placement: (h)ere, page (t)op, page (b)ottom, other (p)age
  \centering
  % file name: F:/Lab Seminar (May 18th, 2011)/Figures/BetheSalpeter.eps
  \includegraphics[width=4in,keepaspectratio]{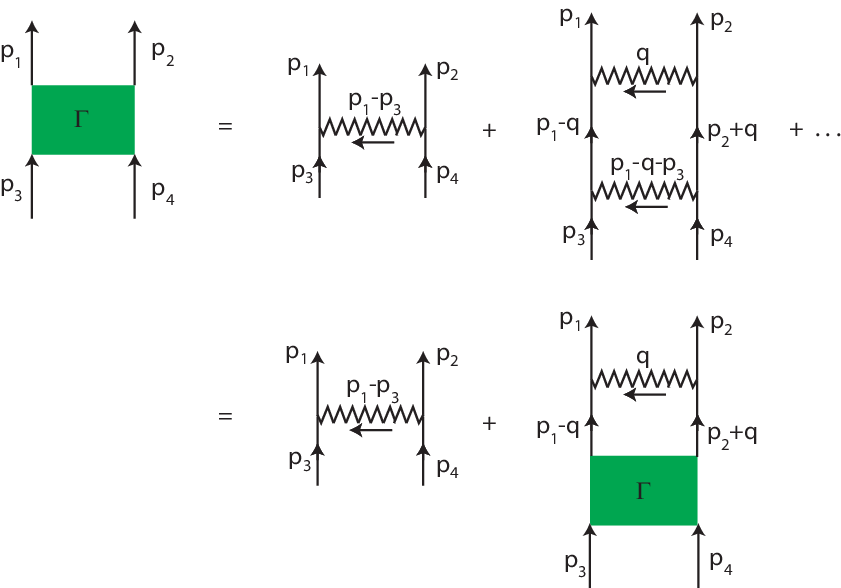}
  \caption{Bethe-Salpeter equation for the T-matrices $\Gamma_\mathcal{F}(p_1,p_2;p_3,p_4)$ in spin channels $\mathcal{F}=0$ and 2 [see Eq.~\eqref{eq: Bethe Salpeter}]. The squares represent the T-matrices, while the free propagators, which describe spinless non-interacting Green's functions $G^0(p)$, are represented by solid lines with arrows. The wavy lines show the interatomic interactions $V_\mathcal{F}(\p)$ in spin channels $\mathcal{F}=0$ and 2.}
  \label{fig:Bethe Salpeter}
\end{figure}

The T-matrix $\Gamma_\mathcal{F}(p_1,p_2;p_3,p_4)$ in the spin channel $\mathcal{F}$, which is given by Eq.~\eqref{eq: T-matrices in spin channels F=0;2}, satisfies the Bethe-Salpeter equation \cite{Bethe51}:
\begin{align}
\Gamma_\mathcal{F}(p_1,p_2;p_3,p_4)=V_\mathcal{F}(\p_1-\p_3)+\frac{i}{\hbar}\integralq &V_\mathcal{F}(\Q)G^0(p_1-q)G^0(p_2+q)\nonumber\\
&\times\Gamma_\mathcal{F}(p_1-q,p_2+q;p_3,p_4).
\label{eq: Bethe Salpeter}
\end{align}
This equation is illustrated in Fig.~\ref{fig:Bethe Salpeter}. Because the form of the Bethe-Salpeter equation is the same for two spin channels $\mathcal{F}=0$ and 2, the subscript $\mathcal{F}$ will be omitted below.
 
Let us introduce the four-vector total momentum $\hbar P=\hbar p_1+\hbar p_2=\hbar p_3+\hbar p_4$, where the second equality indicates the conservation of total momentum and energy, and the four-vector relative momentum $\hbar p=(1/2)(\hbar p_1-\hbar p_2)$, $\hbar p'=(1/2)(\hbar p_3-\hbar p_4)$ for a pair of scattering particles. Equation \eqref{eq: Bethe Salpeter} can then be rewritten as
\begin{align}
\Gamma(p,p',P)=&V(\p-\p')+\frac{i}{\hbar}\int \frac{\text{d}q_0}{2\pi}\integralQ V(\Q)G^0(P/2+p-q)\nonumber\\
&\times G^0(P/2-p+q)\Gamma(p-q,p',P),
\label{eq: Bethe Salpeter 2}
\end{align}
or in a form of an infinite series as
\begin{align}
 \Gamma(p,p',P)=&V(\p-\p')+\frac{i}{\hbar}\integralQ V(\Q)V(\p-\Q-\p')\nonumber\\
&\times \int \frac{\text{d}q_0}{2\pi} G^0(P_0/2+p_0-q_0,\mathbf{P}/2+\p-\Q)G^0(P_0/2-p_0+q_0,\mathbf{P}/2-\p+\Q)\nonumber\\
&+\cdots.
\label{eq: independent of p0}
\end{align}
Via a transformation of variables: $q_0=\tilde{q}_0+p_0$, the integral inside the square brackets in Eq.~\eqref{eq: independent of p0} can be rewritten as
\begin{align}
\int \frac{\text{d}\tilde{q}_0}{2\pi} G^0(P_0/2-\tilde{q}_0,\mathbf{P}/2+\p-\Q)G^0(P_0/2+\tilde{q}_0,\mathbf{P}/2-\p+\Q),
\end{align}
which is independent of $p_0$. In a similar manner, the higher-order terms represented by the dots in Eq.~\eqref{eq: independent of p0} are shown to be independent of $p_0$ and $p'_0$ by iteration. Therefore, the T-matrices are independent of $p_0$ and $p'_0$, and can be written as $\Gamma(\p,\p',P)$. 

Next, we introduce a quantity $\chi(\p,\p',P)$ as an integration kernel of $\Gamma(\p,\p',P)$ \cite{Beliaev1, Fetterbook}:
\begin{align}
\Gamma(\p,\p',P)= \integralQ V(\Q)\chi(\p-\Q,\p',P).
\label{eq: define chi in terms of Gamma}
\end{align}
Note that Eq. \eqref{eq: define chi in terms of Gamma} is similar in form to the equation relating the vacuum scattering amplitude $-M\tilde{f}(\K,\K')/(4\pi\hbar^2)$ to the scattering wavefunction $\psi_\K(\p)$ in momentum space:
\begin{align}
\tilde{f}(\K,\K')=\integralQ V(\Q)\psi_\K(\K'-\Q).
\label{eq: relation between scattering amplitude and scattering wave function}
\end{align}
From the Bethe-Salpeter equation [Eq.~\eqref{eq: Bethe Salpeter 2}] for $\Gamma(\p,\p',P)$, we obtain the equation for $\chi(\p,\p',P)$ as
\begin{align}
\chi(\p,\p',P)=&(2\pi)^3\delta(\p-\p')+\frac{i}{\hbar}\int \frac{\text{d}p_0}{2\pi} G^0(P/2+p)G^0(P/2-p)\nonumber\\
&\integralQ V(\Q)\chi(\p-\Q,\p',P).
\label{eq: equation for chi}
\end{align}
Indeed, by substituting Eq.~\eqref{eq: equation for chi} into Eq.~\eqref{eq: define chi in terms of Gamma}, it can be seen that Eq.~\eqref{eq: Bethe Salpeter 2} is satisfied. Calculating the integral with respect to $p_0$ in Eq.~\eqref{eq: equation for chi} by using $G^0(p)=1/(p_0-\eps{\p}+\mu+i\eta)$, we obtain
\begin{align}
\chi(\p,\p',P)=(2\pi)^3\delta(\p-\p')+\frac{1}{\hbar P_0-\frac{\hbar^2\mathbf{P}^2}{4M}+2\mu-\frac{\hbar^2\p^2}{M}+i\eta}\integralQ  V(\Q)\chi(\p-\Q,\p',P).
\label{eq:chi}
\end{align}
Note that Eq. \eqref{eq:chi} for $\chi(\p,\p',P)$ is similar in form to the Schrodinger equation for the scattering wave function $\psi_\K(\p)$ in momentum space:
\begin{align}
\psi_\K(\p)=(2\pi)^3\delta(\p-\K)-\frac{1}{\frac{\hbar^2\p^2}{M}-\frac{\hbar^2\K^2}{M}-i\eta}\integralQ V(\Q)\psi_\K(\p-\Q).
\label{eq: Schrodinger equation in momentum space}
\end{align}
Then, by using Eqs.~\eqref{eq: relation between scattering amplitude and scattering wave function}, \eqref{eq:chi} and \eqref{eq: Schrodinger equation in momentum space}, $\chi(\p,\p',P)$ can be expressed in terms of $\psi_\K(\p)$ and $\tilde{f}(\K',\K)$ as (see, for example, \cite{Fetterbook})
\begin{align}
\chi(\p,\p',P)=\psi_{\p'}(\p)+\integralQ \psi_\Q(\p)\Big(&\frac{1}{\hbar P_0-\frac{\hbar^2\mathbf{P}^2}{4M}+2\mu-\frac{\hbar^2\Q^2}{M}+i\eta}\nonumber\\
&+\frac{1}{\frac{\hbar^2\Q^2}{M}-\frac{\hbar^2\p^{'2}}{M}-i\eta}\Big)\tilde{f}(\p',\Q)^*.
\label{eq: express chi in terms of f}
\end{align}
Substituting Eq.~\eqref{eq: express chi in terms of f} into Eq.~\eqref{eq: define chi in terms of Gamma}, we obtain the expression of the T-matrix $\Gamma(\p,\p',P)$ written in terms of $\tilde{f}(\K,\K')$ as follows:
\begin{align}
\Gamma(\p,\p',P)=\tilde{f}(\p,\p')+\integralQ \tilde{f}(\p,\Q)\Big(&\frac{1}{\hbar P_0-\frac{\hbar^2\mathbf{P}^2}{4M}+2\mu-\frac{\hbar^2\Q^2}{M}+i\eta}\nonumber\\
&+\frac{1}{\frac{\hbar^2\Q^2}{M}-\frac{\hbar^2\p^{'2}}{M}-i\eta}\Big)\tilde{f}(\p',\Q)^*.
\label{eq:Gamma}
\end{align}
From Eq. \eqref{eq:Gamma}, we can see that the T-matrix $\Gamma_\mathcal{F}(p_1,p_2;p_3,p_4)=\Gamma_\mathcal{F}(\p,\p',P)$ can be fully expressed in terms of the vacuum scattering amplitude $-M\tilde{f}_\mathcal{F}(\p,\p')/(4\pi\hbar^2)$ in spin channel $\mathcal{F}$. This scattering amplitude is a well-defined physical quantity even for a singular interaction potential. 

In the discussion of the T-matrix $\Gamma_{jm,j'm'}(p_1,p_2;p_3,p_4)$ in Sec. \ref{subsection: T-matrix}, we have neglected the dependence on the spin of intermediate states via the quadratic Zeeman energy $q_B(j''+m'')$ in the denominator of Eq.~\eqref{eq: effective potential} and we are now in position to justify the validity of that approximation. From Eq.~\eqref{eq: effective potential}, the difference of $\Gamma_{jm,j'm'}(p_1,p_2;p_3,p_4)$ between the cases in which the term $q_B(j''+m'')$ is and is not neglected has the following order of magnitude:
\begin{align}
&V_\mathcal{F}(\p=0)^2\int \frac{\text{d}^3\Q}{(2\pi)^3} \frac{q_B}{\left(2\mu-2\eps{\Q}+i\eta\right)\left(2\mu-2\eps{\Q}-q_B+i\eta\right)}\nonumber\\
\sim &\, |c_1|n\sqrt{na^3} \nonumber\\
\ll &\, c_0n\sqrt{na^3}
\label{eq: neglect spin of intermediate states}
\end{align}
Here, we consider only atoms in the low-momentum region $\eps{\p_1},\eps{\p_2},\eps{\p_3},\eps{\p_4}\ll c_0n$, subject to a small external magnetic field $q_B\sim |c_1|n\ll c_0n$ as discussed in Sec.~\ref{subsection: T-matrix}. We also used $\mu\sim c_0n$, $V_\mathcal{F}(\p=0) \sim c_0$. From Eq.~\eqref{eq: neglect spin of intermediate states}, it can be seen that up to the order of magnitude of $c_0n\sqrt{na^3}$, the approximation used to evaluate the T-matrix $\Gamma_{jm,j'm'}(p_1,p_2;p_3,p_4)$ in Sec. \ref{subsection: T-matrix} is justified.

%***************************
\section{Derivation of Eq.~(\ref{eq: ferro, second-order correction from q-integral})}
\label{appendix: Derivation of second-order correction from q-integral for both ferro and polar phases}
The second-order contribution to the T-matrix $\Gamma_\mathcal{F}(\p,\p',P)$ given by Eq.~\eqref{eq: Gamma0, Gamma2} is calculated to be
\begin{align}
\Gamma_\mathcal{F}^{(2)}(\p,\p',P)=&\,\mathrm{Im}\{\tilde{f}_\mathcal{F}(\p,\p')\}+f_\mathcal{F}^2\integralQ \Bigg(\frac{1}{\hbar P_0-\frac{\hbar^2\mathbf{P}^2}{4M}+2\mu-\frac{\hbar^2\Q^2}{M}+i\eta}\nonumber\\
&+\frac{1}{\frac{\hbar^2\Q^2}{M}-\frac{\hbar^2\p'^2}{M}-i\eta}\Bigg),
\label{eq: second-order correction to the T-matrix}
\end{align}
where we neglected the momentum dependence of the generalized vacuum scattering amplitude $\tilde{f}_\mathcal{F}(\p,\p')$ in the $\Q$-integral, and replaced them with their zero-momentum limit $f_\mathcal{F}$. These replacements are justified by the fact that the $\Q$-integral in the T-matrix contains $f_\mathcal{F}^2$, and is a second-order correction.

From Eqs.~\eqref{eq: ferro, first-order proper self energies 1} and \eqref{eq: T-matrix written in terms of Gamma0,2}, it can be seen that the contributions to the self-energies and chemical potential from the first-order diagrams in Fig.~\ref{fig:first_order_approximation} involve the T-matrices $\Gamma_\mathcal{F}(\p/2,\pm\p/2,p)$, $\Gamma_\mathcal{F}(\p,{\bf 0},0)=\Gamma_\mathcal{F}({\bf 0},\p,0)$, and $\Gamma_\mathcal{F}({\bf 0},{\bf 0},0)$, whose second-order contributions are given by using Eq.~\eqref{eq: second-order correction to the T-matrix} as
\begin{subequations}
\label{eq: appendix, Gamma(2)}
\begin{align}
\Gamma_\mathcal{F}^{(2)}(\p/2,\pm\p/2,p)=&\,\mathrm{Im}\{\tilde{f}_\mathcal{F}(\p/2,\pm\p/2)\}+f_\mathcal{F}^2\integralQ \Bigg(\frac{1}{\hbar p_0-\frac{\hbar^2\mathbf{p}^2}{4M}+2\mu-\frac{\hbar^2\Q^2}{M}+i\eta}\nonumber\\
&+\frac{1}{\frac{\hbar^2\Q^2}{M}-\frac{\hbar^2\p^2}{4M}-i\eta}\Bigg),\label{eq: Gamma(2)(p/2,p/2,p)}\\
\Gamma_\mathcal{F}^{(2)}(\p,{\bf 0},0)=&\,\Gamma_\mathcal{F}^{(2)}({\bf 0},\p,0)=\,f_\mathcal{F}^2\integralQ \Bigg(\frac{1}{2\mu-\frac{\hbar^2\Q^2}{M}+i\eta}+\frac{1}{\frac{\hbar^2\Q^2}{M}}\Bigg),\\
\Gamma_\mathcal{F}^{(2)}({\bf 0},{\bf 0},0)=&\,f_\mathcal{F}^2\integralQ \Bigg(\frac{1}{2\mu-\frac{\hbar^2\Q^2}{M}+i\eta}+\frac{1}{\frac{\hbar^2\Q^2}{M}}\Bigg).
\end{align}
\end{subequations}
Here, we used the fact that the imaginary parts of the on-shell scattering amplitudes $f_\mathcal{F}(\p,\p')$ with $|\p|=|\p'|$ give second-order corrections [see Eq.~\eqref{eq: imaginary part of on-shell scattering amplitude}], while the off-shell scattering amplitudes $f_\mathcal{F}(\p,{\bf 0})$ and $f_\mathcal{F}({\bf 0},\p)$ are real numbers \cite{Beliaev1}.

The $\Q$-integral in Eq.~\eqref{eq: Gamma(2)(p/2,p/2,p)} can be rewritten in a form that is useful for the calculations in Sec. \ref{section: Second-order approximation} by making a transformation of variables $\Q=\Q'+\p/2$, with which we have
\begin{subequations}
\label{eq: variable transformation in q-integral of T-matrices}
\begin{align}
\K\equiv \Q-\p=&\,\Q'-\p/2,\\
\eps{\Q}+\eps{\K}=&\,\frac{\hbar^2\Q'^2}{M}+\frac{\hbar^2\p^2}{4M},\\
\eps{\p}-\eps{\Q}-\eps{\K}=&\,\frac{\hbar^2\p^2}{4M}-\frac{\hbar^2\Q'^2}{M},\\
\int \text{d}^3\Q=&\int \text{d}^3\Q'.
\end{align}
\end{subequations}
The $\Q$-integral in Eq.~\eqref{eq: Gamma(2)(p/2,p/2,p)} then gives
\begin{align}
&\,\integralQ \Bigg(\frac{1}{\hbar p_0-\frac{\hbar^2\mathbf{p}^2}{4M}+2\mu-\frac{\hbar^2\Q^2}{M}+i\eta}+\frac{1}{\frac{\hbar^2\Q^2}{M}-\frac{\hbar^2\p^2}{4M}-i\eta}\Bigg)\nonumber\\
=&\,\integralQ \Bigg(\frac{1}{\hbar p_0+2\mu-\eps{\Q}-\eps{\K}+i\eta}-\frac{1}{\eps{\p}-\eps{\Q}-\eps{\K}+i\eta}\Bigg).
\label{eq: q-integral after variable transformation}
\end{align}
By substituting the lowest-order chemical potential in Eq.~\eqref{eq: ferro, mu first-order apprx} into Eqs.~\eqref{eq: appendix, Gamma(2)} and \eqref{eq: q-integral after variable transformation}, and using Eqs.~\eqref{eq: ferro, first-order proper self energies 1} and \eqref{eq: T-matrix written in terms of Gamma0,2}, we obtain Eq.~\eqref{eq: ferro, second-order correction from q-integral}. Here, for the second-order correction under consideration, the spin-exchange interaction $c_1(\p,\p')\equiv [f_2(\p,\p')-f_0(\p,\p')]/3$ is neglected because of its small contribution compared with the spin-conserving one $c_0(\p,\p')$. 

%****************************
\section{Contributions to the self-energies and chemical potential from the second-order diagrams}
\label{appendix: Contribution of second-order diagrams}
%############################
\subsection{Ferrromagnetic phase}
\label{appendix: Contribution of second-order diagrams, subsection: Ferrromagnetic phase}
In the second-order contributions to the self-energies and chemical potential, the spin-exchange interaction is neglected since it is smaller than the spin-conserving one by a factor of two hundreds. Therefore, the T-matrices in the second-order diagrams in Figs.~\ref{fig:second_order_Sigma11}-\ref{fig:second_order_mu} are reduced to 
\begin{align}
\Gamma_{jm,j'm'}\simeq \,c_0\delta_{jj'}\delta_{mm'},
\end{align}
where $c_0$ is given by Eq.~\eqref{eq: definition of c0 in terms of a0 and a2}. On the other hand, the propagators in Figs.~\ref{fig:second_order_Sigma11}-\ref{fig:second_order_mu}, which are used to evaluate the second-order self-energies and chemical potential, are given by the first-order Green's functions in Eq.~\eqref{eq: ferro, first-order Green's functions in convenient form}. Then, the contributions to the self-energy $\Sigma^{11}_{jj'}(p)$ from the second-order diagrams in Fig.~\ref{fig:second_order_Sigma11} are given as follows:
\begin{align}
\text{(a1)}=&\,\frac{i}{\hbar^2}n_0c_0^2\sum_m\integralq G_{mm}(q)G_{mm}(q-p)\delta_{j,1}\delta_{j',1}\nonumber\\
=&\,\frac{n_0c_0^2}{\hbar^2}\integralQ \Bigg(\frac{A_{1,\Q} B_{1,\K}}{p_0-\omega_{1,\Q}-\omega_{1,\K}+i\eta}-\frac{A_{1,\K} B_{1,\Q}}{p_0+\omega_{1,\Q}+\omega_{1,\K}-i\eta}\Bigg)\delta_{j,1}\delta_{j',1}\nonumber\\
=&\,\frac{n_0c_0^2}{2\hbar^2}\integralQ \Bigg(\frac{\left\{A_{1,\Q}, B_{1,\K}\right\}}{p_0-\omega_{1,\Q}-\omega_{1,\K}+i\eta}-\frac{\left\{A_{1,\K}, B_{1,\Q}\right\}}{p_0+\omega_{1,\Q}+\omega_{1,\K}-i\eta}\Bigg)\delta_{j,1}\delta_{j',1}, \label{eq: ferro, diagram (a1) for Sigma11}
\end{align}
where $\K\equiv \Q-\p$ and $\left\{A_{j,\Q},B_{j,\K}\right\}\equiv A_{j,\Q} B_{j,\K}+A_{j,\K} B_{j,\Q}$. Here, $G_{0}(q)G_{0}(q-p)$ and $G_{-1}(q)G_{-1}(q-p)$ give zero contributions to the $q_0$-integral in the first line of Eq.~\eqref{eq: ferro, diagram (a1) for Sigma11}, and in deriving the last line we used the fact that the value of the integral in the second line does not change under the exchange of variables $\Q$ and $\K$. Next,
\begin{align}
\text{(a2)}=&\,\frac{i}{\hbar^2}n_0c_0^2\integralq G_{11}(q)G_{11}(q-p)\delta_{j,1}\delta_{j',1}\nonumber\\
=&\,\text{(a1)},
\end{align}
\begin{align}
\text{(a3)}=&\,\text{(a2)}=\,\text{(a1)},
\end{align}
\begin{align}
\text{(a4)}=&\,\frac{i}{\hbar^2}n_0c_0^2\integralq G_{jj}(q)G_{11}(q-p)\delta_{j,j'}\nonumber\\
=&\,\text{(a1)}+\frac{n_0c_0^2}{\hbar^2} \integralQ \frac{B_{1,\K}}{p_0-\omega_{0,\Q}-\omega_{1,\K}+i\eta}\delta_{j,0}\delta_{j',0}\nonumber\\
&+\frac{n_0c_0^2}{\hbar^2} \integralQ \frac{B_{1,\K}}{p_0-\omega_{-1,\Q}-\omega_{1,\K}+i\eta}\delta_{j,-1}\delta_{j',-1},
\end{align}
\begin{align}
\text{(b1)}=&\,\frac{i}{\hbar^2}n_0c_0^2\integralq G^{12}_{11}(q)G^{21}_{11}(q-p)\delta_{j,1}\delta_{j',1}\nonumber\\
=&\,\frac{n_0c_0^2}{\hbar^2} \integralQ C_{1,\Q} C_{1,\K}\Bigg(\frac{1}{p_0-\omega_{1,\Q}-\omega_{1,\K}+i\eta}-\frac{1}{p_0+\omega_{1,\Q}+\omega_{1,\K}-i\eta}\Bigg)\delta_{j,1}\delta_{j',1},
\end{align}
\begin{align}
\text{(b2)}=&\,\text{(b1)},
\end{align}
\begin{align}
\text{(b3)}=&\,\text{(b1)},
\end{align}
\begin{align}
\text{(b4)}=&\,\text{(b1)},
\end{align}
\begin{align}
\text{(c1)}=&\,\frac{i}{\hbar^2}n_0c_0^2\integralq G_{11}(q)G^{12}_{11}(q-p)\delta_{j,1}\delta_{j',1}\nonumber\\
=&\,\frac{n_0c_0^2}{2\hbar^2}\integralQ \Bigg(-\frac{\left\{A_{1,\Q}, C_{1,\K}\right\}}{p_0-\omega_{1,\Q}-\omega_{1,\K}+i\eta}+\frac{\left\{B_{1,\Q} ,C_{1,\K}\right\}}{p_0+\omega_{1,\Q}+\omega_{1,\K}-i\eta}\Bigg)\delta_{j,1}\delta_{j',1},
\end{align}
\begin{align}
\text{(c2)}=&\,\text{(c1)},
\end{align}
\begin{align}
\text{(c3)}=&\,\text{(c1)},
\end{align}
\begin{align}
\text{(c4)}=&\,\,\frac{i}{\hbar^2}n_0c_0^2\integralq G_{jj}(q) G^{12}_{11}(q-p)\delta_{jj'}\nonumber\\
=&\,\text{(d1)}+\frac{n_0c_0^2}{\hbar^2} \integralQ \frac{(-C_{1,\K})}{p_0-\omega_{0,\Q}-\omega_{1,\K}+i\eta}\delta_{j,0}\delta_{j',0}\nonumber\\
&+\frac{n_0c_0^2}{\hbar^2} \integralQ \frac{(-C_{1,\K})}{p_0-\omega_{-1,\Q}-\omega_{1,\K}+i\eta}\delta_{j,-1}\delta_{j',-1},
\end{align}
\begin{align}
\text{(d1)}=&\,\frac{i}{\hbar^2}n_0c_0^2\integralq G_{11}(q)G^{21}_{11}(q-p)\delta_{j,1}\delta_{j',1}\nonumber\\
=&\,\text{(c1)},
\end{align}
\begin{align}
\text{(d2)}=&\,\text{(d1)}=\,\text{(c1)},
\end{align}
\begin{align}
\text{(d3)}=&\,\text{(d1)}=\,\text{(c1)},
\end{align}
\begin{align}
\text{(d4)}=&\,\frac{i}{\hbar^2}n_0c_0^2\integralq G_{jj}(q) G^{21}_{11}(q-p)\delta_{jj'}\nonumber\\
=&\,\text{(c4)},
\end{align}
\begin{align}
\text{(e1)}=&\,\frac{i}{\hbar^2}n_0c_0^2\integralq \Big[ G_{jj}(q)G_{11}(p-q)-G^0_{j}(q)G^0_{1}(p-q) \Big]\delta_{jj'} \nonumber\\
=&\,\frac{n_0c_0^2}{\hbar} \integralQ \Bigg(\frac{A_{1,\Q} A_{1,\K}}{\hbar\left(p_0-\omega_{1,\Q}-\omega_{1,\K}\right)+i\eta}-\frac{B_{1,\Q} B_{1,\K}}{\hbar\left(p_0+\omega_{1,\Q}+\omega_{1,\K}\right)-i\eta}\nonumber\\
&-\frac{1}{\hbar p_0-\eps{\Q}-\eps{\K}+2(c_0+c_1)n_0+i\eta}\Bigg)\delta_{j,1}\delta_{j',1}\nonumber\\
&+\frac{n_0c_0^2}{\hbar} \integralQ \Bigg(\frac{A_{1,\K}}{\hbar\left(p_0-\omega_{0,\Q}-\omega_{1,\K}\right)+i\eta}\nonumber\\
&-\frac{1}{\hbar p_0-\eps{\Q}-\eps{\K}+2(c_0+c_1)n_0+q_B+i\eta}\Bigg)\delta_{j,0}\delta_{j',0}\nonumber\\
&+\frac{n_0c_0^2}{\hbar} \integralQ \Bigg(\frac{A_{1,\K}}{\hbar\left(p_0-\omega_{-1,\Q}-\omega_{1,\K}\right)+i\eta}\nonumber\\
&-\frac{1}{\hbar p_0-\eps{\Q}-\eps{\K}+2(c_0+c_1)n_0+i\eta}\Bigg)\delta_{j,-1}\delta_{j',-1}.
\label{eq: diagram c1}
\end{align}
Here, we should subtract a term containing non-interacting Green's functions given by Eq.~\eqref{eq: non-interacting Green's function} from the contribution of diagram (e1) to avoid double counting of the contribution that has already been taken into account by the definition of the T-matrix and first-order diagrams in Fig.~\ref{fig:first_order_approximation}. Similarly for the diagram (e2), we have
\begin{align}
\text{(e2)}=&\,\frac{i}{\hbar^2}n_0c_0^2\integralq \Big[ G_{11}(q)G_{11}(p-q)-G^0_{1}(q)G^0_{1}(p-q) \Big]\delta_{j,1}\delta_{j',1} \nonumber\\
=&\frac{n_0c_0^2}{\hbar} \integralQ \Bigg(\frac{A_{1,\Q} A_{1,\K}}{\hbar\left(p_0-\omega_{1,\Q}-\omega_{1,\K}\right)+i\eta}-\frac{B_{1,\Q} B_{1,\K}}{\hbar\left(p_0+\omega_{1,\Q}+\omega_{1,\K}\right)-i\eta}\nonumber\\
&-\frac{1}{\hbar p_0-\eps{\Q}-\eps{\K}+2(c_0+c_1)n_0+i\eta}\Bigg)\delta_{j,1}\delta_{j',1}.
\label{eq: diagram c2}
\end{align}
Next, 
\begin{align}
\text{(f1)}=&\,\frac{i}{\hbar}c_0 \sum_m\integralq G_{mm}(q) e^{i\eta q_0} \delta_{jj'}\nonumber\\
=&\,\frac{c_0}{\hbar}\integralQ B_{1,\Q}\, \delta_{jj'},
\end{align}
where we have introduced the convergence factor $e^{i\eta q_0}$ with $\eta\to+0$, which results from the normal order of field operators in physical observables. Similarly, we have
\begin{align}
\text{(f2)}=&\,\frac{i}{\hbar}c_0 \integralq G_{jj}(q) e^{i\eta q_0} \delta_{jj'}\nonumber\\
=&\,\frac{c_0}{\hbar}\integralQ B_{1,\Q}\, \delta_{j,1}\delta_{j',1}.
\label{eq: ferro, diagram (h2) for Sigma11}
\end{align}
By summing up Eq.~\eqref{eq: ferro, first contribution to second-order Sigma11} and Eqs.~\eqref{eq: ferro, diagram (a1) for Sigma11}-\eqref{eq: ferro, diagram (h2) for Sigma11}, we obtain Eqs.~\eqref{eq: ferro, second-order diagrams' contribution, Sigma 11-11}-\eqref{eq: ferro, second-order diagrams' contribution, Sigma 11,-1-1} for $\Sigma^{11(2)}_{jj'}(p)$.

Next, the second-order contributions to the self-energy $\Sigma^{12}_{jj'}(p)$ from the second-order diagrams in Fig.~\ref{fig:second_order_Sigma12} are given as follows:
\begin{align}
\text{(a1)}=&\,\frac{i}{\hbar^2}n_0c_0^2\sum_m\integralq G_{mm}(q)G_{mm}(q-p)\delta_{j,1}\delta_{j',1}\nonumber\\
=&\,\frac{n_0c_0^2}{\hbar^2}\integralQ \Bigg(\frac{A_{1,\Q} B_{1,\K}}{p_0-\omega_{1,\Q}-\omega_{1,\K}+i\eta}-\frac{A_{1,\K} B_{1,\Q}}{p_0+\omega_{1,\Q}+E\omega_{1,\K}-i\eta}\Bigg)\delta_{j,1}\delta_{j',1}\nonumber\\
=&\,\frac{n_0c_0^2}{2\hbar^2}\integralQ \Bigg(\frac{\left\{A_{1,\Q}, B_{1,\K}\right\}}{p_0-\omega_{1,\Q}-\omega_{1,\K}+i\eta}-\frac{\left\{A_{1,\K}, B_{1,\Q}\right\}}{p_0+\omega_{1,\Q}+\omega_{1,\K}-i\eta}\Bigg)\delta_{j,1}\delta_{j',1}, \label{eq: ferro, diagram (a1) for Sigma12}
\end{align}
\begin{align}
\text{(a2)}=&\,\frac{i}{\hbar^2}n_0c_0^2\integralq G_{11}(q)G_{11}(q-p)\delta_{j,1}\delta_{j',1}\nonumber\\
=&\,\text{(a1)},
\end{align}
\begin{align}
\text{(a3)}=&\,\text{(a2)}=\,\text{(a1)},
\end{align}
\begin{align}
\text{(a4)}=&\,\text{(a2)}=\,\text{(a1)},
\end{align}
\begin{align}
\text{(b1)}=&\,\frac{i}{\hbar^2}n_0c_0^2\integralq G^{12}_{11}(q)G^{21}_{11}(q-p)\delta_{j,1}\delta_{j',1}\nonumber\\
=&\,\frac{n_0c_0^2}{\hbar^2} \integralQ C_{1,\Q} C_{1,\K}\Bigg(\frac{1}{p_0-\omega_{1,\Q}-\omega_{1,\K}+i\eta}-\frac{1}{p_0+\omega_{1,\Q}+\omega_{1,\K}-i\eta}\Bigg)\delta_{j,1}\delta_{j',1},
\end{align}
\begin{align}
\text{(b2)}=&\,\text{(b1)},
\end{align}
\begin{align}
\text{(b3)}=&\,\text{(b1)},
\end{align}
\begin{align}
\text{(b4)}=&\,\text{(b1)},
\end{align}
\begin{align}
\text{(c1)}=&\,\frac{i}{\hbar^2}n_0c_0^2\integralq G_{11}(q)G^{12}_{11}(q-p)\delta_{j,1}\delta_{j',1}\nonumber\\
=&\,\frac{n_0c_0^2}{2\hbar^2}\integralQ \Bigg(-\frac{\left\{A_{1,\Q}, C_{1,\K}\right\}}{p_0-\omega_{1,\Q}-\omega_{1,\K}+i\eta}+\frac{\left\{B_{1,\Q} ,C_{1,\K}\right\}}{p_0+\omega_{1,\Q}+\omega_{1,\K}-i\eta}\Bigg)\delta_{j,1}\delta_{j',1},
\end{align}
\begin{align}
\text{(c2)}=&\,\text{(c1)},
\end{align}
\begin{align}
\text{(c3)}=&\,\frac{i}{\hbar^2}n_0c_0^2\integralq G_{11}(q-p)G^{12}_{11}(q)\delta_{j,1}\delta_{j',1}\nonumber\\
=&\,\frac{n_0c_0^2}{2\hbar^2}\integralQ \Bigg(-\frac{\left\{B_{1,\K}, C_{1,\Q}\right\}}{p_0-\omega_{1,\Q}-\omega_{1,\K}+i\eta}+\frac{\left\{A_{1,\K} ,C_{1,\Q}\right\}}{p_0+\omega_{1,\Q}+\omega_{1,\K}-i\eta}\Bigg)\delta_{j,1}\delta_{j',1},
\end{align}
\begin{align}
\text{(c4)}=&\,\text{(c3)},
\end{align}
\begin{align}
\text{(c5)}=&\,\text{(c1)},
\end{align}
\begin{align}
\text{(c6)}=&\,\text{(c1)},
\end{align}
\begin{align}
\text{(c7)}=&\,\text{(c3)},
\end{align}
\begin{align}
\text{(c8)}=&\,\text{(c3)},
\end{align}
\begin{align}
\text{(d1)}=&\,\frac{i}{\hbar^2}n_0c_0^2\integralq G^{12}_{11}(q-p)G^{12}_{11}(q)\delta_{j,1}\delta_{j',1} \nonumber\\
=&\,\text{(b1)},
\end{align}
\begin{align}
\text{(d2)}=&\,\text{(d1)}=\,\text{(b1)},
\end{align}
\begin{align}
\text{(e)}=&\,\frac{i}{\hbar}c_0 \integralq \left[G^{12}_{11}(q) e^{i\eta q_0}-\frac{c_0n_0}{\hbar} G^0_{1}(q)G^0_{1}(-q)\right] \delta_{j,1}\delta_{j',1}\nonumber\\
=&\,\frac{c_0}{\hbar}\integralQ \left(-C_{1,\Q}+\frac{c_0n_0}{2\eps{\Q}-2(c_0+c_1)n_0-i\eta}\right) \delta_{j,1}\delta_{j',1}.
\label{eq: ferro, diagram (i) for Sigma12}
\end{align}
Here, we should subtract a term containing non-interacting Green's functions given by Eq.~\eqref{eq: non-interacting Green's function} from the contribution of diagram (e) to avoid double counting of the contribution that has already been taken into account by the definition of the T-matrix and first-order diagrams in Fig.~\ref{fig:first_order_approximation}. We also have introduced the convergence factor $e^{i\eta q_0}$ with $\eta\to+0$, which results from the normal order of field operators in physical observables. By summing up Eq.~\eqref{eq: ferro, first contribution to second-order Sigma12} and Eqs.~\eqref{eq: ferro, diagram (a1) for Sigma12}-\eqref{eq: ferro, diagram (i) for Sigma12}, we obtain Eq.~\eqref{eq: ferro, second-order diagrams' contribution, Sigma 12,11} for $\Sigma^{12(2)}_{11}(p)$.

It can be shown by changing the direction of momentum from $p$ to $-p$ that the contributions to $\Sigma^{21}_{jj'}(p)$ from the second-order diagrams in Fig.~\ref{fig:second_order_Sigma21} are equal to Eqs.~\eqref{eq: ferro, diagram (a1) for Sigma12}-\eqref{eq: ferro, diagram (i) for Sigma12}. In fact, it can be shown that $\Sigma^{21}_{jj'}(p)=\Sigma^{12}_{jj'}(p)$ to all orders (see, for example,~\cite{Fetterbook}). Finally, the contributions to the chemical potential $\mu$ from the second-order diagrams in Fig.~\ref{fig:second_order_mu} are given as follows:
\begin{align}
\text{(a1)}=&\,\frac{i}{\hbar}c_0 \sum_m\integralq G_{mm}(q) e^{i\eta q_0}\nonumber\\
=&\,\frac{c_0}{\hbar}\integralQ B_{1,\Q},
\label{eq: ferro, diagram (a1) for mu}
\end{align}
\begin{align}
\text{(a2)}=&\,\frac{i}{\hbar}c_0 \integralq G_{11}(q) e^{i\eta q_0} \nonumber\\
=&\,\text{(a1)},
\end{align}
\begin{align}
\text{(b)}=&\,\frac{i}{\hbar}c_0 \integralq \left[G^{12}_{11}(q) e^{i\eta q_0}-\frac{c_0n_0}{\hbar} G^0_{1}(q)G^0_{1}(-q)\right]\nonumber\\
=&\,\frac{c_0}{\hbar}\integralQ \left(-C_{1,\Q}+\frac{c_0n_0}{2\eps{\Q}-2(c_0+c_1)n_0-i\eta}\right) .
\label{eq: ferro, diagram (b) for mu}
\end{align}
By summing up Eq.~\eqref{eq: ferro, first contribution to mu} and Eqs.~\eqref{eq: ferro, diagram (a1) for mu}-\eqref{eq: ferro, diagram (b) for mu}, we obtain Eq.~\eqref{eq: ferro, second-order diagrams' contribution, mu} for $\mu^{(2)}$.

%##########################
\subsection{Polar phase}
\label{appendix: Contribution of second-order diagrams, subsection: Polar phase}
In a manner similar to the case of ferromagnetic phase, the contributions to the self-energy $\Sigma^{11}_{jj'}(p)$ from the second-order diagrams in Fig.~\ref{fig:second_order_Sigma11} are given as follows:
\begin{align}
\text{(a1)}=&\,\frac{i}{\hbar^2}n_0c_0^2\sum_m\integralq G_{mm}(q)G_{mm}(q-p)\delta_{j,0}\delta_{j',0}\nonumber\\
=&\,\frac{n_0c_0^2}{\hbar^2}\integralQ\Bigg[2\Bigg(\frac{A_{1,\Q} B_{1,\K}}{p_0-\omega_{1,\Q}-\omega_{1,\K}+i\eta}-\frac{A_{1,\K} B_{1,\Q}}{p_0+\omega_{1,\Q}+\omega_{1,\K}-i\eta}\Bigg)\nonumber\\
&+\Bigg(\frac{A_{0,\Q} B_{0,\K}}{p_0-\omega_{0,\Q}-\omega_{0,\K}+i\eta}-\frac{A_{0,\K} B_{0,\Q}}{p_0+\omega_{0,\Q}+\omega_{0,\K}-i\eta}\Bigg)\Bigg]\delta_{j,0}\delta_{j',0}\nonumber\\
=&\,\frac{n_0c_0^2}{\hbar^2}\integralQ\Bigg[\Bigg(\frac{\left\{A_{1,\Q}, B_{1,\K}\right\}}{p_0-\omega_{1,\Q}-\omega_{1,\K}+i\eta}-\frac{\left\{A_{1,\K}, B_{1,\Q}\right\}}{p_0+\omega_{1,\Q}+\omega_{1,\K}-i\eta}\Bigg)\nonumber\\
&+\frac{1}{2}\Bigg(\frac{\left\{A_{0,\Q}, B_{0,\K}\right\}}{p_0-\omega_{0,\Q}-\omega_{0,\K}+i\eta}-\frac{\left\{A_{0,\K} ,B_{0,\Q}\right\}}{p_0+\omega_{0,\Q}+\omega_{0,\K}-i\eta}\Bigg)\Bigg]\delta_{j,0}\delta_{j',0}, \label{eq: diagram (a1) for Sigma11}
\end{align}
\begin{align}
\text{(a2)}=&\,\frac{i}{\hbar^2}n_0c_0^2\integralq G_{00}(q)G_{00}(q-p)\delta_{j,0}\delta_{j',0}\nonumber\\
=&\,\frac{n_0c_0^2}{2\hbar^2} \integralQ \Bigg(\frac{\left\{A_{0,\Q}, B_{0,\K}\right\}}{p_0-\omega_{0,\Q}-\omega_{0,\K}+i\eta}-\frac{\left\{A_{0,\K} ,B_{0,\Q}\right\}}{p_0+\omega_{0,\Q}+\omega_{0,\K}-i\eta}\Bigg) \delta_{j,0}\delta_{j',0},
\end{align}
\begin{align}
\text{(a3)}=&\,\text{(a2)},
\end{align}
\begin{align}
\text{(a4)}=&\,\frac{i}{\hbar^2}n_0c_0^2\integralq G_{jj}(q)G_{00}(q-p)\delta_{j,j'}\nonumber\\
=&\,\frac{n_0c_0^2}{\hbar^2} \integralQ\Bigg(\frac{A_{j,\Q} B_{0,\K}}{p_0-\omega_{j,\Q}-\omega_{0,\K}+i\eta}-\frac{A_{0,\K} B_{j,\Q}}{p_0+\omega_{j,\Q}+\omega_{0,\K}-i\eta}\Bigg)\delta_{jj'},
\end{align}
\begin{align}
\text{(b1)}=&\,\frac{i}{\hbar^2}n_0c_0^2\integralq \Big[G^{12}_{00}(q)G^{21}_{00}(q-p)+G^{12}_{1,-1}(q)G^{21}_{1,-1}(q-p)\nonumber\\
&+G^{12}_{-1,1}(q)G^{21}_{-1,1}(q-p)\Big]\delta_{j,0}\delta_{j',0}\nonumber\\
=&\,\frac{n_0c_0^2}{\hbar^2} \integralQ \Bigg[C_{0,\Q} C_{0,\K}\Bigg(\frac{1}{p_0-\omega_{0,\Q}-\omega_{0,\K}+i\eta}-\frac{1}{p_0+\omega_{0,\Q}+\omega_{0,\K}-i\eta}\Bigg)\nonumber\\
&+2C_{1,\Q} C_{1,\K} \Bigg(\frac{1}{p_0-\omega_{1,\Q}-\omega_{1,\K}+i\eta}-\frac{1}{p_0+\omega_{1,\Q}+\omega_{1,\K}-i\eta}\Bigg)\Bigg]\delta_{j,0}\delta_{j',0},
\end{align}
\begin{align}
\text{(b2)}=&\,\frac{i}{\hbar^2}n_0c_0^2\integralq G^{12}_{00}(q) G^{21}_{00}(q-p)\delta_{j,0}\delta_{j',0}\nonumber\\
=&\,\frac{n_0c_0^2}{\hbar^2} \integralQ C_{0,\Q} C_{0,\K}\Bigg(\frac{1}{p_0-\omega_{0,\Q}-\omega_{0,\K}+i\eta}-\frac{1}{p_0+\omega_{0,\Q}+\omega_{0,\K}-i\eta}\Bigg)\delta_{j,0}\delta_{j',0},
\end{align}
\begin{align}
\text{(b3)}=&\,\text{(b2)},
\end{align}
\begin{align}
\text{(b4)}=&\,\text{(b2)},
\end{align}
\begin{align}
\text{(c1)}=&\,\frac{i}{\hbar^2}n_0c_0^2\integralq G_{00}(q)G^{12}_{00}(q-p)\delta_{j,0}\delta_{j',0}\nonumber\\
=&\frac{n_0c_0^2}{2\hbar^2}\integralQ \Bigg(-\frac{\left\{A_{0,\Q}, C_{0,\K}\right\}}{p_0-\omega_{0,\Q}-\omega_{0,\K}+i\eta}+\frac{\left\{B_{0,\Q} ,C_{0,\K}\right\}}{p_0+\omega_{0,\Q}+\omega_{0,\K}-i\eta}\Bigg)\delta_{j,0}\delta_{j',0},
\end{align}
\begin{align}
\text{(c2)}=&\,\text{(c1)},
\end{align}
\begin{align}
\text{(c3)}=&\,\text{(c1)},
\end{align}
\begin{align}
\text{(c4)}=&\,\frac{i}{\hbar^2}n_0c_0^2\integralq G_{jj}(q) G^{12}_{00}(q-p)\delta_{jj'}\nonumber\\
=&\,\frac{n_0c_0^2}{\hbar^2} \integralQ \Bigg(-\frac{A_{j,\Q} C_{0,\K}}{p_0-\omega_{j,\Q}-\omega_{0,\K}+i\eta}+\frac{B_{j,\Q} C_{0,\K}}{p_0+\omega_{j,\Q}+\omega_{0,\K}-i\eta}\Bigg)\delta_{jj'},
\end{align}
\begin{align}
\text{(d1)}=&\,\frac{i}{\hbar^2}n_0c_0^2\integralq G_{00}(q)G^{21}_{00}(q-p)\delta_{j,0}\delta_{j',0}\nonumber\\
=&\,\text{(c1)},
\end{align}
\begin{align}
\text{(d2)}=&\,\text{(d1)}=\,\text{(c1)},
\end{align}
\begin{align}
\text{(d3)}=&\,\text{(d1)}=\,\text{(c1)},
\end{align}
\begin{align}
\text{(d4)}=&\,\frac{i}{\hbar^2}n_0c_0^2\integralq G_{jj}(q) G^{21}_{00}(q-p)\delta_{jj'}\nonumber\\
=&\,\text{(c4)},
\end{align}
\begin{align}
\text{(e1)}=&\,\frac{i}{\hbar^2}n_0c_0^2\integralq \Big[ G_{jj}(q)G_{00}(p-q)-G^0_{j}(q)G^0_{0}(p-q) \Big]\delta_{jj'} \nonumber\\
=&\,\frac{n_0c_0^2}{\hbar} \integralQ \Bigg(\frac{A_{j,\Q} A_{0,\K}}{\hbar\left(p_0-\omega_{j,\Q}-\omega_{0,\K}\right)+i\eta}-\frac{B_{j,\Q} B_{0,\K}}{\hbar\left(p_0+\omega_{j,\Q}+\omega_{0,\K}\right)-i\eta}\nonumber\\
&-\frac{1}{\hbar p_0-\eps{\Q}-\eps{\K}+2c_0n_0-q_Bj^2+i\eta}\Bigg)\delta_{jj'},
\end{align}
\begin{align}
\text{(e2)}=&\,\frac{i}{\hbar^2}n_0c_0^2\integralq \Big[ G_{00}(q)G_{00}(p-q)-G^0_{0}(q)G^0_{0}(p-q) \Big]\delta_{j,0}\delta_{j',0} \nonumber\\
=&\,\frac{n_0c_0^2}{\hbar} \integralQ \Bigg(\frac{A_{0,\Q} A_{0,\K}}{\hbar\left(p_0-\omega_{0,\Q}-\omega_{0,\K}\right)+i\eta}-\frac{B_{0,\Q} B_{0,\K}}{\hbar\left(p_0+\omega_{0,\Q}+\omega_{0,\K}\right)-i\eta}\nonumber\\
&-\frac{1}{\hbar p_0-\eps{\Q}-\eps{\K}+2c_0n_0+i\eta}\Bigg)\delta_{j,0}\delta_{j',0},
\end{align}
\begin{align}
\text{(f1)}=&\,\frac{i}{\hbar}c_0 \sum_m\integralq G_{mm}(q) e^{i\eta q_0} \delta_{jj'}\nonumber\\
=&\,\frac{c_0}{\hbar}\integralQ \left( 2B_{1,\Q}+B_{0,\Q} \right) \delta_{jj'},
\end{align}
\begin{align}
\text{(f2)}=&\,\frac{i}{\hbar}c_0 \integralq G_{jj}(q) e^{i\eta q_0} \delta_{jj'}\nonumber\\
=&\,\frac{c_0}{\hbar}\integralQ B_{j,\Q}\, \delta_{jj'}.
\label{eq: diagram (h2) for Sigma11}
\end{align}
By summing up Eq.~\eqref{eq: polar, first contribution to second-order Sigma11} and Eqs.~\eqref{eq: diagram (a1) for Sigma11}-\eqref{eq: diagram (h2) for Sigma11}, we obtain Eqs.~\eqref{eq: polar, Sigma11(2) 11} and \eqref{eq: Sigma11(2) 00} for $\Sigma^{11(2)}_{11}(p)$ and $\Sigma^{11(2)}_{00}(p)$, respectively.

Next, the contributions to the self-energy $\Sigma^{12}_{jj'}(p)$ from the second-order diagrams in Fig.~\ref{fig:second_order_Sigma12} are given as follows:
\begin{align}
\text{(a1)}=&\,\frac{i}{\hbar^2}n_0c_0^2\sum_m\integralq G_{mm}(q)G_{mm}(q-p)\delta_{j,0}\delta_{j',0}\nonumber\\
=&\,\frac{n_0c_0^2}{\hbar^2}\integralQ\Bigg[2\Bigg(\frac{A_{1,\Q} B_{1,\K}}{p_0-\omega_{1,\Q}-\omega_{1,\K}+i\eta}-\frac{A_{1,\K} B_{1,\Q}}{p_0+\omega_{1,\Q}+\omega_{1,\K}+i\eta}\Bigg)\nonumber\\
&+\Bigg(\frac{A_{0,\Q} B_{0,\K}}{p_0-\omega_{0,\Q}-\omega_{0,\K}+i\eta}-\frac{A_{0,\K} B_{0,\Q}}{p_0+\omega_{0,\Q}+\omega_{0,\K}+i\eta}\Bigg)\Bigg]\delta_{j,0}\delta_{j',0}\nonumber\\
=&\,\frac{n_0c_0^2}{\hbar^2}\integralQ\Bigg[\Bigg(\frac{\left\{A_{1,\Q}, B_{1,\K}\right\}}{p_0-\omega_{1,\Q}-\omega_{1,\K}+i\eta}-\frac{\left\{A_{1,\K}, B_{1,\Q}\right\}}{p_0+\omega_{1,\Q}+\omega_{1,\K}+i\eta}\Bigg)\nonumber\\
&+\frac{1}{2}\Bigg(\frac{\left\{A_{0,\Q}, B_{0,\K}\right\}}{p_0-\omega_{0,\Q}-\omega_{0,\K}+i\eta}-\frac{\left\{A_{0,\K} ,B_{0,\Q}\right\}}{p_0+\omega_{0,\Q}+\omega_{0,\K}+i\eta}\Bigg)\Bigg]\delta_{j,0}\delta_{j',0}, \label{eq: diagram (a1) for Sigma12}
\end{align}
\begin{align}
\text{(a2)}=&\,\frac{i}{\hbar^2}n_0c_0^2\integralq G_{00}(q)G_{00}(q-p)\delta_{j,0}\delta_{j',0}\nonumber\\
=&\,\frac{n_0c_0^2}{2\hbar^2} \integralQ \Bigg(\frac{\left\{A_{0,\Q}, B_{0,\K}\right\}}{p_0-\omega_{0,\Q}-\omega_{0,\K}+i\eta}-\frac{\left\{A_{0,\K} ,B_{0,\Q}\right\}}{p_0+\omega_{0,\Q}+\omega_{0,\K}+i\eta}\Bigg) \delta_{j,0}\delta_{j',0},
\end{align}
\begin{align}
\text{(a3)}=&\,\text{(a2)},
\end{align}
\begin{align}
\text{(a4)}=&\,\text{(a2)},
\end{align}
\begin{align}
\text{(b1)}=&\,\frac{i}{\hbar^2}n_0c_0^2\integralq \Big[G^{12}_{00}(q)G^{21}_{00}(q-p)+G^{12}_{1,-1}(q)G^{21}_{1,-1}(q-p)\nonumber\\
&+G^{12}_{-1,1}(q)G^{21}_{-1,1}(q-p)\Big]\delta_{j,0}\delta_{j',0}\nonumber\\
=&\,\frac{n_0c_0^2}{\hbar^2} \integralQ \Bigg[C_{0,\Q} C_{0,\K}\Bigg(\frac{1}{p_0-\omega_{0,\Q}-\omega_{0,\K}+i\eta}-\frac{1}{p_0+\omega_{0,\Q}+\omega_{0,\K}-i\eta}\Bigg)\nonumber\\
&+2C_{1,\Q} C_{1,\K} \Bigg(\frac{1}{p_0-\omega_{1,\Q}-\omega_{1,\K}+i\eta}-\frac{1}{p_0+\omega_{1,\Q}+\omega_{1,\K}-i\eta}\Bigg)\Bigg]\delta_{j,0}\delta_{j',0},
\end{align}
\begin{align}
\text{(b2)}=&\,\frac{i}{\hbar^2}n_0c_0^2\integralq G^{12}_{00}(q) G^{21}_{00}(q-p)\delta_{j,0}\delta_{j',0}\nonumber\\
=&\,\frac{n_0c_0^2}{\hbar^2} \integralQ C_{0,\Q} C_{0,\K}\Bigg(\frac{1}{p_0-\omega_{0,\Q}-\omega_{0,\K}+i\eta}-\frac{1}{p_0+\omega_{0,\Q}+\omega_{0,\K}-i\eta}\Bigg)\delta_{j,0}\delta_{j',0},
\end{align}
\begin{align}
\text{(b3)}=&\,\text{(b2)},
\end{align}
\begin{align}
\text{(b4)}=&\,\frac{i}{\hbar^2}n_0c_0^2\integralq \Big[G^{12}_{1,-1}(q) G^{21}_{00}(q-p) \left(\delta_{j,1}\delta_{j',-1}+\delta_{j,-1}\delta_{j',1}\right)\nonumber\\
&+G^{12}_{00}(q) G^{21}_{00}(q-p) \delta_{j,0}\delta_{j',0}\Big] \nonumber\\
=&\,\frac{n_0c_0^2}{\hbar^2}\integralQ \Bigg[ C_{1,\Q} C_{0,\K} \Bigg(\frac{1}{p_0-\omega_{1,\Q}-\omega_{0,\K}+i\eta}-\frac{1}{p_0+\omega_{1,\Q}+\omega_{0,\K}-i\eta}\Bigg)\nonumber\\
&\times (\delta_{j,1}\delta_{j',-1}+\delta_{j,-1}\delta_{j',1})+C_{0,\Q} C_{0,\K} \Bigg(\frac{1}{p_0-\omega_{0,\Q}-\omega_{0,\K}+i\eta}\nonumber\\
&-\frac{1}{p_0+\omega_{0,\Q}+\omega_{0,\K}-i\eta}\Bigg)\delta_{j,0}\delta_{j',0}\Bigg],
\end{align}
\begin{align}
\text{(c1)}=&\,\frac{i}{\hbar^2}n_0c_0^2\integralq G_{00}(q)G^{12}_{00}(q-p)\delta_{j,0}\delta_{j',0}\nonumber\\
=&\,\frac{n_0c_0^2}{2\hbar^2}\integralQ \Bigg(-\frac{\left\{A_{0,\Q}, C_{0,\K}\right\}}{p_0-\omega_{0,\Q}-\omega_{0,\K}+i\eta}+\frac{\left\{B_{0,\Q} ,C_{0,\K}\right\}}{p_0+\omega_{0,\Q}+\omega_{0,\K}-i\eta}\Bigg)\delta_{j,0}\delta_{j',0},
\end{align}
\begin{align}
\text{(c2)}=&\,\frac{i}{\hbar^2}n_0c_0^2\integralq \Big[G_{00}(q) G^{12}_{1,-1}(q-p) (\delta_{j,1}\delta_{j',-1}+\delta_{j,-1}\delta_{j',1})\nonumber\\
&+G_{00}(q) G^{12}_{00}(q-p) \delta_{j,0}\delta_{j',0}\Big] \nonumber\\
=&\,\frac{n_0c_0^2}{\hbar^2}\integralQ \Bigg[\Bigg(-\frac{A_{0,\Q} C_{1,\K}}{p_0-\omega_{1,\Q}-\omega_{0,\K}+i\eta}+\frac{B_{0,\Q} C_{1,\K}}{p_0+\omega_{1,\Q}+\omega_{0,\K}-i\eta}\Bigg)\nonumber\\
&\times (\delta_{j,1}\delta_{j',-1}+\delta_{j,-1}\delta_{j',1})+\frac{1}{2}\Bigg(-\frac{\left\{A_{0,\Q},C_{0,\K}\right\}}{p_0-\omega_{0,\Q}-\omega_{0,\K}+i\eta}+\frac{\left\{B_{0,\Q},C_{0,\K}\right\}}{p_0+\omega_{0,\Q}+\omega_{0,\K}-i\eta}\Bigg)\nonumber\\
&\times \delta_{j,0}\delta_{j',0}\Bigg],
\end{align}
\begin{align}
\text{(c3)}=&\,\frac{i}{\hbar^2}n_0c_0^2\integralq G_{00}(q-p) G^{12}_{00}(q) \delta_{j,0}\delta_{j',0} \nonumber\\
=&\,\frac{n_0c_0^2}{2\hbar^2}\integralQ \Bigg(-\frac{\left\{B_{0,\K},C_{0,\Q}\right\}}{p_0-\omega_{0,\Q}-\omega_{0,\K}+i\eta}+\frac{\left\{A_{0,\K},C_{0,\Q}\right\}}{p_0+\omega_{0,\Q}+\omega_{0,\K}-i\eta}\Bigg)\delta_{j,0}\delta_{j',0},
\end{align}
\begin{align}
\text{(c4)}=&\,\frac{i}{\hbar^2}n_0c_0^2\integralq \Big[ G_{00}(q-p) G^{12}_{1,-1}(q) (\delta_{j,1}\delta_{j',-1}+\delta_{j,-1}\delta_{j',1})\nonumber\\
&+ G_{00}(q-p) G^{12}_{00}(q)\delta_{j,0}\delta_{j',0}\Big] \nonumber\\
=&\,\frac{n_0c_0^2}{\hbar^2} \integralQ \Bigg[\Bigg(-\frac{B_{0,\K} C_{1,\Q}}{p_0-\omega_{1,\Q}-\omega_{0,\K}+i\eta}+\frac{A_{0,\K} C_{1,\Q}}{p_0+\omega_{1,\Q}+\omega_{0,\K}-i\eta}\Bigg)\nonumber\\
&\times(\delta_{j,1}\delta_{j',-1}+\delta_{j,-1}\delta_{j',1})+\frac{1}{2} \Bigg(-\frac{\left\{B_{0,\K},C_{0,\Q}\right\}}{p_0-\omega_{0,\Q}-\omega_{0,\K}+i\eta}+\frac{\left\{A_{0,\K},C_{0,\Q}\right\}}{p_0+\omega_{0,\Q}+\omega_{0,\K}-i\eta}\Bigg)\nonumber\\
&\times\delta_{j,0}\delta_{j',0}\Bigg],
\end{align}
\begin{align}
\text{(c5)}=&\,\text{(c1)},
\end{align}
\begin{align}
\text{(c6)}=&\,\text{(c1)},
\end{align}
\begin{align}
\text{(c7)}=&\,\text{(c3)},
\end{align}
\begin{align}
\text{(c8)}=&\,\text{(c3)},
\end{align}
\begin{align}
\text{(d1)}=&\,\frac{i}{\hbar^2}n_0c_0^2\integralq \Big[G^{12}_{00}(q-p)G^{12}_{1,-1}(q)(\delta_{j,1}\delta_{j',-1}+\delta_{j,-1}\delta_{j',1})\nonumber\\
&+ G^{12}_{00}(q-p) G^{12}_{00}(q)\delta_{j,0}\delta_{j',0} \nonumber\\
=&\,\text{(b4)},
\end{align}
\begin{align}
\text{(d2)}=&\,\frac{i}{\hbar^2}n_0c_0^2\integralq G^{12}_{00}(q)G^{12}_{00}(q-p)\delta_{j,0}\delta_{j',0} \nonumber\\
=&\,\text{(b2)},
\end{align}
\begin{align}
\text{(e)}=&\,\frac{i}{\hbar}c_0 \integralq \Big\{G^{12}_{1,-1}(q) e^{i\eta q_0}(\delta_{j,1}\delta_{j',-1}+\delta_{j,-1}\delta_{j',1})\nonumber\\
&+ \Big[G^{12}_{00}(q)e^{i\eta q_0}-c_0n_0G^0_{0}(q)G^0_{0}(-q)\Big]\delta_{j,0}\delta_{j',0}\Big\}\nonumber\\
=&\,\frac{c_0}{\hbar}\integralQ \Bigg[ -C_{1,\Q}(\delta_{j,1}\delta_{j',-1}+\delta_{j,-1}\delta_{j',1})\nonumber\\
&+\Bigg(-C_{0,\Q}+\frac{c_0n_0}{2\eps{\Q}-2c_0n_0-i\eta}\Bigg)\delta_{j,0}\delta_{j',0}\Bigg].
\label{eq: diagram (i) for Sigma12}
\end{align}
By summing up Eq.~\eqref{eq: polar, first contribution to second-order Sigma12} and Eqs.~\eqref{eq: diagram (a1) for Sigma12}-\eqref{eq: diagram (i) for Sigma12}, we obtain Eqs.~\eqref{eq: Sigma12,1,-1(2)} and \eqref{eq: Sigma12,00(2)} for $\Sigma^{12(2)}_{1,-1}(p)$ and $\Sigma^{12(2)}_{00}(p)$, respectively.

As in the case of ferromagnetic phase, it can be shown that $\Sigma^{21}_{jj'}(p)=\Sigma^{12}_{jj'}(p)$ by changing the direction of the momentum from $p$ to $-p$ and using the spin symmetry for the polar phase. Finally, the contributions to the chemical potential $\mu$ from the second-order diagrams in Fig.~\ref{fig:second_order_mu} are given as follows:
\begin{align}
\text{(a1)}=&\,\frac{i}{\hbar}c_0 \sum_m\integralq G_{mm}(q) e^{i\eta q_0} \nonumber\\
=&\,\frac{c_0}{\hbar}\integralQ \left( 2B_{1,\Q}+B_{0,\Q} \right),
\label{eq: diagram (a1) for mu}
\end{align}
\begin{align}
\text{(a2)}=&\,\frac{i}{\hbar}c_0 \integralq G_{00}(q) e^{i\eta q_0} \nonumber\\
=&\,\frac{c_0}{\hbar}\integralQ B_{0,\Q},
\end{align}
\begin{align}
\text{(b)}=&\,\frac{i}{\hbar}c_0 \integralq \Big[G^{12}_{00}(q)e^{i\eta q_0}-c_0n_0G^0_{0}(q)G^0_{0}(-q)\Big]\nonumber\\
=&\,\frac{c_0}{\hbar}\integralQ \Bigg(-C_{0,\Q}+\frac{c_0n_0}{2\eps{\Q}-2c_0n_0-i\eta}\Bigg).
\label{eq: diagram (b) for mu}
\end{align}
By summing up Eq.~\eqref{eq: polar, first contribution to mu} and Eqs.~\eqref{eq: diagram (a1) for mu}-\eqref{eq: diagram (b) for mu}, we obtain Eq.~\eqref{eq: mu(2)} for $\mu^{(2)}$.

%****************************
\section{Imaginary parts of self-energies}
\label{appendix: Imaginary part of self energies}
%############################
\subsection{Ferrromagnetic phase: $\Sigma^{11(2)}_{00}(p)$}
\label{appendix: Imaginary part of self energies, subsection: Ferrromagnetic phase}
By making a transformation of variables $\Q\equiv \p/2+\Q'$, we have
\begin{align}
\K=\,\Q-\p=&\,\Q'-\p/2, \\
\eps{\p}-\eps{\Q}-\eps{\K}=&\,2\left(\eps{\p/2}-\eps{\Q'}\right), \\
\integralQ=&\int \frac{\text{d}^3\Q'}{(2\pi)^3}.
\end{align}
The imaginary part of the last term in the second line of Eq.~\eqref{eq: ferro, second-order diagrams' contribution, Sigma 11-00} can then be rewritten as
\begin{align}
&\,i\,\mathrm{Im}\left\{ n_0\left(\frac{f_0^2+2f_2^2}{3}\right)\integralQ \frac{1}{\eps{\p}-\eps{\Q}-\eps{\K}+i\eta}\right\}\nonumber\\
=&\,n_0\left(\frac{f_0^2+2f_2^2}{3}\right)\integralQ (-i\pi)\delta(\eps{\p}-\eps{\Q}-\eps{\K})\nonumber\\
=&\,-\frac{i\pi\ n_0}{2}\left(\frac{f_0^2+2f_2^2}{3}\right) \int \frac{\text{d}^3\Q'}{(2\pi)^3}\,\delta \left(\eps{\p/2}-\eps{\Q'}\right)\nonumber\\
=&\,-\frac{in_0M^{3/2}}{2\sqrt{2}\pi\hbar^3}\left(\frac{f_0^2+2f_2^2}{3}\right)  \int \limits_{0}^{\infty} \text{d}\eps{\Q'}\sqrt{\eps{\Q'}}\delta\left(\eps{\p/2}-\eps{\Q'}\right) \nonumber\\
=&\,-\frac{i|\p|Mn_0}{8\pi\hbar^2}\left(\frac{f_0^2+2f_2^2}{3}\right).
\label{eq: calculate integral of delta function}
\end{align}
This cancels with the first term in Eq.~\eqref{eq: ferro, second-order diagrams' contribution, Sigma 11-00}. Therefore, by limiting our consideration to a small external magnetic field $q_B\sim |c_1|n\ll c_0n$, and ignoring any difference of the order smaller than $c_0n\sqrt{na^3}$, the imaginary part of $\Sigma^{11(2)}_{00}(p)$ is reduced to
\begin{align}
\mathrm{Im}\Sigma^{11(2)}_{00}(p)=&\,\frac{n_0c_0^2}{\hbar^2}\,\mathrm{Im}\left\{ \integralQ \frac{A_{1,\K}+B_{1,\K}-2C_{1,\K}}{p_0-\omega_{0,\Q}-\omega_{1,\K}+i\eta}\right\} \nonumber\\
=&\,\frac{n_0c_0^2}{\hbar^2} \integralQ \frac{(-\pi\eps{\K})}{\hbar\omega_{1,\K}}\delta(p_0-\omega_{0,\Q}-\omega_{1,\K}).
\end{align}
We then have
\begin{align}
\mathrm{Im}\Sigma^{11(2)}_{00}(p)\Bigg|_{p_0=\omega_{0,\p}}=&\frac{n_0c_0^2}{\hbar^2} \integralQ \frac{(-\pi\eps{\K})}{\hbar\omega_{1,\K}}\delta(\omega_{0,\p}-\omega_{0,\Q}-\omega_{1,\K})\nonumber\\
=&\frac{n_0c_0^2}{\hbar^2} \int \frac{\text{d}^3\K}{(2\pi)^3} \frac{(-\pi\eps{\K})}{\hbar\omega_{1,\K}}\delta(\omega_{0,\p}-\omega_{0,\p+\K}-\omega_{1,\K})\nonumber\\
=&\frac{n_0c_0^2M^{3/2}}{\hbar^5} \int\limits_{0}^{\infty} \text{d}\eps{\K} \frac{\sqrt{\eps{\K}}}{2\sqrt{2}\pi^2} \int\limits_{-1}^{1} \text{d}(\cos\theta) \frac{(-\pi\eps{\K})}{\hbar\omega_{1,\K}} \delta(\omega_{0,\p}-\omega_{0,\p+\K}-\omega_{1,\K}).
\label{eq: ferro, Im Sigma 11,00}
\end{align}
Here, $\theta$ is the angle between $\p$ and $\K$. The argument of the Dirac delta function is
\begin{align}
\omega_{0,\p}-\omega_{0,\p+\K}-\omega_{1,\K}=&\,\frac{\eps{\p}-\eps{\p+\K}}{\hbar}-\omega_{1,\K} \nonumber\\
=&-\frac{\hbar|\p||\K|\cos\theta}{M}-\frac{\hbar\K^2}{2M}-\omega_{1,\K} \nonumber\\
=&\frac{-2\sqrt{\eps{\p}\eps{\K}}\cos\theta-\eps{\K}-\sqrt{\eps{\K}[\eps{\K}+2(c_0+c_1)n_0]}}{\hbar} \nonumber\\
=&-\sqrt{\eps{\K}}\left[\sqrt{\eps{\K}+2(c_0+c_1)n_0}+\sqrt{\eps{\K}}+2\sqrt{\eps{\p}}\cos\theta\right]/\hbar.
\label{eq: ferro, evaluate delta function}
\end{align}
For the low-momentum region under consideration $\eps{\p}\ll c_0n_0$, the expression inside the square brackets of the last line in Eq.~\eqref{eq: ferro, evaluate delta function} is always positive for any value of $\theta\in(0,\pi)$. Therefore, the argument of the Dirac delta function vanishes only at $\eps{\K}=0$, and the value of the integral in the last line of Eq.~\eqref{eq: ferro, Im Sigma 11,00} is, to within a multiplying factor, given by
\begin{align}
&\lim_{\eps{\K}\to 0}\sqrt{\eps{\K}}\frac{\eps{\K}}{\hbar\omega_{1,\K}}\frac{1}{\frac{\partial(\omega_{0,\p}-\omega_{0,\p+\K}-\omega_{1,\K})}{\partial\eps{\K}}}\nonumber\\
=&\hbar \lim_{\eps{\K}\to 0}\sqrt{\eps{\K}}\frac{\eps{\K}}{\sqrt{\eps{\K}[\eps{\K}+2(c_0+c_1)n_0]}}\frac{1}{\left[\frac{2\eps{\K}+2(c_0+c_1)n_0}{\sqrt{\eps{\K}[\eps{\K}+2(c_0+c_1)n_0]}}+1+\frac{\sqrt{\eps{\p}}\cos\theta}{\sqrt{\eps{\K}}}\right]}\nonumber\\
=&0.
\end{align}
This implies that 
\begin{align}
\mathrm{Im}\Sigma^{11(2)}_{00}(p)\Bigg|_{p_0=\omega_{0,\p}}=0.
\label{eq: imaginary part, second order Sigma11-00}
\end{align}

Similarly, we have
\begin{align}
\frac{\partial\,\mathrm{Im}\Sigma^{11(2)}_{00}(p)}{\partial p_0}\Bigg|_{p_0=\omega_{0,\p}}=&\,\frac{n_0c_0^2}{\hbar^2} \integralQ \frac{(-\pi\eps{\K})}{\hbar\omega_{1,\K}}\delta'(\omega_{0,\p}-\omega_{0,\Q}-\omega_{1,\K})\nonumber\\
=&\, \frac{n_0c_0^2}{\hbar^2}\int \frac{\text{d}^3\K}{(2\pi)^3} \frac{(-\pi\eps{\K})}{\hbar\omega_{1,\K}}\delta'(\omega_{0,\p}-\omega_{0,\p+\K}-\omega_{1,\K})\nonumber\\
=&\,\frac{n_0c_0^2M^{3/2}}{\hbar^5} \int\limits_{0}^{\infty} \text{d}\eps{\K} \frac{\sqrt{\eps{\K}}}{2\sqrt{2}\pi^2} \int\limits_{-1}^{1} \text{d}(\cos\theta) \frac{(-\pi\eps{\K})}{\hbar\omega_{1,\K}} \delta'(\omega_{0,\p}-\omega_{0,\p+\K}-\omega_{1,\K}),
\end{align}
where $\delta'(x)$ is the first derivative of the Dirac delta function. Using the identity 
\begin{align}
\delta'[f(x)]=&\frac{\left(\delta[f(x)]\right)'}{f'(x)}\nonumber\\
=&\frac{\left[\delta(x-x_0)/f'(x_0)\right]'}{f'(x)}\nonumber\\
=&\frac{\delta'(x-x_0)}{f'(x_0)f'(x)},
\label{eq: delta function's identity}
\end{align}
where $x_0$ is the zero point of function $f(x)$, we have
\begin{align}
&\,\frac{\partial\mathrm{Im}\Sigma^{11(2)}_{00}(p)}{\partial p_0}\Bigg|_{p_0=\omega_{0,\p}}\nonumber\\
\propto&\, \lim_{\eps{\K}\to 0}\frac{1}{\left[\frac{2\eps{\K}+2(c_0+c_1)n_0}{\sqrt{\eps{\K}[\eps{\K}+2(c_0+c_1)n_0]}}+1+\frac{\sqrt{\eps{\p}}\cos\theta}{\sqrt{\eps{\K}}}\right]}\nonumber\\
&\times \int\limits_{0}^{\infty} \text{d}\eps{\K} \int\limits_{-1}^{1} \text{d}(\cos\theta) \frac{(-1)(\eps{\K})^{3/2}}{\hbar\omega_{1,\K}} \frac{1}{\left[\frac{2\eps{\K}+2(c_0+c_1)n_0}{\sqrt{\eps{\K}[\eps{\K}+2(c_0+c_1)n_0]}}+1+\frac{\sqrt{\eps{\p}}\cos\theta}{\sqrt{\eps{\K}}}\right]}\delta'(\eps{\K})\nonumber\\
=&\,\lim_{\eps{\K}\to 0}\frac{1}{\left[\frac{2\eps{\K}+2(c_0+c_1)n_0}{\sqrt{\eps{\K}[\eps{\K}+2(c_0+c_1)n_0]}}+1+\frac{\sqrt{\eps{\p}}\cos\theta}{\sqrt{\eps{\K}}}\right]}\nonumber\\
&\times (-1) \int\limits_{0}^{\infty} \text{d}\eps{\K}\int\limits_{-1}^{1} \frac{\partial}{\partial \eps{\K}}\Bigg[\frac{(-1)(\eps{\K})^{3/2}}{\hbar\omega_{1,\K}} \frac{1}{\left[\frac{2\eps{\K}+2(c_0+c_1)n_0}{\sqrt{\eps{\K}[\eps{\K}+2(c_0+c_1)n_0]}}+1+\frac{\sqrt{\eps{\p}}\cos\theta}{\sqrt{\eps{\K}}}\right]}\Bigg]\delta(\eps{\K})\nonumber\\
\propto&\, \lim_{\eps{\K}\to 0} \Bigg\{\frac{1}{\left[\frac{2\eps{\K}+2(c_0+c_1)n_0}{\sqrt{\eps{\K}[\eps{\K}+2(c_0+c_1)n_0]}}+1+\frac{\sqrt{\eps{\p}}\cos\theta}{\sqrt{\eps{\K}}}\right]}\nonumber\\
&\times \frac{\partial}{\partial \eps{\K}}\Bigg[\frac{(\eps{\K})^{3/2}}{\hbar\omega_{1,\K}} \frac{1}{\left[\frac{2\eps{\K}+2(c_0+c_1)n_0}{\sqrt{\eps{\K}[\eps{\K}+2(c_0+c_1)n_0]}}+1+\frac{\sqrt{\eps{\p}}\cos\theta}{\sqrt{\eps{\K}}}\right]}\Bigg]\Bigg\}\nonumber\\
=&\,0.
\label{eq: ferro, d/dp0 Im Sigma11-00}
\end{align}
Here, the multiplication factor outside the integrals in Eq.~\eqref{eq: ferro, d/dp0 Im Sigma11-00} corresponds to $f'(x_0)$ in Eq.~\eqref{eq: delta function's identity}.

From Eqs.~\eqref{eq: imaginary part, second order Sigma11-00} and \eqref{eq: ferro, d/dp0 Im Sigma11-00}, we have
\begin{align}
\mathrm{Im}\Sigma^{11(2)}_{00}(p)=0+\mathcal{O}\left[(p_0-\omega_{0,\p})^2\right].
\end{align}

%###########################
\subsection{Polar phase: $\Sigma^{11(2)}_{11}(p)$}
\label{appendix: Imaginary part of self energies, subsection: Polar phase}
The imaginary part of $\Sigma^{11(2)}_{11}(p)$ is given by
\begin{align}
\mathrm{Im}\Sigma^{11(2)}_{11}(p)=&\,\frac{n_0c_0^2}{\hbar^2} \integralQ \frac{(-\pi\eps{\K})}{\hbar\omega_{0,\K}}\left[A_{1,\Q}\delta(p_0-\omega_{1,\Q}-\omega_{0,\K})+B_{1,\Q}\delta(p_0+\omega_{1,\Q}+\omega_{0,\K})\right],
\end{align}
\begin{align}
&\,\mathrm{Im}\Sigma^{11(2)}_{11}(p)\Bigg|_{p_0=\omega_{1,\p}}\nonumber\\
=&\,\frac{n_0c_0^2}{\hbar^2} \integralQ \frac{(-\pi\eps{\K})}{\hbar\omega_{0,\K}}\left[A_{1,\Q}\delta(\omega_{1,\p}-\omega_{1,\Q}-\omega_{0,\K})+B_{1,\Q}\delta(\omega_{1,\p}+\omega_{1,\Q}+\omega_{0,\K})\right]\nonumber\\
=&\, \frac{n_0c_0^2}{\hbar^2}\int \frac{\text{d}^3\K}{(2\pi)^3} \frac{(-\pi\eps{\K})}{\hbar\omega_{0,\K}}\left[A_{1,\Q}\delta(\omega_{1,\p}-\omega_{1,\Q}-\omega_{0,\K})+B_{1,\Q}\delta(\omega_{1,\p}+\omega_{1,\Q}+\omega_{0,\K})\right]\nonumber\\
=&\,\frac{n_0c_0^2M^{3/2}}{\hbar^5} \int\limits_{0}^{\infty} \text{d}\eps{\K} \frac{\sqrt{\eps{\K}}}{2\sqrt{2}\pi^2} \int\limits_{-1}^{1} \text{d}(\cos\theta) \frac{(-\pi\eps{\K})}{\hbar\omega_{0,\K}}\nonumber\\
&\times \Big[A_{1,\p+\K}\delta(\omega_{1,\p}-\omega_{1,\p+\K}-\omega_{0,\K})+B_{1,\p+\K}\delta(\omega_{1,\p}+\omega_{1,\p+\K}+\omega_{0,\K})\Big]\nonumber\\
=&\,\frac{n_0c_0^2M^{3/2}}{\hbar^5} \int\limits_{0}^{\infty} \text{d}\eps{\K} \frac{\sqrt{\eps{\K}}}{2\sqrt{2}\pi^2} \int\limits_{-1}^{1} \text{d}(\cos\theta) \frac{(-\pi\eps{\K})}{\hbar\omega_{0,\K}} A_{1,\p+\K}\delta(\omega_{1,\p}-\omega_{1,\p+\K}-\omega_{0,\K}).
\label{eq: polar, Im Sigma 11,11}
\end{align}
Here, $\theta$ is the angle between $\p$ and $\K$, and in deriving the last line of Eq.~\eqref{eq: polar, Im Sigma 11,11} we used the fact that the argument of the Dirac delta function $\delta(\omega_{1,\p}+\omega_{1,\p+\K}+\omega_{0,\K})$ is always positive. We consider only the low-momentum region $\eps{\p}\ll|c_1|n_0$ and the external parameter region $q_B+2c_1n_0\sim |c_1|n_0$. For $\K$ such that $|\p+\K|>|\p|$, we have $\eps{\p+\K}>\eps{\p}$, $\omega_{1,\p+\K}>\omega_{1,\p}$, and, in turn, $\omega_{1,\p}-\omega_{1,\p+\K}-\omega_{0,\K}<0$. In contrast, for $|\p+\K|\leq|\p|$, we have $\eps{\p+\K}\leq \eps{\p}\ll|c_1|n_0$ and $|\K|\sim |\p|$, $\eps{\K}\sim\eps{\p}\ll|c_1|n_0\ll c_0n_0$. The argument of the delta function in Eq.~\eqref{eq: polar, Im Sigma 11,11} is then reduced to
\begin{align}
&\,\omega_{1,\p}-\omega_{1,\p+\K}-\omega_{0,\K}\nonumber\\
=&\,\frac{\sqrt{q_B(q_B+2c_1n_0)}}{\hbar}\left\{1+\frac{1}{2}\left(\frac{1}{q_B}+\frac{1}{q_B+2c_1n_0}\right)\eps{\p}+\mathcal{O}\left[\left(\frac{\eps{\p}}{|c_1|n}\right)^2\right]\right\}\nonumber\\
&-\frac{\sqrt{q_B(q_B+2c_1n_0)}}{\hbar}\left\{1+\frac{1}{2}\left(\frac{1}{q_B}+\frac{1}{q_B+2c_1n_0}\right)\eps{\p+\K}+\mathcal{O}\left[\left(\frac{\eps{\p}}{|c_1|n}\right)^2\right]\right\}-\omega_{0,\K}\nonumber\\
\simeq&\,\frac{q_B+c_1n_0}{\sqrt{q_B(q_B+2c_1n_0)}}\frac{(\eps{\p}-\eps{\p+\K})}{\hbar}-\omega_{0,\K}\nonumber\\
=&\,-\sqrt{\eps{\K}}\left[\frac{q_B+c_1n_0}{\sqrt{q_B(q_B+2c_1n_0)}}\left(\sqrt{\eps{\K}}+2\sqrt{\eps{\p}}\cos\theta\right)+\sqrt{\eps{\K}+2c_0n_0}\right]/\hbar.
\label{eq: polar, evaluate delta function}
\end{align}
Because $\eps{\p}\ll c_0n_0$, the expression in the square bracket of the last line of Eq.~\eqref{eq: polar, evaluate delta function} is always positive for any value of $\theta\in(0,\pi)$. Therefore, the integral in the last line of Eq.~\eqref{eq: ferro, Im Sigma 11,00} is, to within a multiplication factor, given by
\begin{align}
&\lim_{\eps{\K}\to 0}\sqrt{\eps{\K}}\frac{\eps{\K}}{\hbar\omega_{0,\K}}A_{1,\p+\K}\frac{1}{\frac{\partial(\omega_{1,\p}-\omega_{1,\p+\K}-\omega_{0,\K})}{\partial\eps{\K}}}\nonumber\\
=&\hbar\, \lim_{\eps{\K}\to 0}\sqrt{\eps{\K}}\frac{\eps{\K}}{\sqrt{\eps{\K}(\eps{\K}+2c_0n_0)}}\frac{\hbar\omega_{1,\p+\K}+\eps{\p+\K}+c_1n_0+q_B}{2\omega_{1,\p+\K}}\nonumber\\
&\times\frac{(-1)}{\left[\frac{\sqrt{q_B(q_B+2c_1n_0)}}{q_B+c_1n_0}\left(1+\frac{\sqrt{\eps{\p}}\cos\theta}{\sqrt{\eps{\K}}}\right)+\frac{\eps{\K}+c_0n_0}{\sqrt{\eps{\K}(\eps{\K}+2c_0n_0)}}\right]}\nonumber\\
=&0.
\end{align}
This implies that 
\begin{align}
\mathrm{Im}\Sigma^{11(2)}_{11}(p)\Bigg|_{p_0=\omega_{1,\p}}=0.
\label{eq: polar, imaginary part, second order Sigma11-11}
\end{align}

Similarly, we have
\begin{align}
&\,\frac{\partial\mathrm{Im}\Sigma^{11(2)}_{11}(p)}{\partial p_0}\Bigg|_{p_0=\omega_{1,\p}}\nonumber\\
=&\,\frac{n_0c_0^2}{\hbar^2} \integralQ \frac{(-\pi\eps{\K})}{\hbar\omega_{0,\K}}A_{1,\Q}\,\delta'(\omega_{1,\p}-\omega_{1,\Q}-\omega_{0,\K})\nonumber\\
=&\,\frac{n_0c_0^2}{\hbar^2}\int \frac{\text{d}^3\K}{(2\pi)^3} \frac{(-\pi\eps{\K})}{\hbar\omega_{0,\K}}A_{1,\p+\K}\,\delta'(\omega_{1,\p}-\omega_{1,\p+\K}-\omega_{0,\K})\nonumber\\
=&\,\frac{n_0c_0^2M^{3/2}}{\hbar^5} \int\limits_{0}^{\infty} \text{d}\eps{\K} \frac{\sqrt{\eps{\K}}}{2\sqrt{2}\pi^2} \int\limits_{-1}^{1} \text{d}(\cos\theta) \frac{(-\pi\eps{\K})}{\hbar\omega_{0,\K}}A_{1,\p+\K}\,\delta'(\omega_{1,\p}-\omega_{1,\p+\K}-\omega_{0,\K}).
\end{align}
Using the identity \eqref{eq: delta function's identity}, we have
\begin{align}
&\,\frac{\partial\mathrm{Im}\Sigma^{11(2)}_{11}(p)}{\partial p_0}\Bigg|_{p_0=\omega_{1,\p}}\nonumber\\
\propto&\, \lim_{\eps{\K}\to 0} \Bigg\{\frac{1}{\left[\frac{\sqrt{q_B(q_B+2c_1n_0)}}{q_B+c_1n_0}\left(1+\frac{\sqrt{\eps{\p}}\cos\theta}{\sqrt{\eps{\K}}}\right)+\frac{\eps{\K}+c_0n_0}{\sqrt{\eps{\K}(\eps{\K}+2c_0n_0)}}\right]}\nonumber\\
&\times \frac{\partial}{\partial \eps{\K}}\Bigg[\frac{(\eps{\K})^{3/2}}{\hbar\omega_{1,\K}}A_{1,\p+\K} \frac{1}{\left[\frac{\sqrt{q_B(q_B+2c_1n_0)}}{q_B+c_1n_0}\left(1+\frac{\sqrt{\eps{\p}}\cos\theta}{\sqrt{\eps{\K}}}\right)+\frac{\eps{\K}+c_0n_0}{\sqrt{\eps{\K}(\eps{\K}+2c_0n_0)}}\right]}\Bigg]\Bigg\}\nonumber\\
=&\,0.
\label{eq: polar, d/dp0 Im Sigma11-11}
\end{align}

From Eqs.~\eqref{eq: polar, imaginary part, second order Sigma11-11} and \eqref{eq: polar, d/dp0 Im Sigma11-11}, we obtain
\begin{align}
\mathrm{Im}\Sigma^{11(2)}_{11}(p)=0+\mathcal{O}\left[(p_0-\omega_{1,\p})^2\right].
\end{align}

%****************************
\section{Real parts of self-energies}
\label{appendix: Real part of self energies}
%############################
\subsection{Ferrromagnetic phase: $\Sigma^{11(2)}_{00}(p)$}
\label{appendix: Real part of self energies, subsection: Ferrromagnetic phase}
By limiting our consideration to a small external magnetic field $q_B\sim |c_1|n\ll c_0n$, and ignoring any difference of the order smaller than $|c_0|n\sqrt{na^3}$, which is justified at the second-order approximation, the real part of $\Sigma^{11(2)}_{00}(p)$ given by Eq.~\eqref{eq: ferro, second-order diagrams' contribution, Sigma 11-00} is reduced to
\begin{align}
\mathrm{Re}\,\Sigma^{11(2)}_{00}(p)=&\,\frac{n_0c_0^2}{\hbar} \integralQ \Bigg(-\mathcal{P}\frac{1}{\eps{\p}-\eps{\Q}-\eps{\K}}+\left(A_{1,\K}+B_{1,\K}-2C_{1,\K}\right)\nonumber\\
&\times \mathcal{P}\frac{1}{\hbar\left(p_0-\omega_{0,\Q}-\omega_{1,\K}\right)}\Bigg)+\frac{c_0}{\hbar}\integralQ B_{1,\Q} \nonumber\\
=&\,-\frac{n_0c_0^2}{\hbar} \integralQ \Bigg[\mathcal{P}\frac{1}{\eps{\p}-\eps{\Q}-\eps{\K}}+\frac{1}{4}\left(\frac{1}{\hbar\omega_{1,\Q}}+\frac{1}{\hbar\omega_{1,\K}}\right)\Bigg]\nonumber\\
&+\frac{n_0c_0^2}{\hbar} \integralQ \Bigg[\frac{\eps{\K}}{\hbar\omega_{1,\K}}\mathcal{P}\frac{1}{\hbar\left(p_0-\omega_{0,\Q}-\omega_{1,\K}\right)}\nonumber\\
&+\frac{1}{4}\left(\frac{1}{\hbar\omega_{1,\Q}}+\frac{1}{\hbar\omega_{1,\K}}\right)\Bigg]+\frac{1}{3\pi^2}\frac{n_0c_0}{\hbar}\sqrt{n_0\tilde{a}^3},
\label{eq: ferro, Re second-order Sigma 11-00}
\end{align}
where $\mathcal{P}$ denotes the principle value, and $\tilde{a}$ is defined by Eq.~\eqref{eq: define scattering length a}. Here, we used $A_{1,\K}+B_{1,\K}-2C_{1,\K}=\eps{\K}/(\hbar\omega_{1,\K})$ and 
\begin{align}
\frac{c_0}{\hbar} \integralQ B_{1,\Q}=&\,\frac{c_0}{\pi^2\sqrt{2}\hbar}\left[\frac{(c_0+c_1)n_0M}{\hbar^2}\right]^{3/2}\int\limits_0^\infty  \text{d}x\, \frac{x+1-\sqrt{x(x+2)}}{2\sqrt{x+2}}\nonumber\\
=&\,\frac{c_0}{\pi^2\sqrt{2}\hbar}\left[\frac{(c_0+c_1)n_0M}{\hbar^2}\right]^{3/2}\frac{\sqrt{2}}{3} \nonumber\\
\simeq&\,\frac{1}{3\pi^2}\frac{c_0n_0}{\hbar}\sqrt{n_0\tilde{a}^3},
\end{align}
where we have ignored terms that contain the factor $|c_1|/c_0\ll 1$.

First, we calculate the third line of Eq.~\eqref{eq: ferro, Re second-order Sigma 11-00}. Putting $\Q\equiv\p/2+\Q'$, we have
\begin{subequations}
\label{eq: variable transformation q=p/2+q'}
\begin{align}
\K =&\, \Q-\p=\,\Q'-\p/2,\\
\epsilon^0_\p-\epsilon^0_\Q-\epsilon^0_\K=&\,2(\epsilon^0_{\p/2}-\epsilon^0_{\Q'}),\\
\integralQ \frac{1}{\hbar\omega_{1,\K}}=&\,\int \frac{\text{d}^3\K}{(2\pi)^3} \frac{1}{\hbar\omega_{1,\K}}=\integralQ \frac{1}{\hbar\omega_{1,\Q}},\\
\integralQ \mathcal{P}\frac{1}{\epsilon^0_\p-\epsilon^0_\Q-\epsilon^0_\K}=&\,\frac{1}{2}\int \frac{d^3\Q}{(2\pi)^3}\mathcal{P}\frac{1}{\epsilon^0_{\p/2}-\epsilon^0_{\Q'}}=\,\frac{1}{2}\int \frac{d^3\Q'}{(2\pi)^3}\mathcal{P}\frac{1}{\epsilon^0_{\p/2}-\epsilon^0_{\Q'}}\nonumber\\
=&\,\frac{1}{2}\int \frac{d^3\Q}{(2\pi)^3}\mathcal{P}\frac{1}{\epsilon^0_{\p/2}-\epsilon^0_{\Q}}.\nonumber\\
\end{align}
\end{subequations}
The third line of Eq~\eqref{eq: ferro, Re second-order Sigma 11-00} then becomes
\begin{align}
&\,-\frac{n_0c_0^2}{\hbar} \integralQ \Bigg[\mathcal{P}\frac{1}{\eps{\p}-\eps{\Q}-\eps{\K}}+\frac{1}{4}\left(\frac{1}{\hbar\omega_{1,\Q}}+\frac{1}{\hbar\omega_{1,\K}}\right)\Bigg]\nonumber\\
=&\,-\frac{n_0c_0^2}{2\hbar} \integralQ \left(\mathcal{P}\frac{1}{\epsilon^0_{\p/2}-\epsilon^0_\Q}+\frac{1}{\hbar\omega_{1,\Q}}\right).
\end{align}
By taking a transformation of variables $|\Q| \to \eps{\Q}$, and using the indefinite integration
\begin{align}
\int \text{d}x\, \frac{\sqrt{x}}{a-x}=-2\sqrt{x}-\sqrt{a}\ln \left|\frac{\sqrt{x}-\sqrt{a}}{\sqrt{x}+\sqrt{a}}\right|,
\end{align}
together with the definition of the principle value
\begin{align}
\int \text{d}x\, \mathcal{P}\frac{\sqrt{x}}{a-x}=\lim_{\delta\to 0} \left[\int\limits^{a-\delta}+\int\limits_{a+\delta}\text{d}x\, \frac{\sqrt{x}}{a-x}\right],
\end{align}
we obtain 
\begin{align}
-\frac{n_0c_0^2}{2\hbar} \integralQ \left(\mathcal{P}\frac{1}{\epsilon^0_{\p/2}-\epsilon^0_\Q}+\frac{1}{\hbar\omega_{1,\Q}}\right)=\frac{1}{\pi^2}\frac{n_0c_0}{\hbar}\sqrt{n_0\tilde{a}^3}.
\end{align}

For the remaining term in Eq.~\eqref{eq: ferro, Re second-order Sigma 11-00}, its value in the low-momentum region $\eps{\p}\ll c_0n$ is obtained analytically by making Taylor expansions around $p_0=\omega_{0,\p}$ and $\p={\bf 0}$ as described in Eqs.~\eqref{eq: a Taylor expansion around p0=omega0p} and \eqref{eq: Taylor expansion in low momentum region}. The expansion coefficients are then calculated as follows:
\begin{align}
\mathrm{Re}\Sigma^{11(2)}_{00}(p_0=\omega_{0,\p})=&\,\frac{4}{3\pi^2}\frac{n_0c_0}{\hbar}\sqrt{n_0\tilde{a}^3}+\frac{n_0c_0^2}{\hbar^2} \integralQ \Bigg[\frac{\eps{\K}}{\hbar\omega_{1,\K}}\mathcal{P}\frac{1}{\omega_{0,\p}-\omega_{0,\Q}-\omega_{1,\K}}\nonumber\\
&+\frac{1}{4}\left(\frac{1}{\omega_{1,\Q}}+\frac{1}{\omega_{1,\K}}\right)\Bigg],
\end{align}
\begin{align}
&\,\mathrm{Re}\Sigma^{11(2)}_{00}(p_0=\omega_{0,\p})\Big|_{\p={\bf 0}}\nonumber\\
=&\,\frac{4}{3\pi^2}\frac{n_0c_0}{\hbar}\sqrt{n_0\tilde{a}^3}+\frac{n_0c_0^2}{\hbar}\integralQ \Bigg[\frac{\eps{\Q}}{\hbar\omega_{1,\Q}}\frac{1}{(-\eps{\Q}-\hbar\omega_{1,\Q})}+\frac{1}{2\hbar\omega_{1,\Q}}\Bigg]\nonumber\\
=&\,\frac{4}{3\pi^2}\frac{n_0c_0}{\hbar}\sqrt{n_0\tilde{a}^3}+\frac{n_0c_0^2M^{3/2}}{\pi^2\sqrt{2}\hbar^4}\int\limits_{0}^{\infty} d\eps{\Q}\sqrt{\eps{\Q}}\left[\frac{-\eps{\Q}}{\hbar\omega_{1,\Q}(\eps{\Q}+\hbar\omega_{1,\Q})}+\frac{1}{2\hbar\omega_{1,\Q}}\right]\nonumber\\
=&\,\frac{4}{3\pi^2}\frac{n_0c_0}{\hbar}\sqrt{n_0\tilde{a}^3}+\frac{n_0c_0^2\sqrt{(c_0+c_1)n_0}M^{3/2}}{\pi^2\sqrt{2}\hbar^4}\nonumber\\
&\times\int\limits_{0}^{\infty} dx\sqrt{x}\Bigg[\frac{-x}{\sqrt{x(x+2)}(x+\sqrt{x(x+2)})}+\frac{1}{2\sqrt{x(x+2)}}\Bigg]\nonumber\\
=&\,\frac{4}{3\pi^2}\frac{n_0c_0}{\hbar}\sqrt{n_0\tilde{a}^3}+\frac{n_0c_0^2\sqrt{(c_0+c_1)n_0}M^{3/2}}{\pi^2\sqrt{2}\hbar^4}\frac{\sqrt{2}}{3}\nonumber\\
\simeq&\,\frac{5}{3\pi^2}\frac{n_0c_0}{\hbar}\sqrt{n_0\tilde{a}^3},
\end{align}
\begin{align}
\frac{\partial \mathrm{Re}\Sigma^{11(2)}_{00}(p_0=\omega_{0,\p})}{\partial \omega_{1,\p}}\Bigg|_{\p={\bf 0}}=\,0,
\end{align}
\begin{align}
\frac{\partial^2 \mathrm{Re}\Sigma^{11(2)}_{00}(p_0=\omega_{0,\p})}{\partial (\omega_{1,\p})^2}\Bigg|_{\p={\bf 0}}=&\,-\frac{49n_0c_0^2M^{3/2}}{360\pi^2[(c_0+c_1)n_0]^{3/2}\hbar^2}\nonumber\\
\simeq&\,-\frac{49}{360\pi^2}\sqrt{n_0\tilde{a}^3}\frac{\hbar}{n_0c_0},
\end{align}
\begin{align}
\frac{\partial \mathrm{Re}\Sigma^{11(2)}_{00}(p)}{\partial p_0}\Bigg|_{p_0=\omega_{0,\p}}=\,\frac{n_0c_0^2}{\hbar^2}\integralQ \frac{(-\eps{\K})}{\hbar\omega_{1,\K}}\mathcal{P}\frac{1}{(\omega_{0,\p}-\omega_{0,\Q}-\omega_{1,\K})^2},
\end{align}
\begin{align}
\frac{\partial \mathrm{Re}\Sigma^{11(2)}_{00}(p)}{\partial p_0}\Bigg|_{p_0=\omega_{0,\p},\p={\bf 0}}=&\,n_0c_0^2\integralQ \frac{(-\eps{\Q})}{\hbar\omega_{1,\Q}}\mathcal{P}\frac{1}{(-\eps{\Q}-\hbar\omega_{1,\Q})^2}\nonumber\\
=&\,\frac{n_0c_0^2M^{3/2}}{\sqrt{2}\pi^2\hbar^3}\int\limits_{0}^{\infty} d\eps{\Q}\sqrt{\eps{\Q}}\frac{(-\eps{\Q})}{\hbar\omega_{1,\Q}}\frac{1}{(-\eps{\Q}-\hbar\omega_{1,\Q})^2}\nonumber\\
=&\,-\frac{1}{3\pi^2}\frac{n_0c_0^2M^{3/2}}{\sqrt{(c_0+c_1)n_0}\hbar^3}\nonumber\\
=&\,-\frac{1}{3\pi^2}\sqrt{n_0\tilde{a}^3},
\end{align}
\begin{align}
\frac{\partial}{\partial \omega_{1,\p}}\Bigg(\frac{\partial \mathrm{Re}\Sigma^{11(2)}_{00}(p)}{\partial p_0}\Bigg|_{p_0=\omega_{0,\p}}\Bigg)\Bigg|_{\p={\bf 0}}=\,0,
\end{align}
\begin{align}
\frac{\partial^2}{\partial (\omega_{1,\p})^2}\Bigg(\frac{\partial \mathrm{Re}\Sigma^{11(2)}_{00}(p)}{\partial p_0}\Bigg|_{p_0=\omega_{0,\p}}\Bigg)\Bigg|_{\p={\bf 0}}=&\,-\frac{13n_0c_0^2M^{3/2}}{60\pi^2[(c_0+c_1)n_0]^{5/2}\hbar}\nonumber\\
\simeq&\,-\frac{13}{60\pi^2}\sqrt{n_0\tilde{a}^3}\frac{\hbar^2}{(n_0c_0)^2}.
\end{align}

%############################
\subsection{Polar phase: $\Sigma^{11(2)}_{11}(p)$}
\label{appendix: Real part of self energies, subsection: Polar phase}
Neglecting terms of the order smaller than $c_0n_0\sqrt{n_0a^3}$, which is justified at the second-order approximation, the real part of $\Sigma^{11(2)}_{11}(p)$ is then reduced to
\begin{align}
\mathrm{Re}\Sigma^{11(2)}_{11}(p)=&\,\frac{n_0c_0^2}{\hbar}\integralQ \Bigg[-\mathcal{P}\frac{1}{\eps{\p}-\eps{\Q}-\eps{\K}}+(A_{0,\K}+ B_{0,\K}-2C_{0,\K})\nonumber\\
&\times\Bigg(\mathcal{P}\frac{A_{1,\Q}}{\hbar\left(p_0-\omega_{1,\Q}-\omega_{0,\K}\right)}-\mathcal{P}\frac{B_{1,\Q}}{\hbar\left(p_0+\omega_{1,\Q}+\omega_{0,\K}\right)}\Bigg)\Bigg]\nonumber\\
&+\frac{c_0}{\hbar}\integralQ \left( 3B_{1,\Q}+B_{0,\Q} \right)\nonumber\\
=&\,-\frac{n_0c_0^2}{\hbar}\integralQ \Bigg[\mathcal{P}\frac{1}{\eps{\p}-\eps{\Q}-\eps{\K}}+\frac{1}{4}\left(\frac{1}{\hbar\omega_{1,\Q}}+\frac{1}{\hbar\omega_{0,\K}}\right)\Bigg]+\frac{n_0c_0^2}{\hbar}\integralQ\nonumber\\
&\times \Bigg[(A_{0,\K}+ B_{0,\K}-2C_{0,\K})\Bigg(\mathcal{P}\frac{A_{1,\Q}}{\hbar\left(p_0-\omega_{1,\Q}-\omega_{0,\K}\right)}-\mathcal{P}\frac{B_{1,\Q}}{\hbar\left(p_0+\omega_{1,\Q}+\omega_{0,\K}\right)}\Bigg)\nonumber\\
&+\frac{1}{4}\left(\frac{1}{\hbar\omega_{1,\Q}}+\frac{1}{\hbar\omega_{0,\K}}\right)\Bigg]+\frac{1}{3\pi^2}\frac{c_0n_0}{\hbar}\sqrt{n_0\tilde{a}^3}.
\label{eq: polar, Real part of Sigma11(2)}
\end{align}
Here, we used 
\begin{align}
\frac{c_0}{\hbar} \integralQ B_{0,\Q}=&\,\frac{c_0}{\pi^2\sqrt{2}}\frac{(c_0n_0)^{3/2}M^{3/2}}{\hbar^4}\int\limits_0^\infty \sqrt{x} \text{d}x\, \frac{x+1-\sqrt{x(x+2)}}{2\sqrt{x+2}}\nonumber\\
=&\,\frac{c_0}{\pi^2\sqrt{2}}\frac{(c_0n_0)^{3/2}M^{3/2}}{\hbar^4}\frac{\sqrt{2}}{3} \nonumber\\
=&\,\frac{1}{3\pi^2}\frac{c_0n_0}{\hbar}\sqrt{n_0\tilde{a}^3},\\
\frac{c_0}{\hbar} \integralQ B_{1,\Q} \sim&\,\frac{M^{3/2}}{\hbar^4} \int\limits_{0}^\infty \sqrt{\eps{\Q}}\text{d}\eps{\Q} \frac{-E^1_\Q+\eps{\Q}+c_1n_0+q_B}{2E^1_\Q} \nonumber\\
\sim& \,\frac{c_0(|c_1|n_0)^{3/2}M^{3/2}}{\hbar^4}\nonumber\\
\ll& \frac{c_0n_0}{\hbar}\sqrt{n_0\tilde{a}^3}.
\label{eq: evaluate integral with B1q}
\end{align}

First, we consider the first term in Eq.~\eqref{eq: polar, Real part of Sigma11(2)}:
\begin{align}
&\,-\frac{n_0c_0^2}{\hbar}\integralQ \Bigg[\mathcal{P}\frac{1}{\eps{\p}-\eps{\Q}-\eps{\K}}+\frac{1}{4}\left(\frac{1}{\hbar\omega_{1,\Q}}+\frac{1}{\hbar\omega_{0,\K}}\right)\Bigg]\nonumber\\
=&\,-\frac{n_0c_0^2}{2\hbar} \integralQ \Bigg[\Bigg(\mathcal{P}\frac{1}{\eps{\p}-\eps{\Q}-\eps{\K}}+\frac{1}{2\hbar\omega_{1,\Q}}\Bigg)+\Bigg(\mathcal{P}\frac{1}{\eps{\p}-\eps{\Q}-\eps{\K}}+\frac{1}{2\hbar\omega_{0,\Q}}\Bigg)\Bigg]\nonumber\\
=&\,-\frac{n_0c_0^2}{4\hbar} \Bigg[ \integralQ \Bigg(\mathcal{P}\frac{1}{\eps{\p/2}-\eps{\Q}}+\frac{1}{\hbar\omega_{1,\Q}}\Bigg)+\integralQ \Bigg(\mathcal{P}\frac{1}{\eps{\p/2}-\eps{\Q}}+\frac{1}{\hbar\omega_{0,\Q}}\Bigg)\Bigg]\nonumber\\
\simeq& -\frac{n_0c_0^2M^{3/2}}{4\pi^2\sqrt{2}\hbar^4}\int \text{d}\eps{\Q}\,\sqrt{\eps{\Q}}\Bigg(\mathcal{P}\frac{1}{\eps{\p/2}-\eps{\Q}}+\frac{1}{E^0_\Q}\Bigg)\nonumber\\
=&\,\frac{1}{2\pi^2}\frac{n_0c_0}{\hbar}\sqrt{n_0\tilde{a}^3}.
\label{eq: polar, first term of Re Sigma 11}
\end{align}
Here, as moving from the second line to the third line in Eq.~\eqref{eq: polar, first term of Re Sigma 11} we used a transformation of variables [see Eq.~\eqref{eq: variable transformation q=p/2+q'}]. Furthermore, in the third line the main contributions to the first and the second integrals arise from $\eps{\Q}\sim |c_1|n$ and $\eps{\Q}\sim c_0n$, respectively, which results in the fact that the first integral is smaller than the second one by a factor of the order of $\sqrt{|c_1|/c_0} \ll 1$, and thus, the second integral was neglected. The integral in the next to the last line was directly calculated by using 
\begin{align}
\int\limits_{0}^{\infty} \sqrt{x}\text{d}x \Bigg[\mathcal{P}\frac{1}{a-x}+\frac{1}{\sqrt{x(x+b)}}\Bigg]=&-2\sqrt{x_\infty}+\sqrt{a}\ln\Bigg|\frac{\sqrt{a_+}-\sqrt{a}}{\sqrt{a_+}+\sqrt{a}}\Bigg|+2\sqrt{a_+}-2\sqrt{a_-}\nonumber\\
&-\sqrt{a}\ln\Bigg|\frac{\sqrt{a_-}-\sqrt{a}}{\sqrt{a_-}+\sqrt{a}}\Bigg|+2\sqrt{x_\infty+b}-2\sqrt{b}\nonumber\\
&=\,-2\sqrt{b},
\end{align}
where $x_\infty \equiv \lim_{x\to \infty}, a_{\pm} \equiv a \pm \delta (\delta \to +0)$.

For the remaining term in Eq.~\eqref{eq: polar, Real part of Sigma11(2)}, we can obtain an analytic result for the low-momentum region $\eps{\p}\ll |c_1|n \ll c_0n$ by making Taylor expansions around $p_0=\omega_{1,\p}$ and $\p={\bf 0}$ as described in Eqs.~\eqref{eq: polar, Taylor expansion around p0=omega 1,p} and \eqref{eq: polar, Taylor expansion around p=0}. The coefficients of the expansions are then calculated as follows:
\begin{align}
\mathrm{Re}\Sigma^{11(2)}_{11}(p)\Bigg|_{p_0=\omega_{1,\p}}=&\,\frac{5}{6\pi^2}\frac{n_0c_0}{\hbar}\sqrt{n_0\tilde{a}^3}+\frac{n_0c_0^2}{\hbar^2} \integralQ \Bigg[\frac{\eps{\K}}{\hbar\omega_{0,\K}}\Bigg(\mathcal{P}\frac{A_{1,\Q}}{\omega_{1,\p}-\omega_{1,\Q}-\omega_{0,\K}}\nonumber\\
&-\mathcal{P}\frac{B_{1,\Q}}{\omega_{1,\p}+\omega_{1,\Q}+\omega_{0,\K}}\Bigg)+ \frac{1}{4}\left(\frac{1}{\omega_{1,\Q}}+\frac{1}{\omega_{0,\K}}\right)\Bigg],
\end{align}
\begin{align}
&\,\mathrm{Re}\Sigma^{11(2)}_{11}(p_0=\omega_{1,\p})\Big|_{\p={\bf 0}}\nonumber\\
=&\,\frac{5}{6\pi^2}\frac{n_0c_0}{\hbar}\sqrt{n_0\tilde{a}^3}+\frac{n_0c_0^2}{\hbar^2} \integralQ \Bigg[\frac{\eps{\Q}}{\hbar\omega_{0,\Q}}\Bigg(\frac{A_{1,\Q}}{\omega_{1,\p={\bf 0}}-\omega_{1,\Q}-\omega_{0,\Q}}\nonumber\\
&-\frac{B_{1,\Q}}{\omega_{1,\p=0}+\omega_{1,\Q}+\omega_{0,\Q}}\Bigg)+ \frac{1}{4}\left(\frac{1}{\omega_{1,\Q}}+\frac{1}{\omega_{0,\K}}\right)\Bigg] \nonumber\\
\simeq&\,\frac{5}{6\pi^2}\frac{n_0c_0}{\hbar}\sqrt{n_0\tilde{a}^3}+\frac{n_0c_0^2}{\hbar^2} \integralQ \Bigg[-\frac{\eps{\Q}}{\omega_{0,\Q}(\eps{\Q}+\hbar\omega_{0,\Q})}+\frac{1}{4}\left(\frac{1}{\omega_{1,\Q}}+\frac{1}{\omega_{0,\K}}\right)\Bigg]\nonumber\\
=&\,\frac{5}{6\pi^2}\frac{n_0c_0}{\hbar}\sqrt{n_0\tilde{a}^3}+\frac{n_0c_0^2M^{3/2}}{\pi^2\sqrt{2}\hbar^5}\int\limits_0^\infty \sqrt{\eps{\Q}}\text{d}\eps{\Q}\Bigg[-\frac{\eps{\Q}}{\omega_{0,\Q}(\eps{\Q}+\hbar\omega_{0,\Q})}+\frac{1}{4}\left(\frac{1}{\omega_{1,\Q}}+\frac{1}{\omega_{0,\K}}\right)\Bigg]\nonumber\\
=&\,\frac{5}{6\pi^2}\frac{n_0c_0}{\hbar}\sqrt{n_0\tilde{a}^3}+\frac{n_0c_0^2(n_0c_0)^{1/2}M^{3/2}}{\pi^2\sqrt{2}\hbar^4}\int\limits_0^\infty \sqrt{x} \text{d}x \Bigg[-\frac{x}{\sqrt{x(x+2)}(x+\sqrt{x(x+2)})}\nonumber\\
&+\frac{1}{4}\left(\frac{1}{x}+\frac{1}{\sqrt{x(x+2)}}\right)\Bigg]\nonumber\\
=&\,\frac{5}{3\pi^2}\frac{n_0c_0}{\hbar}\sqrt{n_0\tilde{a}^3}.
\label{eq: polar, real Sigma11-11 at p0=E1p, p=0}
\end{align}
Here, we used the fact that the main contribution to the integral in the second line of Eq.~\eqref{eq: polar, real Sigma11-11 at p0=E1p, p=0} comes from $\eps{\Q}\sim c_0n_0\gg c_1n_0$, and we can approximate $\hbar\omega_{1,\Q}\simeq \eps{\Q}, \hbar \omega_{1,\p=0}\simeq 0, A_{1,\Q}\simeq 1, B_{1,\Q} \simeq 0$. This is because the integral converges at both the upper limit $\eps{\Q} \gg c_0n_0$, and the lower limit $\eps{\Q}\to 0$. Next, we have
\begin{align}
\frac{\partial \mathrm{Re}\Sigma^{11(2)}_{11}(p_0=\omega_{1,\p})}{\partial \omega_{0,\p}}\Bigg|_{\p={\bf 0}}=\,0,
\end{align}
\begin{align}
\frac{\partial^2 \mathrm{Re}\Sigma^{11(2)}_{11}(p_0=\omega_{1,\p})}{\partial (\omega_{0,\p})^2}\Bigg|_{\p={\bf 0}}=&\,\left(- \,\frac{1}{3\pi^2}\frac{q_B+c_1n_0}{\sqrt{q_B(q_B+2c_1n_0)}}+\frac{71}{360\pi^2}\right)\sqrt{n_0\tilde{a}^3}\frac{\hbar}{n_0c_0},
\end{align}
where we used the result of the following integral:
\begin{align}
\int\limits_0^\infty \text{d}x\, \frac{13x^2+11x^3-\sqrt{x^3(2+x)}+29\sqrt{x^5(2+x)}+8\sqrt{x^7(2+x)}}{3(2+x)^{5/2}\left(x+\sqrt{x(2+x)}\right)^3}=\,\frac{71}{90\sqrt{2}}.
\end{align}
Therefore, the real part of the self-energy $\Sigma^{11(2)}_{11}(p_0=\omega_{1,\p})$ can be written as
\begin{align}
\mathrm{Re}\Sigma^{11(2)}_{11}(p_0=\omega_{1,\p})\simeq&\, \frac{5}{3\pi^2}\frac{n_0c_0}{\hbar}\sqrt{n_0\tilde{a}^3}+\left(-\,\frac{1}{6\pi^2}\frac{q_B+c_1n_0}{\sqrt{q_B(q_B+2c_1n_0)}}+\frac{71}{720\pi^2}\right)\nonumber\\
&\times\sqrt{n_0\tilde{a}^3}\frac{\hbar\omega_{0,\p}^2}{n_0c_0}.
\end{align}
In a similar manner, the real part of the self-energy $\Sigma^{11(2)}_{11}(-p)|_{p_0=\omega_{1,\p}}$ can be calculated by replacing $\K\equiv \Q-\p$ with $\K\equiv \Q+\p$, and we obtain
\begin{align}
\mathrm{Re}\Sigma^{11(2)}_{11}(-p)|_{p_0=\omega_{1,\p}}\simeq&\, \frac{5}{3\pi^2}\frac{n_0c_0}{\hbar}\sqrt{n_0\tilde{a}^3}+\left(\frac{1}{6\pi^2}\frac{q_B+c_1n_0}{\sqrt{q_B(q_B+2c_1n_0)}}+\frac{71}{720\pi^2}\right)\nonumber\\
&\times\sqrt{n_0\tilde{a}^3}\frac{\hbar\omega_{0,\p}^2}{n_0c_0}.
\end{align}

Next, we have
\begin{align}
\frac{\partial \mathrm{Re}\Sigma^{11(2)}_{11}(p)}{\partial p_0}\Bigg|_{p_0=\omega_{1,\p},\p={\bf 0}}=&\,-\frac{n_0c_0^2}{\hbar^2}\integralQ \frac{\eps{\Q}}{\hbar\omega_{0,\Q}}\Bigg[\frac{A_{1,\Q}}{(\omega_{1,\p={\bf 0}}-\omega_{1,\Q}-\omega_{0,\Q})^2}\nonumber\\
&-\frac{B_{1,\Q}}{(\omega_{1,\p={\bf 0}}+\omega_{1,\Q}+\omega_{0,\Q})^2}\Bigg].
\label{eq: polar, real d/dp0 of Sigma11-11 at p0=E1p, p=0}
\end{align}
As above, the main contribution to the integral in Eq.~\eqref{eq: polar, real d/dp0 of Sigma11-11 at p0=E1p, p=0} arises from $\eps{\Q}\sim c_0n_0 \gg c_1n_0$. We can then approximate $\hbar\omega_{1,\Q}\simeq \eps{\Q}, \hbar \omega_{1,\p=0}\simeq 0, A_{1,\Q}\simeq 1, B_{1,\Q} \simeq 0$, and the integral can be evaluated straightforwardly as
\begin{align}
\frac{\partial \mathrm{Re}\Sigma^{11(2)}_{11}(p)}{\partial p_0}\Bigg|_{p_0=\omega_{1,\p},\p={\bf 0}}\simeq&\,-\frac{n_0c_0^2M^{3/2}}{\pi^2\sqrt{2}\hbar^4}\int\limits_0^\infty\sqrt{\eps{\Q}}\,\text{d}\eps{\Q}\,\frac{\eps{\Q}}{\omega_{0,\Q}(\eps{\Q}+\hbar\omega_{0,\Q})^2}\nonumber\\
=&\,-\frac{n_0c_0^2M^{3/2}}{\pi^2\sqrt{2}\hbar^3}(n_0c_0)^{-1/2}\int\limits_0^\infty \sqrt{x}dx \frac{x}{\sqrt{x(x+2)}[x+\sqrt{x(x+2)}]^2}\nonumber\\
=&\,-\frac{1}{3\pi^2}\sqrt{n_0\tilde{a}^3}.
\end{align}
Similarly, we have
\begin{align}
\frac{\partial \mathrm{Re}\Sigma^{11(2)}_{11}(-p)}{\partial p_0}\Bigg|_{p_0=\omega_{1,\p},\p={\bf 0}}=\,\frac{1}{3\pi^2}\sqrt{n_0\tilde{a}^3}.
\end{align}
For the derivatives with respect to $\omega_{0,\p}$, we obtain
\begin{align}
\frac{\partial}{\partial \omega_{0,\p}}\Bigg(\frac{\partial \mathrm{Re}\Sigma^{11(2)}_{11}(p)}{\partial p_0}\Bigg|_{p_0=\omega_{1,\p}}\Bigg)\Bigg|_{\p={\bf 0}}=&\,0,
\end{align}
\begin{align}
\frac{\partial}{\partial \omega_{0,\p}}\Bigg(\frac{\partial \mathrm{Re}\Sigma^{11(2)}_{11}(-p)}{\partial p_0}\Bigg|_{p_0=\omega_{1,\p}}\Bigg)\Bigg|_{\p={\bf 0}}=&\,0,
\end{align}
\begin{align}
\frac{\partial^2}{\partial (\omega_{0,\p})^2}\Bigg(\frac{\partial \mathrm{Re}\Sigma^{11(2)}_{11}(p)}{\partial p_0}\Bigg|_{p_0=\omega_{1,\p}}\Bigg)\Bigg|_{\p={\bf 0}}=&\,\left(- \,\frac{1}{3\pi^2}\frac{q_B+c_1n_0}{\sqrt{q_B(q_B+2c_1n_0)}}+\frac{7}{60\pi^2}\right) \nonumber\\
&\times\sqrt{n_0\tilde{a}^3}\frac{\hbar^2}{(n_0c_0)^2},
\end{align}
\begin{align}
\frac{\partial^2}{\partial (\omega_{0,\p})^2}\Bigg(\frac{\partial \mathrm{Re}\Sigma^{11(2)}_{11}(-p)}{\partial p_0}\Bigg|_{p_0=\omega_{1,\p}}\Bigg)\Bigg|_{\p={\bf 0}}=&\,\left(\frac{1}{3\pi^2}\frac{q_B+c_1n_0}{\sqrt{q_B(q_B+2c_1n_0)}}-\frac{7}{60\pi^2}\right)\nonumber\\
&\times \sqrt{n_0\tilde{a}^3}\frac{\hbar^2}{(n_0c_0)^2}.
\end{align}
From the above results, we obtain the real parts of the self-energies $\Sigma^{11(2)}_{11}(\pm p)$ as given in Eqs.~\eqref{eq: real part, second order Sigma11-11, polar} and \eqref{eq: real part, second order Sigma22-11, polar}.

%% References


\bibliographystyle{model1a-num-names}
\bibliography{<your-bib-database>}



\end{document}